\documentclass[twocolumn]{autart}    % Enable this line and disable the 
\usepackage{graphicx}               % document whose style resembles the
\usepackage{amsmath,amsfonts, amssymb,color}
\usepackage{tikz, subcaption, cite}
\usepackage{amsmath,amsfonts, amssymb,color}
\usepackage{epsfig} 
\usepackage{psfrag}
\DeclareMathOperator{\diag}{diag}
\usepackage{tcolorbox}
\newcommand{\mc}{\mathcal}

\newcommand{\bs}{\boldsymbol}
\newcommand{\tb}{\color{blue}}
\newcommand{\tr}{\color{red}}

\usepackage{algorithm}
\usepackage{algpseudocode}
\algrenewcommand\algorithmicforall{\textbf{foreach}}
\algrenewcommand\algorithmicindent{.8em}

\begin{document}

\newtheorem{theorem}{Theorem}
\newtheorem{problem}{Problem}
\newtheorem{condition}{Condition}
\newtheorem{definition}{Definition}
\newtheorem{lemma}{Lemma}
\newtheorem{proposition}{Proposition}
\newtheorem{corollary}{Corollary}
\newtheorem{remark}{Remark}
\newtheorem{assumption}{Assumption}
\newtheorem{example}{Example}

\begin{frontmatter}

\title{A Survey of Graph-Theoretic Approaches for Analyzing the Resilience of Networked Control Systems\thanksref{footnoteinfo}} % Title, preferably not more 

\thanks[footnoteinfo]{This material is based in part upon work supported by NSF CAREER award 1653648.  Corresponding author M.~Pirani. Tel. +1 416 716 9162.}

\author[Author1]{Mohammad Pirani}\ead{mohammad.pirani@utoronto.ca},     % Add the 
\author[Paestum]{Aritra Mitra}\ead{amitra20@seas.upenn.edu}, 
\author[Rome]{Shreyas Sundaram}\ead{sundara2@purdue.edu}            % e-mail address 

\address[Author1]{Department of Electrical and Computer Engineering, University of Toronto, Toronto, ON, Canada.} 

\address[Paestum]{Department of Electrical and Systems Engineering, University of Pennsylvania, Philadelphia, PA, USA} % Please supply                                              
\address[Rome]{Elmore Family School of Electrical and Computer Engineering, Purdue University, West Lafayette, IN, USA}             % full addresses
     % here.

\begin{keyword}                           % Five to ten keywords,  
Networked Control Systems, Graph Theory, Resilient Distributed Algorithms            % chosen from the IFAC 
\end{keyword}                             % keyword list or with the 
                                          % help of the Automatica 
                                          % keyword wizard

\begin{abstract}   
As the scale of networked control systems increases and interactions between different subsystems become more sophisticated, questions of the resilience of such networks increase in importance. The need to redefine  classical system and control-theoretic notions using the language of graphs has recently started to gain attention as a fertile and important area of research. This paper presents an overview of graph-theoretic methods for analyzing the resilience of networked control systems. We discuss various distributed algorithms operating on networked systems and investigate their resilience against adversarial actions by looking at the structural properties of their underlying networks. We present graph-theoretic methods to quantify the attack impact, and reinterpret some system-theoretic notions of robustness from a graph-theoretic standpoint to mitigate the impact of the attacks. Moreover, we discuss miscellaneous problems in the security of networked control systems which use graph-theory as a tool in their analyses. We conclude by introducing some avenues for further research in this field. 
\end{abstract}

\end{frontmatter}

\section{Introduction}
\begin{figure}[t!]
\centering
\includegraphics[scale=.56]{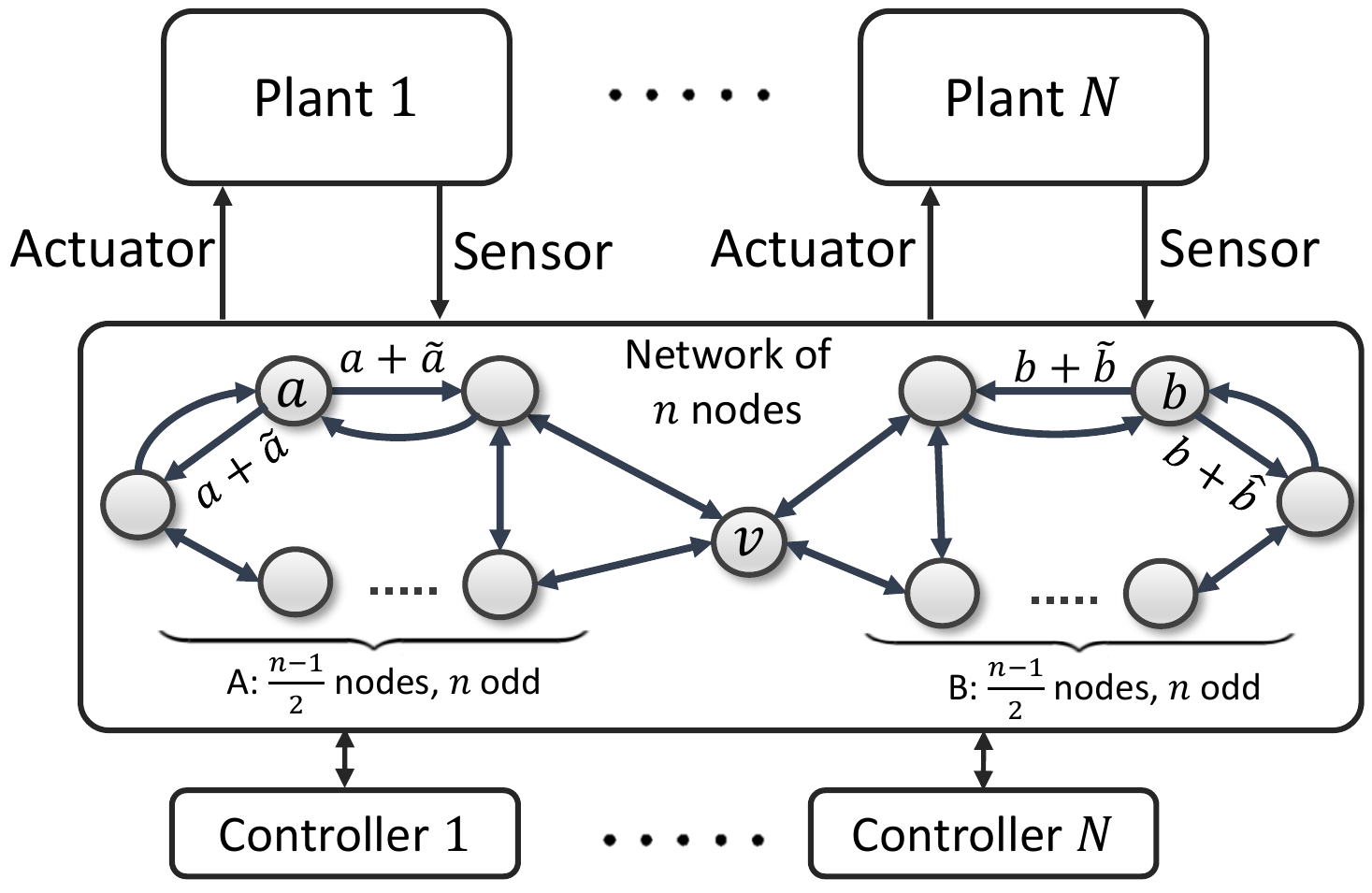}
\caption{A schematic figure of a networked  control system.}
\label{fig:byz}
\end{figure}

Networked control systems (NCS) are spatially
distributed control systems wherein the control loops are closed through a wired or wireless communication network. A schematic of an NCS is shown in Fig.~\ref{fig:byz}. 
The network part of the control loop can be in the form of a sensor network,  a controller network, or an actuator network.  NCSs are often also highly dynamic, with subsystems, actuators, and sensors entering and leaving the network over time. 

As opposed to  monolithic systems where a single decision-maker (human or machine) possesses all available knowledge and information related to the system,  NCSs typically  involve multiple decision-makers, each with access to information that is not available to other decision-makers.  Coordination among the decision-makers in such systems can be achieved through the use of distributed algorithms. These algorithms are  executed  concurrently by each of the decision-makers, incorporating both their local information and any information received from other decision-makers in the network (via communication channels). However,  as the size and the complexity of interconnections in NCSs increase, these distributed algorithms become more prone to failures, degradation, and attacks.  In this survey, we focus our attention on attacks performed by adversarial agents in NCSs.  The subtle difference between a fault and an attack is that in the latter, the attacker uses their knowledge about the system model to target the vulnerable parts of an NCS to either maximize their impact, minimize their visibility, or minimize their effort to attack. As the attacker intelligently optimizes its actions, distributed algorithms have to be carefully designed to withstand adversarial actions, rather than more generic classes of faults considered by 
 classical fault-tolerant control methods. 
Among various approaches to the resilience and security of distributed algorithms,  the goal of this survey paper is to focus on the intersection of systems and control theory, graph theory, and  communication and computation techniques to provide tools to solve security problems in NCSs. Notably, these are problems that are intractable using any set of methods in isolation. 
 
\subsection{Security of Networked Systems}
\label{sec:faultvsattack}
As mentioned above, the main difference between adversarial actions and faults stems from  the ability of the attacker  to carefully target vulnerable parts of the system, particularly by learning about the system (and the deployed algorithms) before the attack. Cyber-attacks are thus classified into different categories based on their knowledge level and their ability to disrupt resources. In addition to system-theoretic properties, the structure of the underlying large-scale network plays a key role in determining  specific attack behaviors. The following simple example shows the role of the graph structure in distinguishing attacks from faults. 
\begin{example}
Consider the NCS shown in Fig.~\ref{fig:byz}. If the network experiences a single fault distributed uniformly at random, over the total number of $n$ nodes in the network, then any given node  becomes faulty with probability $p=\frac{1}{n}$. For large $n$, this value is small. On the other hand, an attacker that wishes to disconnect the graph can do so by targeting node $v$. 
This prevents nodes in set $A$ from receiving true information from nodes in set $B$. 
\end{example}
The dichotomy between random failures and targeted node removal has been studied in the complex networks literature, particularly in the case of scale-free networks, which arise in a large number of applications \cite{pfattach}. These scale-free networks  have a few nodes with large degrees (so called ``hub nodes"); when such  nodes are targeted for removal \cite{barabasi, barabasialbertt}, they cause the network to split into multiple parts.
 
Other than simply removing nodes from the network as discussed above, the attacker can perform more complex actions. One such action is to manipulate the dynamics of a subset of nodes in the network by injecting carefully crafted attack signals (or incorrect data) into their update rules. These attacks create a discrepancy between the information that the targeted nodes send to their neighbors and the true information they are supposed to send. Here, based on the communication medium, another network-theoretic feature of attacks arises which distinguishes them from faults: if the communication is point-to-point (as opposed to wireless broadcast), the attacker has the ability to send inconsistent information to different neighbors whereas benign faults result in incorrect but consistent information being sent through the network. An example is shown in Fig.~\ref{fig:byz} where the node whose true value is $a$ sends a wrong but consistent value of $a+\tilde{a}$ to its neighbors, whereas the node with value $b$ shares wrong and inconsistent information to neighbors, i.e., it sends $b+\Tilde{b}$ to one node and $b+\hat{b}$ to the other. 
%Sending inconsistent information, however, requires specific physical infrastructure. In particular, it can only happen in point-to-point communications, e.g., wired networks, whereas in broadcast networks it is not possible to perform such an attack. 

\subsection{Applications}
Networked control systems have found numerous applications in today's engineering systems, some of which we mention briefly below. 

\subsubsection*{Automotive and Intelligent Transportation Systems}

The concept of connected vehicles, denoted by V2X, effectively transforms  transportation systems into a network of processors.  From this perspective, V2X  refers to (i) each vehicle's wireless communications with its surroundings, including other vehicles, road infrastructure, and the cloud, and (ii) wired communication within each vehicle between several electronic control units (ECU) in a controller area network (CAN). At the higher level, the nodes represent vehicles, road infrastructure, the cloud, and any other component that is able to send information. The wireless communication between those nodes is modeled by edges, e.g., the dedicated short range communication system (DSRC) for V2V communication. For the wired intra-vehicle network, the nodes are ECUs and the edges are buses transmitting data. 
Wireless communications between vehicles and their surroundings are prone to intrusions. Several works have reported different types of attacks on inter-vehicular networks, along with defense mechanisms \cite{biron, BenOthman}. Attacks on the wired intra-vehicle network or on safety critical ECUs (e.g., engine control unit, active steering, or brake system) can have life-threatening consequences.  Moreover,  stealthy attacks on the CAN Bus system, including an attack that embeds malicious code in a car’s telematics unit and completely erases any evidence of its presence after a crash, have been reported \cite{Koscher}. Several other attacks using both wired and wireless communications have been studied in the literature. An example is an attack that can enter the vehicle via a Bluetooth connection through the radio ECU and be disseminated to other safety-critical ECUs \cite{Checkoway}.

\subsubsection*{Smart Buildings and IoT}
Smart buildings are the integration of a vast number of sensors, smart devices, and appliances to control heating, ventilation and air conditioning, lighting, and home security systems through a building automation system (BAS).  In a building automation networked system, the home appliances are the nodes and the wireless communications between them and between each appliance and the center are the edges. When home devices are connected to the internet, they form a key part of the {\it Internet of Things}.  The objectives of building automation are to improve occupant comfort, ensure efficient operation of building systems, and reduce  energy consumption and operating costs. However, the high level of connectivity, automation, and remote accessibility of devices also makes it critical to protect smart buildings against failures and attacks \cite{Stankovic}.

\subsubsection*{Power Systems}

The traditional practice in power grids
is to institute safeguards against physical faults using protective devices \cite{hooshyar}. However,  the emergence of new technologies including smart meters, smart appliances, and renewable energy resources, together with available communication
technologies introduces further vulnerabilities to potential cyber-attacks \cite{Sridhar}. Cyber attacks in power systems can happen at three different levels: (i) generation and transmission level, i.e., automatic generation control (AGC) loops \cite{Huang}, (ii) distribution level, e.g.,  islanded micro-grids \cite{shahidehpour}, and (iii)  market level, e.g., false data injection in electricity
markets \cite{lxie}. Further discussion on the cyber-security of power systems can be found in \cite{WooiTen, Ericsson}.

\subsubsection*{Blockchain}
A blockchain is a growing list of records, called blocks, that are linked together using cryptography. Each block contains a cryptographic hash of the previous block in a tree structure, called a Merkle tree. As each block contains information about the block previous to it, they form a chain, with each additional block reinforcing the ones before it.  Blockchains are considered secure by design and exemplify a distributed computing system with high attack tolerance \cite{lin}. One of the most recognized applications of blockchains is in cryptocurrency, e.g., bitcoins.

There are several other applications for which the security of large-scale networked systems plays a crucial role. Examples include  swarm robotics (with applications ranging from search and rescue missions  to  mining and agricultural systems) \cite{swarm1, swarm2}, water and waste-water networks \cite{slay2007lessons}, nuclear power plants \cite{StuxnetSymantecReport}, and social networks \cite{Marsden, JacksonBook}. Further discussion on applications can be found in \cite{survey}.

\subsection{Early Works on the Resilience of Networked Control Systems}

We provide a brief literature review on the security of networked control systems; starting from  centralized (non-graph-theoretic) approaches and then followed up by graph-theoretic methods. 

\subsubsection{Centralized Resilient Control Techniques}

Centralized fault-tolerant techniques have a long history in the systems and control community \cite{masoum, saif, Darouach, depersis}. These early works focused on detecting and mitigating faults in the system and were not equipped to overcome adversarial actions. Distributed methods for fault-detection that were subsequently developed will be discussed in the next subsection. These later efforts focused on the goal of providing defense mechanisms for control systems against attacks in three layers of {\it attack prevention, attack detection}, and {\it attack endurance}.  We provide a brief overview of these defense layers below. Detailed discussions can be found in \cite{survey, teixeira2015secureCSM}.  

The first layer of defense is to prevent the attack from happening. Cryptography, network coding, model randomization, differential privacy, and moving target defense are among well-known attack prevention mechanisms used for control systems \cite{medard, farokhi, dwork, ny, Motwani, moving}.
In many cases, however, it is not always possible to prevent all attacks since regular users may not be distinguishable from the intruders. In those cases, the second layer comes into play which aims to detect and isolate the attack. Observer-based techniques have been proposed to detect the attacks which compare the state estimates under the healthy and the attacked cases \cite{Pasqualettiii}.  When the control system does not satisfy the required observability conditions, coding-theory, e.g., parity check methods, can be used to detect the attacks \cite{Blahut}.  In some cases, an adversary delivers fake sensor measurements to a system operator to conceal its effect on the plant. Certain types of such attacks, referred to as ``replay attacks'', have been addressed by introducing physical watermarking (by adding a Gaussian signal to the control input) to bait the attacker to reveal itself  \cite{watermark}. The sub-optimality of the resulting control action is the cost paid to detect the attacks in those cases.
In addition to the above model-based techniques, anomaly detection methods have been proposed based on machine learning techniques. For example, Neural Networks (NNs) and Bayesian learning have been studied for anomaly
detection in the context of security \cite{garcia, tsai, yhe}. 
When attack detection is not possible, the system must be at least resilient enough to withstand the attacks or mitigate the impact of the attack.  Probabilistic methods for attack endurance in NCSs for both estimation and control were studied in  \cite{sinopolitac, joao, touri, fawzi}. Redundancy-based approaches are used to bypass the attacks by using the healthy redundant parts \cite{redundant1,redundant2, abbas, momani}. Such  redundancy in large-scale systems can be in the form of adding parts, e.g., extra sensors, or the connection between the parts, i.e., network connectivity. On the other hand, several control-theoretic methods have also been proposed to mitigate the attack impact, including event-triggered control for tackling denial of service attacks \cite{heemels, tesi}. Robust control techniques have also been shown to be useful tools to mitigate the attack impact \cite{zhubasar}.

\subsubsection{Resilient Distributed Techniques}
The theory of distributed algorithms has a long history in computer science  with a variety of applications in telecommunications, scientific computing, distributed information processing, and real-time process control \cite{tsisiklis, tsisikliss}.  In  the control systems community,  distributed control algorithms have been studied for several decades \cite{varayaasurvey, varayamitter, varayaa, tsisikliss, Witsenhausen}, with an explosion of interest in recent decades due to their applications in distributed coordination of multi-agent systems, formation control of mobile robots, state estimation of power-grids, smart cities, and intelligent transportation systems, and distributed energy systems \cite{Jadbaba, ren2, tomlin, Olfati1}.

The earliest works on the security of distributed algorithms can be found in the computer science literature  \cite{Lynch, FISCHER}, typically with the focus on simple network topologies (such as complete graphs).  One of the main approaches to address the resilience and security of distributed systems is to leverage the physical redundancy that the network connectivity provides. Hence, resilient distributed estimation and control algorithms usually use (different types of) network connectivity measures to quantify the resilience against certain adversarial actions \cite{Sundaram2011, Hogan, Pasqualetti123}. To do this, system-theoretic notions, such as controllability (or observability) and detectability are reinterpreted in terms of graph-theoretic quantities with the help of tools such as algebraic graph theory or structured systems theory \cite{Pasqualettiii}. 

In parallel to control-theoretic approaches, several other approaches to the design of resilient distributed algorithms have been developed. Recently, federated learning techniques have found numerous applications in computer networks. The general principle of federated learning is to train local models on local data samples and exchange parameters, e.g., the weights of a deep neural network, between these local models at some frequency to generate a global model \cite{Jakub, Jiang}. With federated learning, only machine learning parameters are exchanged. These parameters can be encrypted before sharing between learning rounds to guarantee privacy. For the cases where these parameters may still leak information about the underlying data samples, e.g., by making multiple specific queries on specific datasets, secure aggregation techniques have been developed \cite{Tong,ByzFed1,Ghosh,pillutla}. 

When a NCS becomes larger in scale, the  notions of attack prevention, detection, and resilience discussed above depend more on the interconnections between components (sensors, actuators, or controllers) in the network. With this in mind, the focus of this survey paper is to  present the theoretical works in the literature on  graph-theoretic interpretations of the security in NCSs. 

\textbf{Related Survey Papers.} There are some recently published survey papers on related topics, including \cite{giraldo}, which provides an overview of security and privacy in a variety of cyber-physical systems (e.g., smart-grids, manufacturing systems, healthcare units, industrial control systems, etc.);  \cite{ishii}, which focuses on resilient consensus problems; \cite{bajwa_surv}, which focuses on distributed statistical inference and machine learning under attacks; and 
\cite{prorok}, which discusses the applications of resilient distributed algorithms to multi-robot systems. Compared to \cite{giraldo} where the exposition is essentially of a qualitative nature, our survey provides a mathematical treatment of security in networked systems, covering the necessary technical background in linear algebra, graph theory, dynamical systems, and structured systems theory. Our paper differs from \cite{ishii}, \cite{bajwa_surv}, and \cite{prorok} in that it has a much broader scope: we provide a detailed discussion of graph-theoretic measures for the resilience of a variety of distributed algorithms, special cases of which include consensus and distributed statistical inference. Moreover, we provide a comprehensive view of the role of various connectivity measures on the resilience levels for distributed algorithms. We also discuss ways to maintain a desired level of resilience when the network loses connectivity.

\section{Mathematical Preliminaries}
\label{sec:not}
\subsection{Graph Theory}
\begin{figure}[t!]
\centering
\includegraphics[scale=.55]{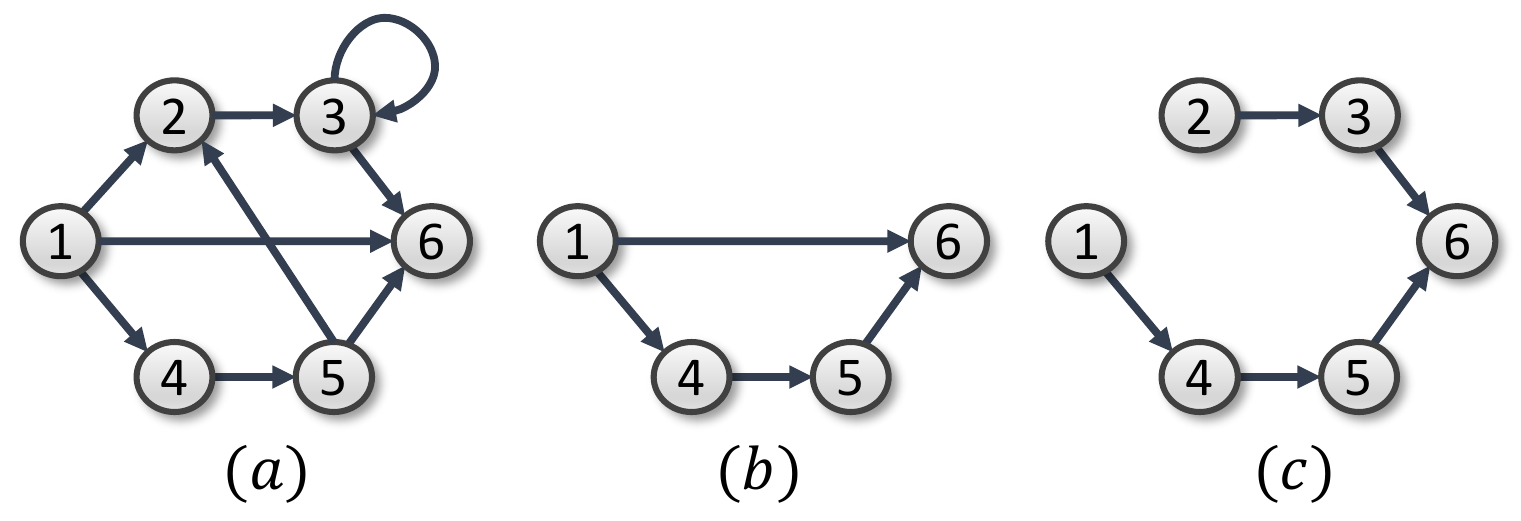}
\caption{(a) An example of a digraph $\mathcal{G}$, (b) an induced subgraph of $\mathcal{G}$, (c) a spanning subgraph of $\mathcal{G}$. }
\label{fig:by9qdyz}
\end{figure}

A weighted graph is a pair $(\mathcal{G},w)$ where  $\mathcal{G}=\{\mathcal{V},\mathcal{E}\}$ is a directed graph in which $\mathcal{V}$ is the set of vertices (or nodes)\footnote{Throughout this paper, we use words node, vertex, and agent interchangeably.} and $\mathcal{E}\subseteq \mathcal{V}\times \mathcal{V}$ is the set of edges and $w:\mathcal{E}\to \mathbb{R}$ is a weight function. In particular, $(j, i)\in \mathcal{E}$ if and only if there exists an edge from $j$ to $i$ with some weight $w_{ji}\neq 0$. Graph $\mathcal{G}$ is undirected if $(j, i) \in \mathcal{E}$ implies $(i, j) \in \mathcal{E}$ and $w_{ij}=w_{ji}$.
The in-neighbors of vertex $i \in \mathcal{V}$  are denoted  $\mathcal{N}_i^{\rm in} \triangleq \{j \in \mathcal{V}~|~(j, i) \in \mathcal{E}, j\neq i\}$. Similarly, the out-neighbors of $i$ are $\mathcal{N}_i^{\rm out} \triangleq \{j \in \mathcal{V}~|~(i, j) \in \mathcal{E}, j\neq i\}$.
The in-degree (or simply degree) of  node $i$ is $d_i=\sum_{j}w_{ji}$. The minimum degree of a graph $\mathcal{G}$ is denoted by $d_{\min}(\mathcal{G}) = \min_{i\in\mathcal{V}}d_i$. If $(i,i)\in \mathcal{E}$, then $i$ is said to have a self loop (but it is not counted in the degree of $i$). A subgraph of $\mathcal{G}$ is a graph $\bar{\mathcal{G}}=\{\bar{\mathcal{V}}, \bar{\mathcal{E}}\}$ with $\bar{\mathcal{V}} \subseteq \mathcal{V}$ and $\bar{\mathcal{E}}\subseteq \mathcal{E}$. A subgraph is induced if it is obtained from $\mathcal{G}$ by deleting a set of vertices (and all edges coming into and out of those vertices), but leaving all other edges intact. The subgraph $\mathcal{H}$ is called spanning if it contains all
vertices of $\mathcal{G}$, i.e., $\bar{\mathcal{V}}=\mathcal{V}$. 
An example of a digraph $\mathcal{G}$ together with an induced and a spanning subgraph of $\mathcal{G}$ is shown in Fig.~\ref{fig:by9qdyz}.

%The in-degree for node $i$ is $d_i=|\mathcal{N}_i^{\rm in}|$.  Let $A$ be the adjacency matrix of $\mathcal{G}$, where $A_{ij} > 0$ if $(i,j) \in \mathcal{E}$ and zero otherwise. The Laplacian is given by $L \triangleq D - A$, where $D = \diag(d_1, d_2, \ldots, d_n)$.

\subsubsection{Paths and Cycles in Graphs}
For  subsets $\mathcal{F}, \mathcal{B}\subset \mathcal{V}$, a path from $\mathcal{F}$ to $\mathcal{B}$ is a sequence of vertices $ v_1,  v_2, ...,  v_{t}$ where $ v_1\in \mathcal{F}$,  $ v_{t}\in \mathcal{B}$,
and $(v_j,v_{j+1})\in \mathcal{E}$ for $1 \leq  j \leq t- 1$. A cycle is a path where $ v_t=v_1$. A simple path contains no repeated vertices. A {\it directed acyclic graph} is a digraph with no cycles. For a subset $\mathcal{X}\subset \mathcal{V}$, an $\mathcal{X}$-rooted path (respectively $\mathcal{X}$-topped path) is a path which starts from a node $v\in \mathcal{X}$ (respectively ends at some $v\in \mathcal{X}$) . Two paths are disjoint if they have no common vertices and two paths are internally disjoint if they have no common vertices except for possibly the starting and ending vertices.  A set of paths $P_1, P_2, ..., P_r$ are (internally) vertex disjoint if the paths are pairwise (internally) vertex disjoint. For example, the two paths $P_1:1,2,3,6$ and $P_2:1,4,5,6$ in Fig.~\ref{fig:by9qdyz} (a) are internally vertex disjoint paths between nodes 1 and 6. Given two subsets $\mathcal{X}_1,\mathcal{X}_2\subset \mathcal{V}$, a set of $r$ vertex disjoint paths, each with
start vertex in $\mathcal{X}_1$ and end vertex in $\mathcal{X}_2$, is called an $r$-linking from $\mathcal{X}_1$ to $\mathcal{X}_2$.\footnote{There are various algorithms to find linkings, such as the Ford-Fulkerson algorithm, which has
run-time polynomial in the number of vertices \cite{West01}.}
The length of a path is the summation of the edge weights in the path. The distance between a pair of nodes $i$ and $j$ is the length of the shortest path between $i$ and $j$. 
The {\it effective resistance}, $\mathfrak{R}_{ij}$, between two vertices $i$ and $j$ in a graph is the equivalent resistance between these two vertices when we treat the resistance of each edge $e$ as $\frac{1}{w_e}$, where $w_e$ is the edge weight.

\subsubsection{Graph Redundancy Measures}
\label{sec:connectivity}

\begin{figure}[t!]
\centering
\includegraphics[scale=.55]{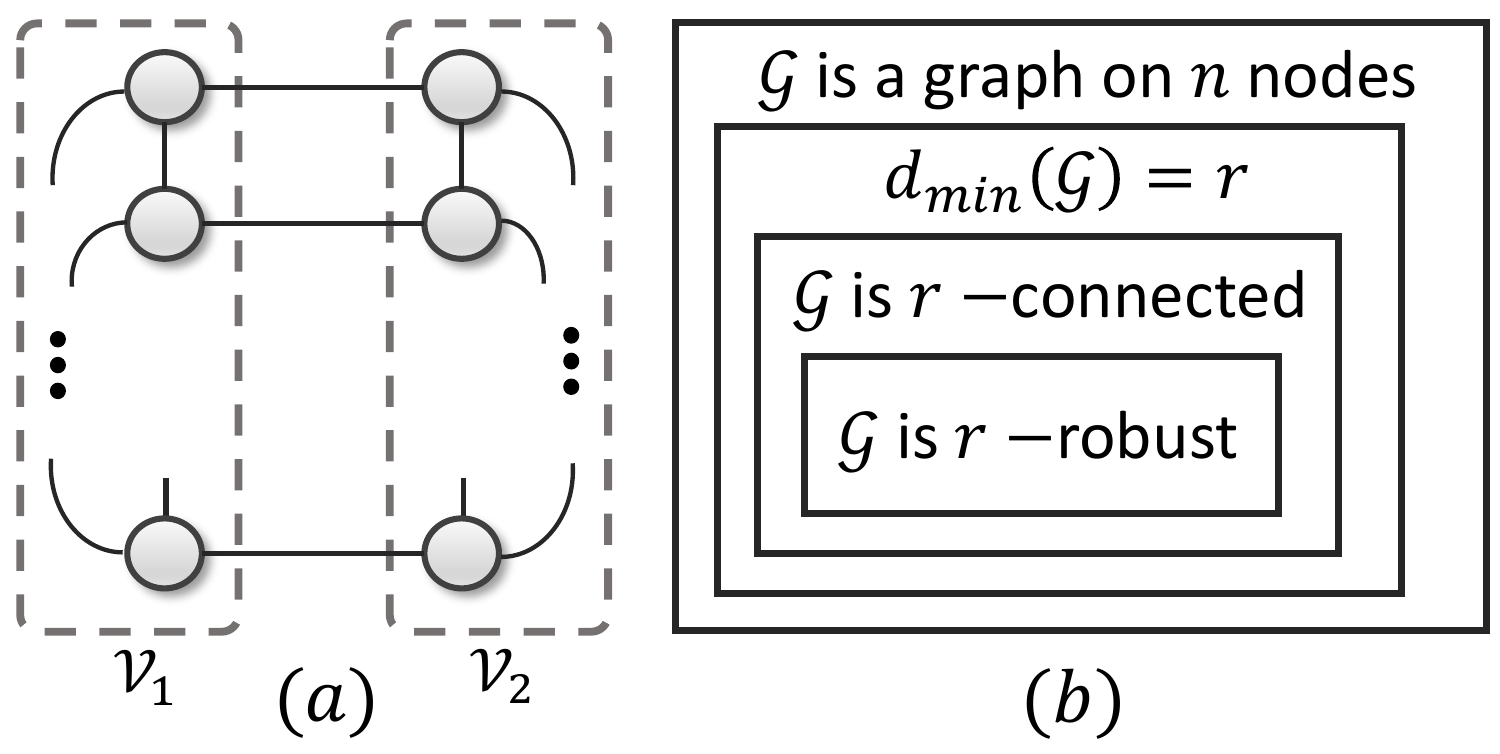}
\caption{(a) Graph $\mathcal{G}=(\mathcal{V},\mathcal{E})$ with $\mathcal{V}=\mathcal{V}_1\cup \mathcal{V}_2$ and $|\mathcal{V}_1|=|\mathcal{V}_2|=\frac{n}{2}$. $\mathcal{V}_1$ and $\mathcal{V}_2$ are complete graphs. There is a one to one connection between nodes in $\mathcal{V}_1$ and nodes in $\mathcal{V}_2$.~(b) Relationships between different graph connectivity measures.}
\label{fig:by9qwfqdyz}
\end{figure}

A graph $\mathcal{G}$ is called {\it strongly connected} if there is a path between each pair of vertices $i,j\in \mathcal{V}$. A graph is said to be disconnected if there exists at least one pair of vertices $i,j\in \mathcal{V}$ such that there is no path from $i$ to $j$. Other than the above binary measures of connectivity, there are several other graph connectivity measures, some of which are mentioned bellow. 
\begin{itemize}
    \item \textbf{Vertex and Edge Connectivity:} A \emph{vertex-cut} in a graph $\mathcal{G}=\{\mathcal{V}, \mathcal{E}\}$ is a subset $\mathcal{S} \subset \mathcal{V}$  of vertices such that removing the vertices in $\mathcal{S}$ (and any resulting associated edges) from the graph causes the remaining graph to be disconnected. A \emph{$(j, i)$-cut} in a graph is a subset $\mathcal{S}_{ij}\subset \mathcal{V}$ such that if the nodes $\mathcal{S}_{ij}$ are removed, the resulting graph contains no path from vertex $j$ to vertex $i$.   Let $\kappa_{ij}$ denote the size of the smallest $(j, i)$-cut between any two vertices $j$ and $i$. The graph $\mathcal{G}$ is said to have \emph{vertex connectivity} $\kappa(\mathcal{G})$ (or to be \emph{$\kappa$-vertex connected}) if $\kappa_{ij}\geq \kappa$ for all $i,j \in \mathcal{V}$. Similarly, the \emph{edge connectivity} $e(\mathcal{G})$ of a graph  $\mathcal{G}$ is the minimum number of edges whose deletion disconnects the graph. The vertex connectivity, edge connectivity, and minimum degree satisfy
\begin{equation}
\kappa(\mathcal{G})\leq e(\mathcal{G})\leq d_{\rm min}(\mathcal{G}).
\label{eqn:veredgemin}
\end{equation}

\item \textbf{Graph Robustness \cite{zhang2012robustness, Hogan}:} For some $r \in \mathbb{N}$, a subset $S$ of nodes in the graph $\mathcal{G}=(\mathcal{V},\mathcal{E})$ is said to be {\it $r$-reachable} if there exists a node $i \in S$ such that $|\mathcal{N}_i^{\rm in}\setminus S| \ge r$.  Graph $\mathcal{G}$ is said to be {\it $r$-robust} if for every pair of nonempty, disjoint subsets $~\mathcal{X}_1, \mathcal{X}_2 \subseteq \mathcal{V}$, either $\mathcal{X}_1$ or $\mathcal{X}_2$ is $r$-reachable. If $\mathcal{G}$ is $r$-robust, then it is at least $r$-connected.

For some $r, s \in \mathbb{N}$, a graph
is said to be $(r,s)$-robust if for all pairs of disjoint nonempty subsets $\mathcal{X}_1,\mathcal{X}_2\subset \mathcal{V}$, at least one of the following conditions
hold:
\begin{itemize}
    \item [(i)] All nodes in $\mathcal{X}_1$ have at least $r$ neighbors outside $\mathcal{X}_1$.
    \item [(ii)] All nodes in $\mathcal{X}_2$ have at least $r$ neighbors outside $\mathcal{X}_2$.
    \item [(iii)] There are at least $s$ nodes in $\mathcal{X}_1\cup \mathcal{X}_2$ that each have at
least $r$ neighbors outside their respective sets.
\end{itemize}
Based on the above definitions, $(r, 1)$-robustness is equivalent to $r$-robustness.

\iffalse
\item \textbf{Isoperimetric Constant:} The {\it edge-boundary} of a set of nodes $S \subset V$ is given by $\partial{S} = \{(i,j) \in E \mid i \in S, j \in V\setminus{S}\}$.  The {\it isoperimetric constant} of $G$ is defined as \cite{ChungSpectral}
 \begin{equation}
 i(\mathcal{G})\triangleq \min_{S \subset V, |S| \le \frac{n}{2}}\frac{|\partial S|}{|S|}.
 \label{eqn:iso}
 \end{equation}
By choosing $S$ as the vertex with the smallest degree we obtain $i(\mathcal{G}) \le d_{min}$. If $i(\mathcal{G})> r-1$  graph $\mathcal{G}$, then the graph is at least $r$-robust. 

\fi

\end{itemize}
The gap between the robustness and node connectivity (and minimum degree) parameters can be arbitrarily large, as illustrated by the graph $\mathcal{G}$ in Fig.~\ref{fig:by9qwfqdyz} (a).  While the minimum degree and node connectivity of the graph $\mathcal{G}$ is $n/2$, it is only $1$-robust (consider subsets $\mathcal{V}_1$ and $\mathcal{V}_2$). 
In Fig.~\ref{fig:by9qwfqdyz} (b), a Venn diagram is used to show the relationship between various graph connectivity measures. 

\subsection{Matrix Terminology}

For a matrix $\bs{M}_{m\times n}$ with $n \leq m$, the singular values are ordered as $\sigma_1(\bs{M}) \le \sigma_2(\bs{M}) \le \ldots \le \sigma_n(\bs{M})$.                       
When $M$ is a square matrix, the real parts of the eigenvalues are ordered as  $\Re \left(\lambda_1(\bs{M})\right) \le \Re \left(\lambda_2(\bs{M})\right) \le \ldots \le \Re \left(\lambda_n(\bs{M})\right)$. Matrix $\bs{M}$ is called nonnegative  if its elements are nonnegative, and it is  a {\it Metzler} matrix if its off-diagonal elements are nonnegative.  We use $\mathbf{e}_i$ to indicate the $i$-th vector of the canonical basis. 

\subsection{Spectral Graph Theory and Linear Systems}

 The  adjacency matrix of a graph of $n$ nodes is denoted by $\bs{A}_{n\times n}$, where $\bs{A}_{ij}=w_{ij}$ if $(j, i) \in \mathcal{E}$ with the edge weight $w_{ij}$ and $\bs{A}_{ij}=0$ otherwise.   The Laplacian matrix of the graph is $\bs{L} \triangleq \bs{D} - \bs{A}$, where $\bs{D} = \diag(d_1, d_2, \ldots, d_n)$.  The real parts of the  Laplacian eigenvalues are nonnegative and are denoted by $0 = \Re\left(\lambda_1(\bs{L})\right) \le \Re\left(\lambda_2(\bs{L})\right)  \le \ldots \le \Re\left(\lambda_n(\bs{L})\right) $.\footnote{We consider Laplacian matrices for graphs with positive edge weights. Graph spectra for negative edge weights have been studied in \cite{chen, hou}.} The second smallest eigenvalue of the Laplacian matrix, $\lambda_2(\bs{L})$, is called the {\it algebraic connectivity} of the graph and is greater than zero if and only if $\mathcal{G}$ is a connected graph. Moreover, we always have \cite{Godsil}
\begin{equation}
\lambda_2(\bs{L})\leq \kappa(\mathcal{G}).
\label{eqn:veredgemin123}
\end{equation}
Given a connected  graph $\mathcal{G}=\{\mathcal{V},\mathcal{E}\}$, an orientation of the graph $\mathcal{G}$ is defined by assigning  a direction (arbitrarily) to each edge in $\mathcal{E}$. For graph $\mathcal{G}$ with $m$ edges, labeled as $e_1, e_2, ..., e_m$, its node-edge incidence matrix $\mathcal{B}(\mathcal{G})\in \mathbb{R}^{n\times m}$ is defined as
$$[\mathcal{B}(\mathcal{G})]_{kl}=
  \begin{cases}
    1       & \quad  \text{if node $k$ is the head of edge $l$},\\
   -1  & \quad \text{if node $k$ is the tail of edge $l$},\\
   0  & \quad \text{otherwise}.\\
  \end{cases}
  $$
The graph Laplacian satisfies $L=\mathcal{B}(\mathcal{G})\mathcal{B}(\mathcal{G})'$.

A discrete-time linear time-invariant system is represented in the  state-space form as follows:
\begin{align}
     &{\boldsymbol{x}}[t+1]=\boldsymbol W\boldsymbol{x}[t]+\boldsymbol B\boldsymbol{u}[t],\nonumber\\
    &\boldsymbol{y}[t]=\boldsymbol C \boldsymbol{x}[t]+\boldsymbol D\boldsymbol{u}[t],
    \label{eqn:struc}
\end{align}
where $\boldsymbol{x}\in \mathbb{R}^n$ is the state vector, $\boldsymbol{u}\in \mathbb{R}^m$ is the vector of $m$ inputs, $\boldsymbol{y}\in \mathbb{R}^q$ is the vector of $q$ outputs, and $\boldsymbol W\in \mathbb{R}^{n\times n}$, $\boldsymbol B\in \mathbb{R}^{n\times m}$, $\boldsymbol C\in \mathbb{R}^{q\times n}$, and $\boldsymbol D\in \mathbb{R}^{q\times m}$, are called state, input, output, and feed-forward matrices, respectively. Similarly,  the state-space model of a continuous-time linear system is given
by 
\begin{align}
     &\dot{\boldsymbol{x}}=\boldsymbol W\boldsymbol{x}+\boldsymbol B\boldsymbol{u},\nonumber\\
    &\boldsymbol{y}=\boldsymbol C\boldsymbol{x}+\boldsymbol D\boldsymbol{u}.
    \label{eqn:struccont}
\end{align}
A state space form of a linear system is compactly represented as $(\boldsymbol W,\boldsymbol B,\boldsymbol C,\boldsymbol D)$ or $(\boldsymbol W,\boldsymbol B,\boldsymbol C)$  for cases where there is no feed-forward term.
A linear system is called (internally) positive if  its state and output are non-negative for every non-negative input and every non-negative initial state. A continuous-time linear system $(\boldsymbol W,\boldsymbol B,\boldsymbol C)$  is positive if and only if $\boldsymbol W$ is a Metzler matrix and $\boldsymbol B$ and  $\boldsymbol C$ are non-negative element-wise \cite{Farina}.  Moreover, for such a positive system with transfer function $G(s)=\boldsymbol C(s\boldsymbol I_n-\boldsymbol W)^{-1}\boldsymbol B$, the system $\mathcal{H}_{\infty}$ norm is obtained from the DC gain of the system, i.e.,  $\|G\|_{\infty}=\sigma_{n}(G(0))$, where  $\sigma_{n}$ is the maximum singular value of matrix $G(0)$.

\subsection{Structured Systems Theory}
\label{sec:str}
Consider the linear time-invariant system \eqref{eqn:struc}. With this system, associate the matrices $\boldsymbol W_{\lambda} \in \{0,\lambda\}^{n\times n}$, $\boldsymbol B_{\lambda} \in \{0,\lambda\}^{n\times m}$, $\boldsymbol C_{\lambda} \in \{0,\lambda\}^{q\times n}$, and $\boldsymbol D_{\lambda} \in \{0,\lambda\}^{q\times m}$. Specifically, an entry in these matrices is zero if the corresponding entry in the system matrices is equal to zero, and the matrix entry is a free parameter (denoted by $\lambda$) otherwise. This type of representation of \eqref{eqn:struc} shows the structure of the linear system regardless of the specific values of the elements in the matrices. Thus, it is called a {\it structured system} and can be equivalently represented by a directed graph $\mathcal{G}=\{\mathcal{X}, \mathcal{U}, \mathcal{Y}, \mathcal{E}_{\mathcal{X}\mathcal{X}},  \mathcal{E}_{\mathcal{X}\mathcal{Y}}, \mathcal{E}_{\mathcal{U}\mathcal{X}},\mathcal{E}_{\mathcal{U}\mathcal{Y}}\}$, where
\begin{itemize}
\item $\mathcal{X} \triangleq \{x_1, x_2, ..., x_n\}$ is the set of states;

\item $\mathcal{Y} \triangleq \{y_1,y_2,...,y_q\}$ is the set of measurements;

\item  $\mathcal{U} \triangleq \{u_1, u_2, ..., u_{m}\}$ is the set of inputs;

\item $\mathcal{E}_{\mathcal{X}\mathcal{X}}=\{(x_j,x_i)|\boldsymbol{W}_{ij}\neq 0\}$ is the set of edges corresponding to interconnections between the state vertices;
\item $\mathcal{E}_{\mathcal{U}\mathcal{X}}=\{(u_j,x_i)|\boldsymbol{B}_{ij}\neq 0\}$ is the set of edges corresponding to connections between the input vertices and the
state vertices;
\item $\mathcal{E}_{\mathcal{X}\mathcal{Y}}=\{(x_j,y_i)|\boldsymbol{C}_{ij}\neq 0\}$ is the set of edges corresponding to connections between the state vertices and the
output vertices;
\item $\mathcal{E}_{\mathcal{U}\mathcal{Y}}=\{(u_j,y_i)|\boldsymbol{D}_{ij}\neq 0\}$ is the set of edges corresponding to connections between the input vertices and the
output vertices.
\end{itemize}

A structured system is said to have a certain property, e.g., controllability or invertibility, if that
property holds for at least one numerical choice of free parameters $\lambda$
in the system. The following  theorem introduces graphical conditions for structural controllability and observability of linear systems.

\begin{thm} [\cite{Dion}]
The  pair $(\boldsymbol W, \boldsymbol B)$ (resp. $(\boldsymbol W, \boldsymbol C)$) is structurally
controllable (resp. observable) if and only if the graph $\mathcal{G}=\{\mathcal{X},\mathcal{U},\mathcal{E}_{\mathcal{X},\mathcal{X}}, \mathcal{E}_{\mathcal{U},\mathcal{X}}\}$ satisfies both of the following
properties:
\begin{enumerate}
    \item[(i)]  Every state vertex $x_i \in \mathcal{X}$ can be reached by a path from (resp. has a path to) some input vertex (resp. some output vertex).
    \item[(ii)] $\mathcal{G}$ contains a subgraph that is a disjoint union of cycles and $\mathcal{U}$-rooted paths (resp. $\mathcal{Y}$-topped paths), which covers all of the state vertices.
\end{enumerate}\label{thm:strcon}
\end{thm}

In Sec.  \ref{sec:attackdetection}, we will revisit structured systems by discussing structural conditions for the system to be invertible.

\begin{example}
The graph shown in Fig.~\ref{fig:by9qstrucqdyz} (a) is structurally controllable as it satisfies both conditions in Theorem \ref{thm:strcon}. However, it is not structurally observable: condition (ii) does not hold since the cycle  and $\mathcal{Y}$-topped paths are not disjoint. Graph (b) is  structurally controllable and observable. The set of disjoint $\mathcal{U}$-rooted paths (respectively $\mathcal{Y}$-topped paths) and cycles is $\mc{P}=\big\{\{u, x_1, x_4, x_5\},\{x_2\},\{x_3\}\big\}$ (respectively $\mc{C}=\big\{\{x_1, x_4, x_5, y\},\{x_2\},\{x_3\}\big\}$).
\begin{figure}[t!]
\centering
\includegraphics[scale=.55]{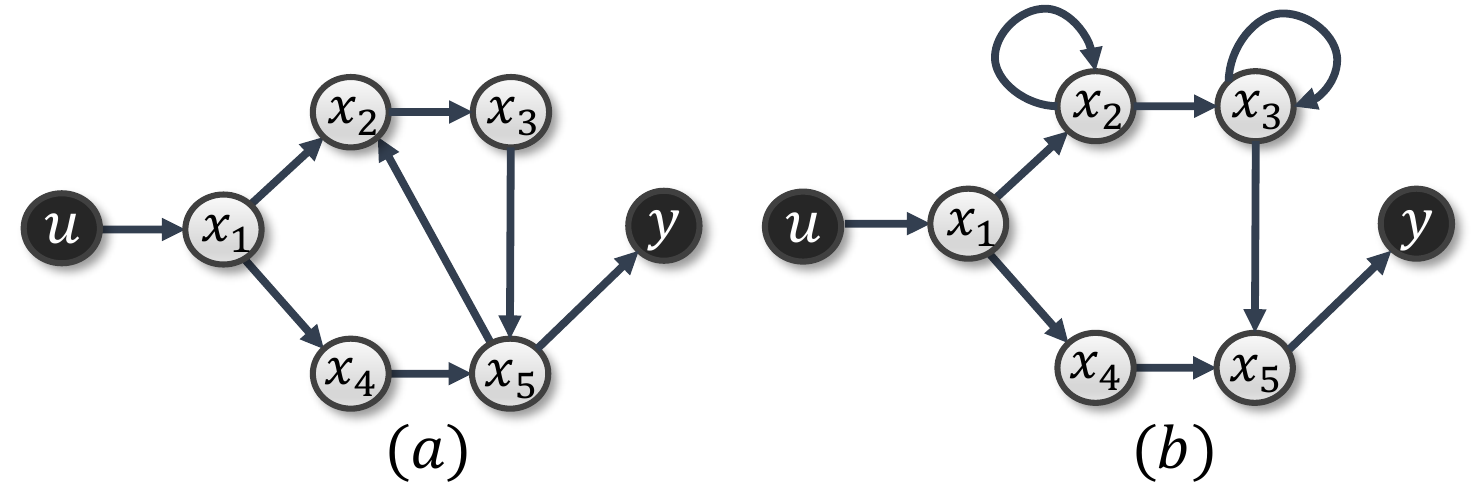}
\caption{(a) A structurally controllable but not observable graph, (b) A structurally controllable and observable graph.}
\label{fig:by9qstrucqdyz}
\end{figure}
\end{example}

\section{Notions of Resilience}
\label{sec:attackmodel}

In this section, we discuss the notions of resilience in NCSs which are sought in various distributed algorithms. To classify the resilience against each type of adversary, the first step is to distinguish a regular node from an adversarial one in NCSs. 

In distributed algorithms, each agent  is given an updating rule which is a function of its own states (and local information) and the states and information obtained from its neighbors. In many distributed algorithms, the agents only need to know their own updating rule and their neighbor set. However, in some problems, the agents not only need to know their neighbor sets, they must also be aware of the entire network topology (and maybe the updating policy of other agents). Based on this, the computational burden and the required storage limits for distributed algorithms may be different. 
Regular agents (nodes) in a NCS are those who obey the prescribed updating rule. The objective of a regular node can be either to calculate an exact desired value (e.g., a function of the state of other nodes) or an approximation of that value. The required precision of calculation depends on many factors including the cost of computing the exact value.
Deviations from normal behaviour can be considered as either a fault or an adversarial action (attack). Faults are those which happen unintentionally and (often) randomly with a given distribution. Communication dropouts, disturbances and noise, and failures due to unmodeled physics are all categorized as faults. Attacks, as discussed in Section \ref{sec:faultvsattack}, are  failures which intentionally happen to the system and can be viewed as decisions of an intelligent intruder. From the network's perspective, the attacker (or an adversarial agent) is  one who intentionally disregards the prescribed updating rule: the attacker updates its state and sends it to its neighbors in an arbitrary (and potentially worst case) manner. Since this type of deliberate adversarial behaviour is the focus of the current paper,  we further classify them in the following definition.

\begin{definition}[\textbf{Malicious vs. Byzantine}]
An adversarial agent is called malicious if it updates its state in an arbitrary manner. Thus, it sends incorrect but consistent values to all of its out-neighbors at each time-step. An adversarial agent is Byzantine if it can update its state arbitrarily and is capable of sending inconsistent values to different neighbors at each time-step.
\end{definition}

Both malicious and Byzantine agents  are allowed to know the entire network topology, the local information of all agents, and the algorithms executed by all agents. Furthermore, both malicious and Byzantine agents are allowed to collude amongst themselves to select their actions.  
 An example of malicious and Byzantine agents was discussed in Section \ref{sec:faultvsattack}.
Based on the above definition, every malicious node is Byzantine but not vice versa.  Malicious attacks may be appropriate for wireless broadcast models of communication or when the state of an agent is directly sensed by its neighbors (e.g., via cameras), whereas Byzantine attacks follow the wired point-to-point model of communication. 

In return for providing so much power to the adversarial agents, it is typical to assume a bound on the number of such agents. The following definitions  quantify the maximum number of tolerable attacks in a given network.

\begin{definition}[\textbf{$f$-total and $f$-local sets}]
For $f \in \mathbb{N}$, a set $\mathcal{C}\subset \mathcal{V}$ is said to be $f$-total if it contains at most $f$ nodes in the network, i.e., $|\mathcal{C}|\leq f$. A set $\mathcal{C}\subset \mathcal{V}$ is  $f$-local if it contains at most $f$ nodes in the neighborhood of each node outside that set, i.e., $|\mathcal{N}_i^{\rm in}\cap \mathcal{C}|\leq f$ for all $i\in \mathcal{V}\setminus \mathcal{C}$. 
\end{definition}

\begin{definition}[\textbf{$f$-local adversarial model}]
For $f \in \mathbb{N}$, a set $\mathcal{F}$ of adversarial nodes is $f$-locally bounded if $\mathcal{F}$ is an $f$-local set. 
\end{definition}

The set of adversarial nodes $\{a,b\}$ in Fig.~\ref{fig:byz} is a 2-total and 1-local set. Thus, it is 1-locally bounded. Note that every $f$-total set is also an $f$-local set but not vice versa. The $f$-total adversarial model is predominant in the literature on resilient distributed algorithms  \cite{Lynch, Lamport, Bouzid}. However, in order to allow the number of adversarial agents to potentially scale with the network, several of the algorithms discussed in this survey  allow the adversarial set to be $f$-local. 

Based on the above discussions, the notion of resilience can be stated as follows.
\begin{tcolorbox}
\textbf{Resilient Distributed Algorithm:}
Under a given adversarial model (e.g., $f$-locally bounded or $f$-total, malicious or Byzantine), a distributed algorithm operating on network $\mathcal{G}$ is called {\it resilient} if each regular node in $\mathcal{G}$ can compute its desired value (within some specified tolerance)  despite the actions of  the adversarial agents in $\mathcal{G}$. 
\end{tcolorbox}

Thus, various specific notions of resilience can be considered,  based on the above definition and depending on the type and number of adversaries, and the desired value computed by the regular agents.  

\section{Connectivity: The Earliest Measure of Resilience}
\label{sec:distfunccalc}

In this section, we first discuss the role of network connectivity in reliable information dissemination over networks. Then, with the help of structured systems theory, we tie together the traditional graph property of connectivity with system-theoretic notions to find conditions for reliable calculation of node values in a network.

\subsection{Connectivity as a Measure of Resilience}

In the following example,  we see how the existence of redundant paths between a  pair of nodes can facilitate reliable transmission of information between those nodes. 

\begin{example}\label{exp:vw98}
For the graph shown in Fig.~\ref{fig:by167rsvdyz} (a), suppose that $v_1$ tries to obtain the true value of $v_4$, i.e., $\psi_4[0]=2$. This value can be transmitted to $v_1$ through $v_2$ and $v_3$.\footnote{This can be done using a flooding algorithm, i.e., each node reads and stores their neighbors' values and broadcasts them to their out-neighbors in the next time step.} Suppose that $v_2$ is malicious and pretends that $\psi_4[0]$ has a value other than its true value. In this case, as $v_1$ receives inconsistent information  from  $v_2$ and  $v_3$, it can not conclude which value is the true one. In this case, a redundant path can serve as a tie breaker and help  $v_1$ to obtain the true value, Fig.~\ref{fig:by167rsvdyz} (b). 
\begin{figure}[t!]
\centering
\includegraphics[scale=.5]{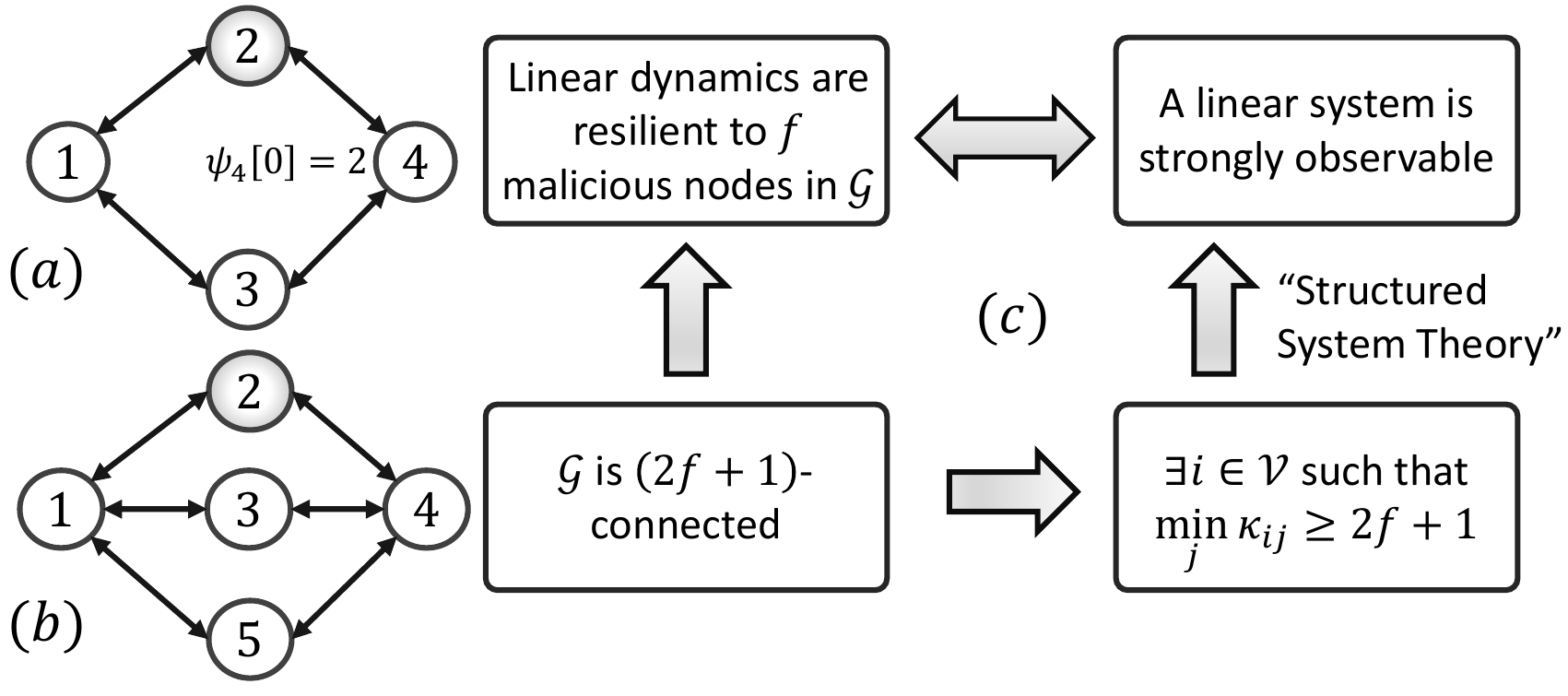}
\caption{(a) A graph with (a) two disjoint paths between nodes $v_1$ and $v_4$, and (b)  three disjoint paths between nodes $v_1$ and $v_5$. (c) Schematics of interconnections between system and graph properties.}
\label{fig:by167rsvdyz}
\end{figure}
\end{example}

One can generalize the observation given in Example \ref{exp:vw98} by saying that if there are $f$ adversarial nodes in the network, there should be $2f+1$ disjoint paths between any given pair of nodes in order to make sure that  information can be transmitted reliably between those nodes. 
The number of disjoint paths between node pairs is related to the vertex connectivity via Menger's theorem \cite{west}. 
\begin{thm}[\textbf{Menger's Theorem}]
Graph $\mathcal{G}$ has vertex connectivity $r$ if and only if there are $r$ internally vertex disjoint paths between each pair of nodes in $\mathcal{G}$. 
\end{thm}

This fundamental observation that $2f+1$ connectivity is required to overcome $f$ adversarial agents is classical in the computer science literature \cite{Lynch}.  
In the following subsection, we describe how the same result (namely that a connectivity of $2f+1$ is required to reliably exchange information in networks despite malicious agents) arises in the context of linear iterative dynamics for information dissemination in networks. In the process, this will introduce the use of zero-dynamics (and strong observability) together with structured systems theory as a means to analyze the resilience of linear dynamics on graphs. 

\subsection{A System-Theoretic Perspective on  Resilient Exchange of Information}
\label{subsec:func_calc}
In this subsection, we discuss reliable calculation of node values in a network in the presence of adversaries. In this setting, each agent $i\in \mathcal{V}$ tries to gather the values (measurements, positions, votes, or other data) of all other agents, despite the actions of malicious agents in the network. These values can be later used to calculate any arbitrary function of the agents' values. Here, we consider a broadcast model of communication where each node transmits the same value to all its neighbors. Hence, the adversarial agents are malicious, but not Byzantine. Our goal is to show that the topology of the network (specifically, its connectivity) completely characterizes the resilience of linear iterative strategies to malicious behavior. To this end, we first formally introduce the model of a distributed system under attack.

\textbf{Distributed System Model Under Attack:} Consider a network of $n$ agents (or processors) whose communication is represented by a time-varying graph $\mathcal{G}=\{\mathcal{V},\mathcal{E}[t]\}$.
Suppose that each agent $i\in \mathcal{V}$  begins with some (possibly private) initial value\footnote{This value is called the state or the opinion of node $i$, depending on the context. }  $\psi_i[0] \in \mathbb{R}$ and updates its value over time according to a prescribed rule, i.e., 
\begin{equation}
\psi_i[t+1]=g_i(\psi_j^i[t], t), \quad j\in \mathcal{N}_i[t], t\in \mathbb{Z}_{\geq 0},
\end{equation}
where $\psi_j^i[t]$ is the state of node $j$ sent to node $i$ at time step $t$ and $\psi_i^i[t]=\psi_i[t]$. The update rule, $g_i$, which is designed {\it a priori}, can be an arbitrary function and may be different for each node. 
For example, for the standard linear consensus protocol \cite{Jadbaba}, this function is simply some linear combination of the values of node $i$'s neighbors:  
\begin{equation}
\psi_i[t+1]=\sum_{j\in \mathcal{N}_i[t]}w_{ij}[t]\psi_j^i[t], 
\label{eqn:lineariteration}
\end{equation}
where $w_{ij}[t]$ is the weight assigned to node $j$'s value by node $i$ at time step $t$.

Recall that node $i$ is a  {\it regular} node if it does not deviate from its prescribed update rule $g_i(\cdot)$. The set of regular nodes is denoted by $\mathcal{R}$. A deviation can stem from a failure, e.g.,  disturbance or noise with a known model, a time delay or signal dropout, or an adversarial action (attack) in the form of arbitrary state updates. Some fundamental differences between faults and attacks  were discussed in Section \ref{sec:faultvsattack}. 
Consider a set $\mathcal{F}=\{{i_1}, {i_2}, ..., {i_f}\}\subset\mathcal{V}$ of malicious nodes. One way to represent an adversarial action at time step $t$ is to use an additive attack signal $\zeta_i[t]$ in the updating rule \eqref{eqn:lineariteration}. In particular, instead of applying the update
equation \eqref{eqn:lineariteration},  each node $i\in \mathcal{F}$ updates its state as
\begin{equation}
\psi_{i}[t+1]=\sum_{j\in \mathcal{N}_{i}[t]}w_{i j}[t]\psi_{j}^i[t]+\zeta_{i}[t].
\label{eqn:lineariterationwd}
\end{equation}
 Here, an agent is malicious in $T$ time steps if $\zeta_{i}[t]\neq 0$ for at least one time step $0\leq t \leq T-1$.  Writing \eqref{eqn:lineariterationwd} in vector form yields
\begin{align}
\boldsymbol{\psi}[t+1]&=\boldsymbol{W}\boldsymbol{\psi}[t]+\underbrace{[\mathbf{e}_{i_1} \hspace{3mm} \mathbf{e}_{i_2} \hspace{3mm} ... \hspace{3mm} \mathbf{e}_{i_f}]}_{\boldsymbol {B}_{\mathcal{F}}}\boldsymbol{\zeta}_{\mathcal{F}}[t],\nonumber\\
\boldsymbol y[t]&=\boldsymbol C\boldsymbol\psi[t]
\label{eqn::OAEGI0}
\end{align}
where $\boldsymbol{\zeta}_{\mathcal{F}}[t]=[\zeta_{i_1}, \zeta_{i_2}, ..., \zeta_{i_f}]$ is the additive error (attack) vector. Matrix $\boldsymbol W$ represents the communications between agents and $\boldsymbol C=[\bs{C}_1'\hspace{2mm}\bs{C}_2'\hspace{2mm}...\hspace{2mm}\bs{C}_q']'$,  where $\bs{C}_i$ determines the measurement of node $i$. 

\textbf{Observability and Connectivity:} The objective for each node $i$ is to recover the vector $\boldsymbol \psi[0]=[\psi_1[0]\hspace{1mm} \psi_2[0], ..., \hspace{1mm} \psi_n[0]]'$. To build intuition for the adversarial setting, let us start by looking at the case without malicious nodes. Moreover, let us consider a time-invariant graph. In this case, based on the linear iterative scheme in Eq. \eqref{eqn:lineariteration}, the task of estimating $\boldsymbol \psi[0]$ at node $i$ can be equivalently viewed as the problem of recovering the initial condition of the linear discrete-time dynamical system
$$ \boldsymbol{\psi}[t+1]=\boldsymbol{W}\boldsymbol{\psi}[t],$$
based on the following observation model:
\begin{equation}
    y_i[t] =\boldsymbol C_i\boldsymbol{\psi}[t].
    \label{eqn:measuremetn}
\end{equation}
Here, $\boldsymbol C_i$ is a $(d_i +1)\times n$ matrix with a single 1 in each row that denotes the states available to node $i$ (these positions correspond to the neighbors of node $i$, including node $i$). From basic linear systems theory, it then follows that $\boldsymbol \psi[0]$ can be recovered by node $i$ if the pair $(\boldsymbol{W}, \boldsymbol{C}_i)$ is observable. In \cite{Sundaramwithoutfault}, it is shown that if the underlying network is connected, then the weight matrix $\boldsymbol {W}$ can be designed in a way that ensures observability of  $(\boldsymbol{W}, \boldsymbol{C}_i)$. 

The above discussion reveals how concepts from systems theory such as observability can be combined with basic graph-theoretic notions such as connectivity to study the process of information diffusion over networks. It is natural to thus wonder whether a marriage of ideas between systems theory and graph theory will continue to be fruitful while analyzing the adversarial setting. The results from \cite{Sundaram2011}, summarized below,  establish that this is indeed the case.

\iffalse
\begin{rem}\label{rem:initialconditions}
Note that the initial conditions are not corrupted since they are equally trusted for malicious and regular agents. Instead, in this problem setup, the goal is to avoid the malicious nodes spreading confusion about the initial values of non-malicious nodes during the system's evolution. 
\end{rem} \textcolor{blue}{Can we add this remark elsewhere? Seems to be disrupting the flow a bit.}
\fi

Suppose that a subset $\mathcal{F}$ of nodes is  malicious and deviates from the update rule \eqref{eqn:lineariteration}. Thus, the new goal is to recover $\boldsymbol\psi[0]$ for the model in  \eqref{eqn::OAEGI0}. As in the non-adversarial case, we start by examining the observation model at node $i$. We note that the set of all values seen by node $i$ during the first $L+1$
time-steps of the linear iteration (for any non-negative integer $L$) is given by
\begin{equation}
    y_i[0:L]=\mathcal{O}_{i,L}\boldsymbol{\psi}[0]+\mathcal{M}_{i,L}^{\mathcal{F}}\boldsymbol{\zeta}_{\mathcal{F}}[0:L-1], 
    \label{eqn:oaisbv}
\end{equation}
where $y_i[0:L]=\big[y'_i[0] \hspace{2mm} y'_i[1]\hspace{2mm} \cdots \hspace{2mm} y'_i[L]\big]'$ and $\boldsymbol{\zeta}_{\mathcal{F}}[0:L-1]=\big[\zeta'_{\mathcal{F}}[0] \hspace{2mm} \zeta'_{\mathcal{F}}[1]\hspace{2mm} \cdots \hspace{2mm} \zeta'_{\mathcal{F}}[L-1]\big]'$. Matrices $\mathcal{O}_{i,L}$ and $\mathcal{M}_{i,L}^{\mathcal{F}}$ are the observability and invertibility matrices, respectively (from the perspective of node $i$), and can be expressed recursively as
\begin{equation}
\mathcal{O}_{i,L}= \begin{bmatrix}
      \boldsymbol C_i     \\[0.3em]
     \mathcal{O}_{i,L-1}\boldsymbol{W}
     \end{bmatrix}, \hspace{3mm} \mathcal{M}_{i,L}^{\mathcal{F}}=\begin{bmatrix}
      \mathbf{0}  &  \mathbf{0}\\[0.3em]
   \mathcal{O}_{i,L-1}\boldsymbol B_{\mathcal{F}} & \mathcal{M}_{i,L-1}^{\mathcal{F}}
     \end{bmatrix},
\end{equation}
where $\mathcal{O}_{i,0}=\boldsymbol C_i$ and $\mathcal{M}_{i,0}^{\mathcal{F}}$ is the empty matrix (with zero
columns). The question of interest is the following: \textit{Under what conditions can node $i$ recover $\boldsymbol{\psi}[0]$ based on a sufficiently large sequence of observations, despite the presence of the unknown inputs $\boldsymbol{\zeta}_{\mathcal{F}}$?}

As it turns out, the answer to the above question is intimately tied to the system-theoretic concept of \textit{strong observability}. In particular, the linear system \eqref{eqn::OAEGI0} is said to be strongly observable w.r.t. node $i$ if ${y}_i[t]=0$ for all $t$ implies $\boldsymbol{\psi}[0]=0$  (regardless of the values of the unknown inputs $\boldsymbol{\zeta}_{\mathcal{F}}[t]$). Moreover, if such a  strong observability condition holds, then this is equivalent to saying that node $i$ will be able to uniquely determine the initial condition $\boldsymbol{\psi}[0]$ based on the knowledge of its output sequence, regardless of the unknown inputs. 

\textbf{Strong Observability and Connectivity:} For the non-adversarial setting, connectivity was enough to ensure observability of the pair $(\boldsymbol{W},\boldsymbol{C}_i), \forall i \in \mathcal{V}$. In a similar vein, we need to now discern how the structure of the underlying network impacts strong observability. To this end, we present a simple argument to demonstrate that if the network is not adequately connected, then system \eqref{eqn::OAEGI0} will not be strongly observable w.r.t. certain nodes in the graph. For simplicity, let $\mathcal{G}$ be undirected. Now suppose the connectivity $\kappa$ of $\mathcal{G}$ is such that $\kappa \leq |\mathcal{F}|$. This implies the existence of a vertex cut $\mathcal{S}_2$ of size at most $|\mathcal{F}|$ that separates the graph into two disjoint parts. Let the vertex sets for these disjoint parts be denoted by $\mathcal{S}_1$ and $\mathcal{S}_3$. After reordering the nodes such that the nodes in $\mathcal{S}_1$ come first, followed by those in $\mathcal{S}_2$ and then $\mathcal{S}_3$, the consensus weight matrix takes the following form:

$$\boldsymbol{W}= \begin{bmatrix} \boldsymbol{W}_{11} & \boldsymbol{W}_{12} & \boldsymbol{0}\\
                                  \boldsymbol{W}_{21} & \boldsymbol{W}_{22} & \boldsymbol{W}_{23}\\
                                  \boldsymbol{0} & \boldsymbol{W}_{32} & \boldsymbol{W}_{33}
                                  \end{bmatrix}. $$
                                  
The structure of the above matrix follows immediately from the fact that the nodes in $\mathcal{S}_1$ can interact with those in $\mathcal{S}_3$ only via the nodes in $\mathcal{S}_2$. Now suppose all the nodes in $\mathcal{S}_2$ are adversarial; this is indeed feasible since $|\mathcal{S}_2| \leq |\mathcal{F}|$. Moreover, let the initial condition $\boldsymbol{\psi}[0]$ be of the form $$\boldsymbol{\psi}'[0] = \begin{bmatrix} \boldsymbol{\psi}_{\mathcal{S}_1}[0] & \boldsymbol{\psi}_{\mathcal{S}_2}[0]& \boldsymbol{\psi}_{\mathcal{S}_3}[0] \end{bmatrix} = \begin{bmatrix} \boldsymbol{0} & \boldsymbol{0} & \boldsymbol{v} \end{bmatrix},$$
where $\boldsymbol{v}$ is a non-zero vector in $\mathbb{R}^{|\mathcal{S}_3|}.$ If the adversarial inputs are of the form $\boldsymbol{\zeta}_{\mathcal{F}}[t]=-\boldsymbol{W}_{23}\boldsymbol{\psi}_{\mathcal{S}_3}[t]$, then it is easy to see that the states of the nodes in $\mathcal{S}_1$ and $\mathcal{S}_2$ remain at zero. Thus, an agent $i$ in $\mathcal{S}_1$ observes a sequence of zeroes. It follows that there is no way for agent $i$ to distinguish the zero initial condition from the non-zero initial condition we considered in this example. Thus, system \eqref{eqn::OAEGI0} is not strongly observable w.r.t. node $i$. 

\textbf{Discussion:} The above argument serves to once again highlight the interplay between control- and graph-theory in the context of information diffusion over networks. Moreover, it suggests that in order for every node to uniquely determine the initial condition, the \textit{connectivity of the network has to somehow scale with the number of adversaries.} Using a more refined argument than the one we presented above, it is possible to show that a connectivity of $2f+1$ is necessary for the problem under consideration \cite{Sundaram2011,Pasqualetti123},  where $f$ is the maximum number of malicious nodes in the network. 

As shown in \cite{Sundaram2011}, a  connectivity of $2f+1$ is also sufficient for  the linear iterative strategy to reliably disseminate information between regular nodes in the network despite the actions of up to $f$ malicious adversaries.   

\begin{thm} [\cite{Sundaram2011}] 
Given a fixed network with $n$ nodes described
by a graph $\mathcal{G}=\{\mathcal{V},\mathcal{E}\}$, let $f$ denote the maximum number
of malicious nodes that are to be tolerated in the network, and let $\kappa_{ij}$ denote the size of the smallest $(i,j)$-cut between any two vertices $j$ and $i$. Then, regardless of the actions of the malicious nodes, node $i$ can uniquely determine all of the
initial values in the network via a linear iterative strategy if
and only if $\min_{j}\kappa_{ij}\geq 2f+1$. Furthermore, if this condition is
satisfied, $i$ will be able to recover the initial values after the nodes run the linear iterative strategy with almost any choice of weights for at most $n$ time-steps.
\label{thm:1disfunccalc}
\end{thm}

\iffalse
We rephrase the necessary condition in Theorem \ref{thm:1disfunccalc} in the following corollary. 

\begin{corollary}\label{cor:cnads9}
Given a fixed network with $n$ nodes described
by a graph $\mathcal{G}=\{\mathcal{V},\mathcal{E}\}$, if $\kappa_{ij}\leq 2f$ for some positive integer $f$, then then there exists a set $\mathcal{F}$ of $f$ malicious nodes along with a vector of additive errors $\boldsymbol \zeta_{\mathcal{F}}[t]$ such that
node $i$ cannot calculate any function that involves node $j$’s initial value (regardless of the number of time-steps for which the iteration is performed).
\end{corollary}
\fi

%\textbf{Discussion:} Note that the lower bound of $2f+1$ on the network connectivity is an \textit{algorithm-independent} fundamental bound for information dissemination, under the adversary model we consider here \cite{Lynch, Dolev, Hromkovic}. This  shows that linear iterative strategies are as powerful as any other strategy for information dissemination in the presence of malicious agents (under the wireless broadcast model of communication).

A key ingredient in the proof of Theorem \ref{thm:1disfunccalc} is 
establishing that if $\mc{G}$ is $(2f+1)$-connected, then the tuple 
 $(\bs{W},\bs{B}_{\mc{F}},\bs{C}_i)$ is strongly observable (i.e., does not possess any zero dynamics) $\forall i\in\mc{V}$, for almost all choices of the weight matrix $\bs{W}$, and under any $f$-total adversarial set $\mc{F}$.\footnote{Note that when $\mc{F}=\emptyset$ (i.e., in the absence of adversaries), we immediately recover that connectivity of $\mc{G}$ implies observability of $(\bs{W},\bs{C}_i), \forall i\in\mc{V}$, for almost all choices of the weight matrix $\bs{W}$.} This interdependence between system and graph-theoretic properties is schematically illustrated in Fig.~\ref{fig:by167rsvdyz}(c). 
 
One important point to note is that the approaches developed in \cite{Sundaram2011} and \cite{Pasqualetti123} to combat adversaries require each regular agent to possess complete knowledge of the network structure, and to perform a large amount of computation to identify the malicious sets. For large-scale networks, this may be infeasible. One may thus ask whether it is possible to resiliently diffuse information across a network when each regular agent only has \textit{local} knowledge of its own neighborhood, and can run only simple computations. In the subsequent sections, we provide an overview of work showing that this is indeed possible. However, as we shall see, the lack of global information will dictate the need for stronger requirements on the network topology (relative to $(2f+1)$-connectivity). 
 
\begin{rem}[\textbf{Byzantine Attacks}]
For point-to-point communications, which are prone to Byzantine attacks, in addition to the network connectivity, which has to be at least $2f+1$, the total number of nodes must satisfy $n\geq 3f+1$. This is because of the fact that if $i$ receives $j$'s value reliably,  it still does not know what $j$ told other nodes
in the network. Thus, there must be a sufficient number of non-Byzantine nodes in the network in order for $i$ to ascertain what $j$ told ‘most’ of the nodes \cite{Dolev}.
\end{rem}

\begin{example}
Consider the graph shown in Fig.~\ref{fig:bdistfunctcqdyz}(a). The objective is for node $ 1$ to calculate the  function $\sum_{i=1}^6\psi_i^2[0]$ despite the presence of a malicious node in the network, i.e, $f=1$.  
\begin{figure}[t!]
\centering
\includegraphics[scale=.48]{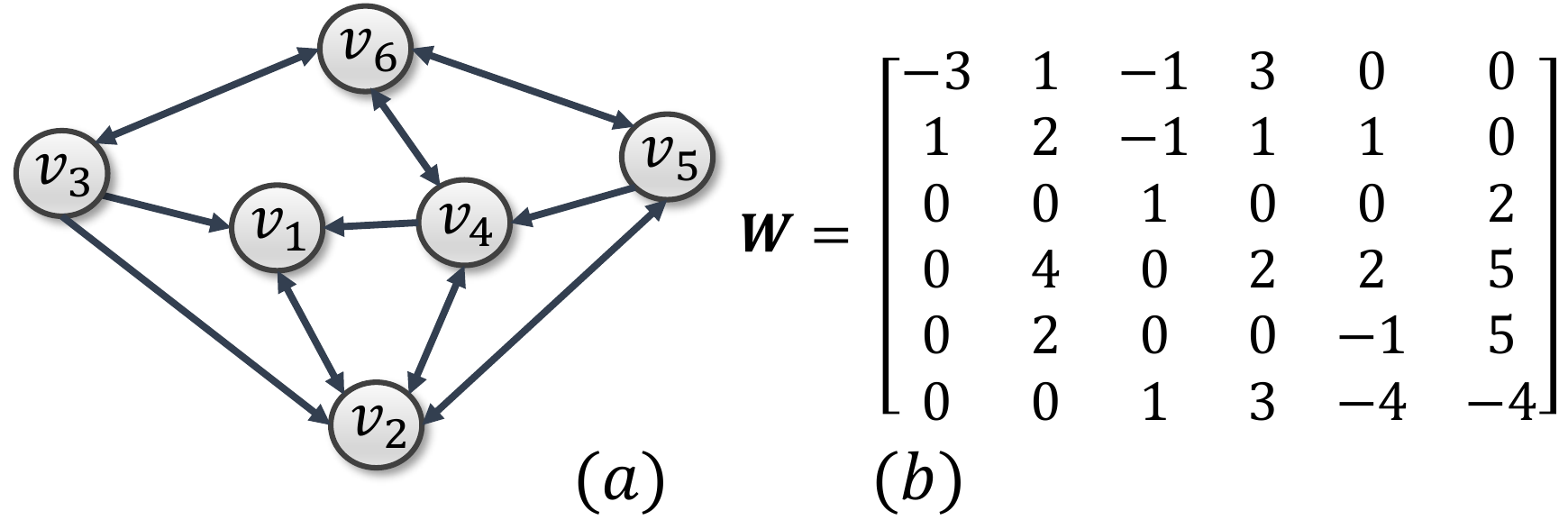}
\caption{(a) Graph discussed in Example, (b) The weight matrix corresponding to graph (a).}
\label{fig:bdistfunctcqdyz}
\end{figure}
In this graph, nodes $ v_2,v_3$, and $v_4$ are neighbors of $v_1$, and $v_5$ and $v_6$ have three internally
vertex-disjoint paths to $v_1$. Thus, $\kappa_{j1}\geq 3$ for all $j$ and, based on Theorem \ref{thm:1disfunccalc}, $v_1$ is able to calculate the desired function after running the linear iteration (with almost any choice of weights) for at most $n=6$ time-steps.  We take each of the edge and self loop weights to be i.i.d. random variables from the set $\{-5, -4, -3,-2,-1, 1, 2, 3, 4, 5\}$ with equal probabilities. These weights produce the matrix shown in Fig.~\ref{fig:bdistfunctcqdyz} (b). Since node $ 1$ has access to its own state and the states of its neighbors, we have $\boldsymbol C_1=[\mathbf{I}_4 \quad \mathbf{0}]$. Based on these values, matrices $\mathcal{O}_{1,L}$ and $\mathcal{M}_{1,L}^{\mathcal{F}}$ are obtained and $v_1$ can calculate the initial state vector $\boldsymbol \psi [0]$ using \eqref{eqn:oaisbv}. Suppose that the initial values of the nodes are $\boldsymbol\psi [0]=[3 \hspace{1.5 mm} -1 \hspace{1.5 mm} 4\hspace{1.5 mm} -4 \hspace{1.5 mm} 7 \hspace{1.5mm} 11]'$ and $v_4$ is a malicious node. At time steps 1 and 2, $v_4$ adds  an additive error of $\zeta_4[0]=-8$ and $\zeta_4[0]=-12$ to its updating rule. The values of all nodes over the first three time-steps of the linear iteration are given by $\boldsymbol \psi[0]=[3  \hspace{1.5 mm} -1 \hspace{1.5 mm} 4 \hspace{1.5 mm} -4 \hspace{1.5 mm}  7 \hspace{1.5 mm} 11]'$,  $\boldsymbol \psi[1]=[-26   \hspace{1.5 mm} 0  \hspace{1.5 mm} 26 \hspace{1.5 mm} 49 \hspace{1.5  mm} 46 \hspace{1.5 mm} -80]'$, and $\boldsymbol \psi[2]=[199 \hspace{1.5 mm} 43 \hspace{1.5 mm} -134 \hspace{1.5 mm} -222 \hspace{1.5  mm} -446 \hspace{1.5 mm} 309]'$. The values seen by $v_1$ at time-step $t$ are given by $y_1[t]=\boldsymbol C_1\boldsymbol \psi[t]$; node $v_1$ can now use $y_1[t]$ to calculate the vector of initial values, despite the efforts of the malicious node. Node $ v_1$ has to find a set $\mathcal{F}_i$ for which $y_1[0:2]$ falls into the column space of $\mathcal{O}_{1,2}$ and $\mathcal{M}_{1,2}^{\mathcal{F}}$. In this example, $v_1$ can figure out that this holds for $j=4$. Then, it finds vectors $\bar{\boldsymbol \psi}$ and $\boldsymbol \zeta_{\mathcal{F}_4}[0:1]$  such that $y_i[0:2]=\mathcal{O}_{1,2}\bar{\boldsymbol \psi}+\mathcal{M}_{1,2}^{\mathcal{F}_4}\boldsymbol{\zeta}_{\mathcal{F}_4}[0:1]$ as $\boldsymbol \zeta = [3 \hspace{1.5 mm} -1 \hspace{1.5 mm} 4\hspace{1.5 mm} -4 \hspace{1.5 mm} 7 \hspace{1.5mm} 11]'$ and $\boldsymbol \zeta_{\mathcal{F}_4}[0:1]=[-8 \hspace{2mm} -12]'$. Node $v_1$ now has access to $\boldsymbol\psi[0]=\bar{\boldsymbol\psi}$ and can calculate $\sum_{i=1}^6\psi_i^2[0]=212$. 

It is worth noting that for the network in Fig.~\ref{fig:bdistfunctcqdyz}(a), we have $\kappa_{26}=2$, since  the set $\mc{F} = \{v_4, v_5\}$ forms a $(2, 6)$-
cut (i.e., removing nodes $ v_4$ and $v_5$ removes all paths from $v_2$ to $v_6$). Thus, node $v_6$ is not
guaranteed to be able to calculate any function of node $v_2$’s value when there is a faulty node in the system. In particular, one can
verify that in the example above, where node $v_4$ is malicious
and updates its values with the errors $\zeta_4[0]=-8$ and $\zeta_4[0]=-12$, the values seen by $v_6$ during the first three time steps of the linear iteration are the same as the values seen by node $v_6$ when $\psi_1[0]=4$, $\psi_1[0]=-3$ and node $ 5$ is
malicious, with $\psi_5[0]=4$ and $\psi_5[0]=6$. In other words node $v_6$ cannot distinguish the case when node $v_4$ is faulty from the case where node $v_5$ is faulty (with different initial values in the network). 
\end{example}

\iffalse
The detector uses the linear filter
\begin{align}
\hat{\boldsymbol{\psi}}[t+1]&=(\boldsymbol{W}-\boldsymbol K\boldsymbol C\boldsymbol{W})\hat{\boldsymbol{\psi}}[t]+\boldsymbol K\boldsymbol{y}[t+1],\nonumber\\
\boldsymbol{z}^a[t]&=\boldsymbol{y}[t]-\boldsymbol C\boldsymbol W\hat{\boldsymbol{\psi}}[t-1],
\label{eqn:4}
\end{align}
where $\boldsymbol K$ is the observer gain matrix, $\boldsymbol C=[\mathbf{e}_{i_1}', \mathbf{e}_{i_2}', ..., \mathbf{e}_{i_m}']'$ is the output matrix representing the locations of $m$ sensors, and $\boldsymbol{z}^a[t]$ is called the residue.

An attack is called {\it undetectable} or {\it perfect} if the residues under normal operation and during an attack are the
same, i.e., $\Delta\boldsymbol{z}[t]\triangleq\boldsymbol{z}[t]-\boldsymbol{z}^a[t]=0$, in which case the centralized detector cannot distinguish an
attack from normal operation. 

We assume that  centralized detector observes sensor measurements and is aware of matrices $\boldsymbol{W}$ and $\boldsymbol C$ in \eqref{eqn::OAEGI0}, but not $\boldsymbol B_{\mathcal{F}}$, i.e., the attacker's strategy.
\fi

\section{ Resilient Distributed  Consensus}
\label{sec:rdistcons}

Distributed consensus is a well studied application of information diffusion in networks.  In distributed consensus, every node in the network
has some information to share with the others, and the
entire network must come to an agreement on an appropriate 
function of that information \cite{Olfati1,Olfati2,Jadbaba, Sundaram}. In the resilient version of distributed consensus, the algorithm has to be modified in such a way that it maintains the consensus value in a desired region despite the actions of adversarial agents who attempt to steer the states outside that region (or disrupt agreement entirely). The desired steady state value can vary according to the application of interest \cite{piranicdc, Ghaderi13}. 

\begin{rem}[\textbf{A Fundamental Limitation}]\label{rem:fundlimit}
In the standard linear consensus dynamics, a single malicious agent (shown in black in Fig.~\ref{fig:cha0s82}) can drive the consensus value towards its own state simply by keeping its value constant,\footnote{These types of adversarial agents are referred to as {\it stubborn} agents in the literature \cite{Ghaderi13, Ozdaglar, ACC}.} as shown in 
Fig.~\ref{fig:cha0s82}. More generally, since the initial values of the nodes are assumed to be known only to the nodes themselves, an adversarial node can simply change its own initial value and participate in the rest of the algorithm as normal. This would allow the adversary to affect the final consensus value (through its modified local value), but never be detected. Thus, perfect calculation of any function of initial values is generally impossible under adversarial behaviour. This is a fundamental limitation of all distributed algorithms for any problem where each agent holds data that is required by others to compute their functions (e.g., as in consensus, function calculation, or distributed optimization), as will be discussed later.
\begin{figure}[t!]
\centering
\includegraphics[scale=.44]{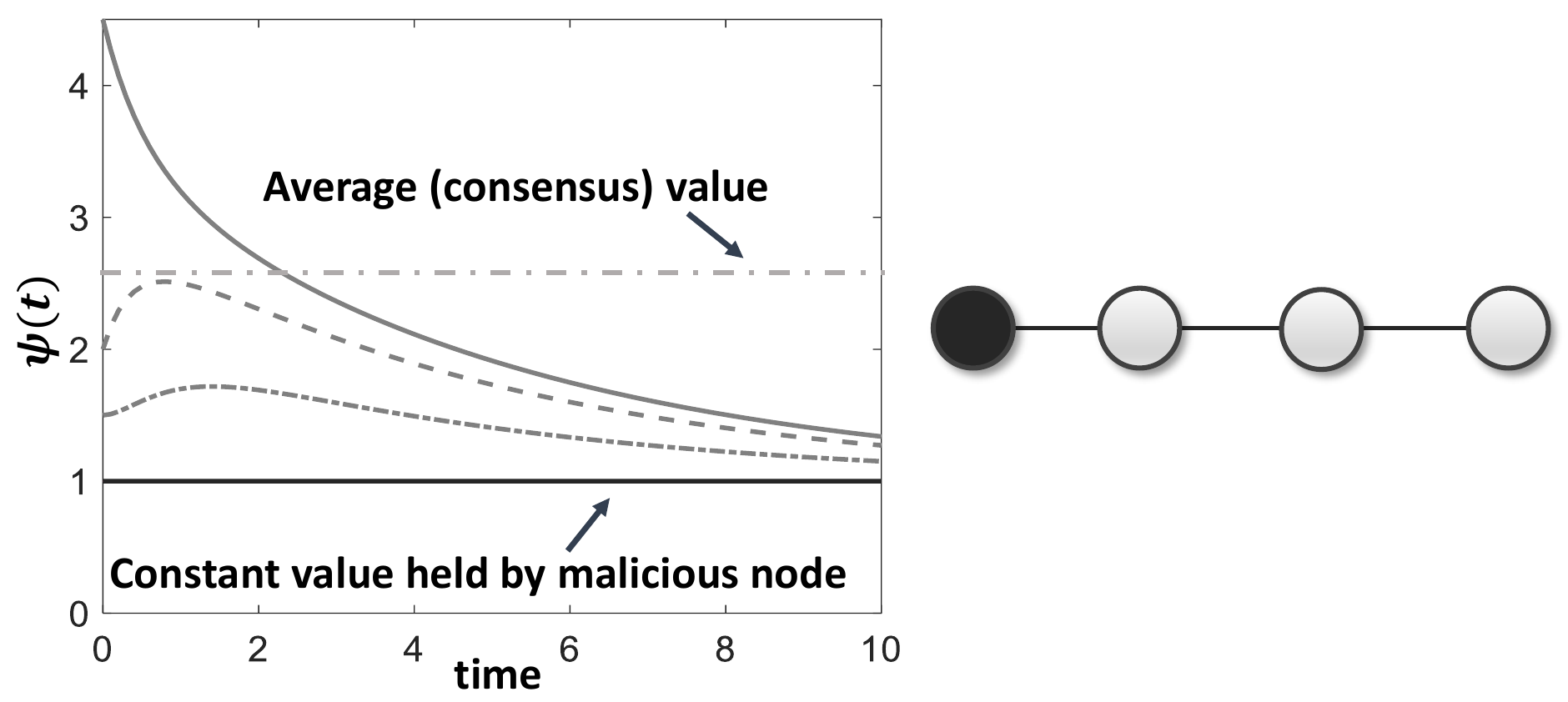}
\caption{State trajectories of agents in the presence of a stubborn agent.}
\label{fig:cha0s82}
\end{figure}
\end{rem}

\iffalse

Linear consensus is a special case of the linear iteration rule \eqref{eqn:lineariteration} with the following additional conditions 
\begin{itemize}
    \item  $\sum_{j=1}^n w_{ij}=1, \quad \forall i \in \mathcal{V},  t\in \mathbb{Z}_{\geq 0}$;
    
    \item $w_{ij}[t]=0$ \quad whenever \quad $j \notin \mathcal{N}_i[t], t\in \mathbb{Z}_{\geq 0}$;  \hspace{8mm} \boldsymbol{(\bigg\ast)}
    
    \item  $w_{ij}[t]\geq \alpha$ for some $\alpha >0$ whenever  $j \in \mathcal{N}_i[t], t\in \mathbb{Z}_{\geq 0}$. 
\end{itemize}
In addition to the above conditions, the underlying graph must have a certain structure for reaching  asymptotic consensus. In particular, for time-invariant networks, the above algorithm reaches asymptotic consensus if and only if the underlying graph $\mathcal{G}$ has a rooted directed spanning tree. For the cases where $\mathcal{G}$ is time varying, there is a sufficient condition in which there is a uniformly bounded sequences of continuous time intervals such that the union of digraphs across each interval has a rooted directed spanning tree \cite{ren2}.

The model of attack signal is the same as  \eqref{eqn::OAEGI0} in which the elements of  matrix $\boldsymbol W$  satisfy conditions ${(\boldsymbol{\ast})}$. We introduce two approaches for resilient  distributed consensus.

\fi

\subsection{Classical Approaches}
\label{sec:jasd98}
Classical results on distributed consensus in the presence of Byzantine agents date back to the computer science literature \cite{Dolev2} showing that the regular agents can always reach a consensus if and only if (1) the number of Byzantine agents is less than $\frac{1}{2}$ of the network connectivity, and (2) less than $\frac{1}{3}$ of the total number of agents. Similar results have been derived to detect and identify malicious nodes in consensus dynamics, as discussed in the previous section  \cite{Sundaram2011, Pasqualetti123}. However, these works require each regular node to have full knowledge of the network topology, and for each regular node to perform complex computations.  The following subsection describes an alternative scalable and ``purely local" method for resilient consensus. 

\iffalse

\subsection{Attack Identification Based Approach}
This method is based on the fact that an attacked detection and isolation module  has to be run concurrently with the consensus dynamics. Based on this idea, a malicious behaviour has to be  detected and identified in order to be  avoided in the  consensus dynamics. Thus, the attack detection scheme discussed in Section XX is specifically applied to consensus dynamics, i.e., matrix $A$ in \eqref{eqn:3} has to satisfy the consensus weight conditions mentioned above.

The communication is assumed to be broadcast and, thus, each agent sends the same values to its neighbors. The goal is to provide graph-theoretic conditions for identification of malicious agents. 
\begin{thm}
Given consensus dynamics \eqref{eqn::OAEGI0} with weights satisfying conditions $(i)$ to $(iii)$, then 
up to $k$ malicious agents can be detected and identified if and only if the network is $2k+1$ connected.
\end{thm}

Similar to attack detection discussed for general linear systems in Section XX, detection and identification of attacks in consensus dynamics are generic, in which the attack is identifiable for almost all choices of weight matrix $\boldsymbol{W}$ satisfying the consensus conditions.

\fi

\iffalse
We need the following definitions.
\begin{definition}
For a linear consensus system of the form \eqref{eqn::OAEGI0}, the input $\boldsymbol{\zeta}(t)$ introduced by a set $|\mathcal{F}|$ of misbehaving agents is undetectable if 
...
and it is said to be unidentifiable if
...
\end{definition}
\fi

\subsection{Purely Local Approaches}

\label{sec:wmsr}

By imposing stronger conditions on the network topology (beyond being just $(2f+1)$-connected), one can formulate algorithms that can handle worst case $f$-local Byzantine attacks with much less computational cost. In this class of algorithms, which were first named {\it approximate agreement} \cite{Dolev2}, each node
disregards the largest and smallest $f$ values received from its neighbors at each iteration and updates its state to be the average of a carefully chosen subset of the remaining values (such quantities are known as {\it trimmed means} in the robust statistics literature \cite{huber72}). These methods were extended to a class of algorithms named  {\it Mean-Subsequence-Reduced (MSR)} algorithms \cite{Azad}. In \cite{leblank}  a continuous-time
variation of the MSR algorithms, named the Adversarial
Robust Consensus Protocol (ARC-P) was proposed. In what follows, we discuss an extension of MSR algorithms, called Weighted-Mean-Subsequence  Reduced (W-MSR) in \cite{Hogan}, which can  handle $f$-local adversarial agents.

The algorithm is as follows.
\begin{itemize}
    \item [1)] Let $\psi_i[t]\in \mathbb{R}$ be the value maintained at each time step by each regular node $i$. At each time-step $t$, each regular node $i$ receives values $\psi_i[t],$ $j\in \mathcal{N}_i$ from all of its neighbors, and ranks them from largest to smallest.
    \item[2)]  If there are $f$ or more values larger than $\psi_i[t]$, normal
node $i$ removes the  $f$ largest values. If there are fewer
than $f$ values larger than $\psi_i[t]$, regular node $i$ removes all of these larger values. This same logic is applied to the smallest values in regular node $i$’s neighborhood. Let $\mathcal{R}_i[t]$ denote the set of nodes whose values were
removed by $i$ at time-step $t$.
\item[3)] Each regular node $i$ updates its value as
\begin{equation}
\psi_i[t+1]=w_{ii}[t]\psi_i[t]+ \sum_{j\in \mathcal{N}_i[t]\setminus{\mathcal{R}_i[t]}}w_{ij}[t]\psi_j[t], 
\label{eqn:lineariterconsensusation}
\end{equation}
\end{itemize}
where $w_{ii}[t]$ and $w_{ij}[t]$ satisfy the following conditions:
\begin{itemize}
    \item  $\sum_{j=1}^n w_{ij}=1, \quad \forall i \in \mathcal{V},  t\in \mathbb{Z}_{\geq 0}$;
    
    \item $w_{ij}[t]=0 \hbox{ whenever } j \notin \mathcal{N}_i[t], t\in \mathbb{Z}_{\geq 0}$;  \hfill 
    
    \item  $w_{ij}[t]\geq \alpha$ for some $\alpha >0$ whenever  $j \in \mathcal{N}_i[t], t\in \mathbb{Z}_{\geq 0}$. 
\end{itemize}

We call the largest number of values that each node could
throw away the {\it parameter} of the algorithm (it is equal to $2f$ in the above algorithm). Before presenting conditions for resilient consensus, in the following example we show that large network connectivity is no longer sufficient to guarantee consensus under W-MSR algorithms.
\begin{example}
 In the graph shown in Fig.~\ref{fig:by9qwfqdyz} (a), suppose that the initial value of nodes in set $\mathcal{V}_1$ and set $\mathcal{V}_2$ are zero and 1, respectively. For  $f=1$, if each node disregards the largest and smallest values in its neighborhood, then the value of nodes in sets $\mathcal{V}_1$ and $\mathcal{V}_2$ remains the same as their initial value for all $t\geq 0$. As a result, consensus will not be achieved even though there are no malicious nodes. This lack of consensus is despite the fact that the connectivity of the graph is $\frac{n}{2}$, and arises due to the fact that the local state-dependent filtering in W-MSR causes all nodes between sets $\mathcal{V}_1$ and $\mathcal{V}_2$ to be disconnected at each iteration. 
\end{example}

Although the network connectivity is no longer an appropriate metric for analyzing the resilience of W-MSR dynamics, the notion of graph robustness from \cite{zhang2012robustness, Hogan}  (see Section \ref{sec:connectivity}) turns out to be the key concept. We first start with the following concept. Denote the maximum and minimum values of the normal
nodes at time-step $t$ as $M[t]$ and $m[t]$.

\begin{definition}[\textbf{$f$-local safe}] Under the $f$-local adversarial model, the W-MSR algorithm is said to be $f$-local safe if both of the following conditions are satisfied: (i) all regular nodes reach consensus for any choice of initial values, and (ii) the regular nodes’ values (including the final consensus value) are always in the range $\big[m[0],M[0]\big]$. 
\end{definition}

The following result provides conditions under which the
W-MSR algorithm guarantees (or fails) to be $f$-local safe. 

\begin{thm} [\cite{Hogan}] 
Under the $f$-local Byzantine adversary model, the W-MSR algorithm with parameter $2f$ is $f$-local safe if the
network $\mathcal{G}$ is $(2f + 1)$-robust. Furthermore, for any $f>0$, there exists a $2f$-robust
network which fails to reach consensus based on the W-MSR
algorithm with parameter $2f$.
\label{thm:j9f8dsh}
\end{thm}

%{\tb Mentione that those results are extended to second order conesnsus by Mehran.}

As discussed in Section \ref{sec:connectivity}, the robustness condition used in Theorem \ref{thm:j9f8dsh} is much stronger than the network connectivity condition which was required in classical distributed consensus algorithms. However, this  stronger  condition can be considered as the price to be paid for a computationally tractable resilient consensus algorithm which is able to tolerate worst-case Byzantine attacks. Furthermore, under this condition, one gains the ability to tolerate $f$-local adversaries (rather than $f$-total adversaries) ``for free". 

\iffalse
\begin{example}
An example of a $2f$-robust graph mentioned in Theorem  \ref{thm:j9f8dsh} which fails to reach a consensus is shown in Fig.~\ref{fig:cha0s82} (right). There, $S_1, S_2$ and $S_3$ are all complete components with $|S_1|=|S_3|=2f$, $|S_2|=4f$. Each node in $S_1$ connects to $2f$ nodes of $S_2$ and each node in $S_3$ connects to
the other $2f$ nodes of $S_2$, and all these connections are undirected. Nodes $a$ and $b$ have incoming edges
from all nodes in $S_1$ and $S_3$, respectively.  It can be verified that this graph is $2f$-robust. 
We choose $f$ nodes of $S_1$ and also $f$ nodes of $S_3$ to
be malicious; note that this constitutes an $f$-local set of malicious nodes. Then we assign node $a$
with initial value $m$, node $b$ with initial value $M$ and the other regular nodes with initial values
$c$, such that $m<c<M$. Malicious nodes in $S_1$ and $S_3$ will keep their values unchanged at $m$
and $M$, respectively. We can see that, by using the W-MSR algorithm, the values of nodes $a$
and $b$ will never change and thus consensus cannot be reached.
\end{example}
\fi

Several other variations of the above approach, including extensions to second order consensus with asynchronous time delay and applications to formation control of mobile robots, are discussed in \cite{dibajiautomatica, dibajitac, Kelsey}. Applications of resilient consensus in multi-robot systems using Wi-Fi communication is studied in \cite{gil}. 
Resilient flocking in multi-robot systems requires the extension of the above techniques to time varying networks, which is studied in \cite{saldana}. There, it is shown that if the required network robustness condition is not satisfied at all times, the network can still reach resilient consensus  if the union of communication graphs over a bounded period of time satisfies $(2f+1)$-robustness. Moreover, a control policy to attain such resilient behavior in the context of perimeter surveillance with a team of robots was proposed. 

\subsection{Resilient Vector Consensus}
\label{subsec:resilient_vector_consensus}

\begin{figure}[t!]
\centering
\includegraphics[scale=.59]{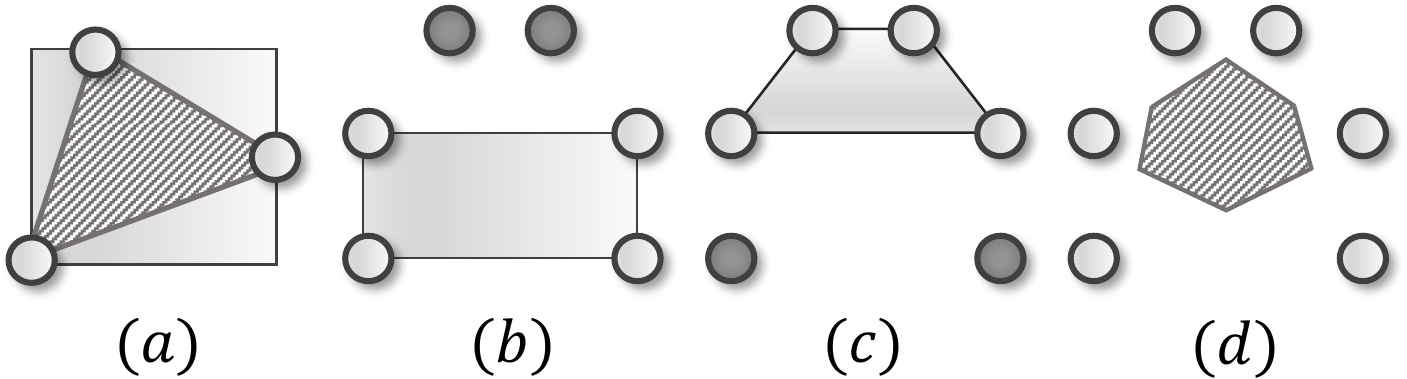}
\caption{(a) The box over-estimating the convex-hull of nodes in $\mathbb{R}^2$. (b),(c) show non-intersecting convex hull of the regular nodes. (d) the dark region is the centerpoint region of all six nodes. }
\label{fig:cha0scovexhull}
\end{figure}

The W-MSR algorithm described in the previous section considered the case where agents maintain and exchange scalar quantities and  remove ``extreme" values at each iteration. However, the extension to  multi-dimensional vectors requires further considerations since there may not be a total ordering among vectors. One option is to simply run W-MSR on each component of the vector separately. If the graph is $(2f+1)$-robust, this would guarantee that all regular agents reach a consensus in a hypercube formed by the initial vectors of the regular agents (despite the action of any $f$ local set of Byzantine agents). Keeping the agents' states within the convex hull of the initial vectors, however, requires some subtleties, as we will discuss in this section. We know that  the convex hull of a set of vectors is a subset of the region (the box) formed by the convex hull of their components separately. This is shown in Fig.~\ref{fig:cha0scovexhull} (a) in which the triangle is the convex hull of the three points in $\mathbb{R}^2$ and the grey rectangle is the box formed by calculating the convex hull of each component of the three vectors separately. Thus, the component-wise convex hull gives an overestimate of the actual convex hull of the vectors. 
 
First attempts to address Byzantine resilient vector consensus to the convex hull of the initial values of the regular agents were provided in \cite{vaidya, vaidya2},  and were further developed in the context of rendezvous  multi robot systems by \cite{park}. To describe these approaches, we first need to extend the notion of $f$-local safe points.

\begin{definition}[\cite{park}]
Given a set of $n$ nodes in $\mathbb{R}^d$
of which at most $f$ are adversarial, a point $p$ that is guaranteed to lie in the interior of the convex hull of $(n-f)$ regular points is called an $f$-safe point.  
\end{definition}

Based on the above definition, the resilient vector consensus algorithm relies on the computation of $f$-safe points by each node. In particular, 
\begin{itemize}
    \item  Let $\bs {\psi}_i[t]\in \mathbb{R}^{d}$ be the value maintained at each time step by each regular node $i$.  In the iteration $t$, a regular node $i$ gathers the state values of its neighbors $\mathcal{N}_i[t]$. 
    \item Each regular node $i$ computes an $f_i$-safe point, denoted by $s_i[t]$, of points corresponding to its neighbors’ states.
    \item Each regular agent $i$ then updates its state by moving toward the safe point $s_i[t]$, i.e.,
    \begin{equation}
       \boldsymbol \psi_i[t+1]= \alpha_i[t] s_i[t]+\left(1-\alpha_i[t]\right)\boldsymbol \psi_i[t],
    \end{equation}
where $\alpha_i[t]\in (0,1)$ is a dynamically chosen parameter whose value depends on the application. 
\end{itemize}

It was shown in \cite{park} that if all regular agents follow the above routine, they are guaranteed to converge to some point in the convex hull of their initial states. The following proposition shows conditions of the existence of $f$-safe points. 

\begin{prop}[\cite{park}]
Given $n$ points in $\mathbb{R}^d$, where $d\in \{1,2,..., 8\}$, and at most $f$ points belong to adversaries, then there exists an $f$-safe point if $n\geq (f+1)(d+1)$. It also holds for $d>8$ if Reay’s conjecture is true \cite{Reay}.
\label{prop:vna089h}
\end{prop}

In \cite{waseem2},  it is shown that $n\geq (f+1)(d+1)$ is also a necessary condition for the existence of an $f$-safe point. An example is shown in Fig.~\ref{fig:cha0scovexhull} (b) and (c). The malicious nodes are shown with darker colors.  Here, for $f=2$ malicious nodes, there is no 2-safe point (an interior point in the intersection of convex hull of four regular nodes).

The main question is how to find these $f$-safe points from a given set of points in  $\mathbb{R}^d$. In \cite{vaidya, park}, this is done via the {\it Tverberg partitioning algorithm}, which partitions points into subsets such that the convex hull of the partitions have a non-empty intersection, provided that the number of nodes is sufficiently large; see \cite{Tverberg2} for more details. A similar approach is recently adopted in \cite{yilinmo}. However, finding Tverberg partitions is computationally hard in practice (although it is not proved that the problem is NP hard). To achieve fast algorithms, one has to pay the price of reducing the number of parts in the partition, i.e., the number of malicious nodes.  Linear time approximation algorithms to find Tverberg points, i.e., $f$-safe nodes, have been proposed in  \cite{Mulzer}, provided that $f\leq \lceil \frac{n}{2^d}\rceil-1$.  On the other hand, Tverberg partitioning  provides strong conditions for $f$-safe points, i.e., the outcome of the Tverberg partitioning algorithm are $f$-safe points, but the reverse is not true.  A less conservative  approach to find $f$-safe points via the notion of a centerpoint was developed in \cite{waseem}, as explained below.

\begin{definition}
Given a set $X$ of $n$ points in $\mathbb{R}^d$, a centerpoint $p$ is a point, not necessarily
from X, such that any closed half-space\footnote{A closed half-space in $\mathbb{R}^d$ is a set of the form $\{x \in \mathbb{R}^d : a' x \geq  b\}$ for some $a\in \mathbb{R}^d\setminus\{0\}$.}of $\mathbb{R}^d$ containing $p$ also contains at least $\frac{n}{d+1}$ points from $X$.
\end{definition}

By the Centerpoint Theorem, every finite set of points in $\mathbb{R}^d$ has a centerpoint \cite{Matousek}. It is shown that for a set of $n$ points in $\mathbb{R}^d$ and $f\leq \lceil\frac{n}{d+1}\rceil-1$, the region of
$f$-safe points is {\it equivalent to} the centerpoint region \cite{waseem}.  
The centerpoint of six points in shown in Fig.~\ref{fig:cha0scovexhull} (d). The following result provides conditions for  finding  $f$-safe points. 

 \begin{prop}[\cite{waseem}]
Given $n$ points in $\mathbb{R}^d$ for which at most $f$ points belong to adversaries, then an $f$-safe point can be computed (using centerpoint) if 
\begin{align}
f&\leq \lceil\frac{n}{d+1}\rceil-1 \quad d=2,3,\nonumber\\
f&\leq \lceil\frac{n}{d^2}\rceil-1 \hspace{8mm} d>3.
\end{align}
\label{prop:n0va8gb}
\end{prop}

\iffalse
finding $f$-safe points using Tverberg partitioning, the complexity of finding $f$-safe points using via centerpoints is polynomial in $n$.
Finding safe points using Tverberg partitioning, Proposition \ref{prop:vna089h} requires $n$ to scale linearly with $d$, but in  \ref{prop:n0va8gb} it scales quadratically with $d$. This can be considered as the price to pay for using centerpoint theorem which is less computational compared to Tverberg partitioning.
\fi

\section{Resilient Distributed Optimization}
\label{sec:rdist_opt}
While the consensus problem discussed in the previous section considered the scenario where each agent has a static initial value, a more general setting is that of distributed optimization.
In this setting, each agent $i \in \mathcal{V}$ has a convex function $g_i:\mathbb{R}\to \mathbb{R}$ (with bounded subgradients) which is only available to agent $i$.\footnote{We discuss the case of multi-dimensional functions later in this section.} The objective is
for the agents to solve the following global optimization problem in a distributed manner:
\begin{equation}\label{eqn:1do}
\underset{\psi}{\text{min}}\hspace{2mm} g(\psi)=\frac{1}{n}\sum_{i=1}^ng_i(\psi).
\end{equation}
A common approach to solve this problem is to use a synchronous iterative consensus-based protocol in which agents use a combination of consensus dynamics and gradient flow to find a minimizer of $g(\psi)$ \cite{nedicc}. More specifically, at every time-step $t\in \mathbb{N}$, each agent $i$ maintains an estimate $\psi_i(t)\in \mathbb{R}$ of the solution to \eqref{eqn:1do}, and updates it based on the information received from its neighbors, as follows
\begin{equation}\label{eqn:2do}
\psi_i[t+1]=w_{ii}[t]\psi_i[t]+\sum_{j\in \mathcal{N}_i[t]} w_{ij}[t]\psi_j[t]-\alpha_t d_i[t].
\end{equation}
In the above update rule, $d_i[t]$ is a subgradient of $g_i$ evaluated at $w_{ii}[t]\psi_i[t]+\sum_{j\in \mathcal{N}_i[t]} w_{ij}[t]\psi_j[t]$,  and $\alpha_t$ is the step size sequence corresponding to the influence of the subgradient on the update rule at each time-step. Dynamics \eqref{eqn:2do} can be represented in the following vector form as
\begin{equation}
   \boldsymbol \psi[t+1]=\boldsymbol W\boldsymbol\psi[t]-\alpha_t\boldsymbol d[t],
\end{equation}
where $\boldsymbol W$ is a doubly-stochastic matrix. The following result shows that the update rule \eqref{eqn:2do} allows the
nodes in the network to distributively solve the global optimization problem \eqref{eqn:1do}. 
\iffalse
We can easily observe that 
\begin{align}\label{eqn:3do}
    x[t+1]&=A[t]A[t-1]...A[0]x[0]\nonumber\\
    &-\sum_{s=1}' A[t]A[t-1]...A[s] \alpha_{s-1}d[s-1]-\alpha_td[t],
\end{align}
Here, we define $\Phi(t,s)\triangleq A[t]A[t-1]...A[s]$ for $t\geq s$ as  the state matrix for which  $\Phi(t,s)\triangleq 0$ when $t< s$. Hence, \eqref{eqn:2do}
becomes
\begin{equation}
x[t+1]=\Phi(t,0)x[0]-\sum_{s=1}' \Phi(t,s) \alpha_{s-1}d[s-1]-\alpha_td[t].
\end{equation}
\fi
\iffalse
\newline
\textbf{Assumption 1 (Lower Bounded Weights):} There exists $\eta>0$ such that for all $t\in \mathbb{N}$ and $i\in \mathcal{V}$, if $j\in \{i\}\cup \mathcal{N}_i$, then $w_{ij} \geq \eta$.
\newline
\textbf{Assumption 2 (Double Stochasticity):} For all $t\in \mathbb{N}$ and $i\in \mathcal{V}$, the weights satisfy $w_{ii}[t]+\sum_{j\in \mathcal{N}^{\rm in}[t]} w_{ij}[t]=1$.
\fi
\begin{prop}[\cite{nedicc}]
Suppose that $\mathcal{G}$ is strongly connected at each time step and that the subgradients of each of
the local functions $g_i$ are bounded. For the update rule \eqref{eqn:2do} with step sizes satisfying $\sum_{t\in \mathbb{N}}\alpha_t=\infty$ and $\sum_{t\in \mathbb{N}}\alpha_t^2<\infty$, we have 
\begin{equation}
    \lim_{t\to \infty} \|\psi_i[t]-\psi^*\|=0,
\end{equation}
for all $i\in \mathcal{V}$, where $\psi^*$ is the global minimizer of $g(\psi)$.
\end{prop}

Our objective here is to summarize the vulnerabilities of such protocols to adversarial agents, and to provide an overview of secure distributed optimization algorithms that have provable safety guarantees despite the presence of such agents. Recent works have focused on some gradient-based metrics to detect and identify malicious agents in a distributed optimization algorithm \cite{Scagione}. However, similar to previous sections, our focus is on graph-theoretic methods. As before, we assume that adversarial nodes can update their states in a completely arbitrary manner. 

\begin{rem}[\textbf{Fundamental Limitation Revisited}]
Following the fundamental limitation discussed in Remark \ref{rem:fundlimit}, one can easily argue that it is generally impossible to compute $\psi^*$ when there are adversarial agents in the network, since one can never infer their local functions accurately.  
As an example, suppose that node $ n$ is adversarial and wishes the states to converge to $\bar{\psi}\in \mathbb{R}$. It simply chooses a function $\bar{g}(\psi)$ such that the minimizer of $\sum_{i=1}^{n-1}g_i(\psi)+\bar{g}_n(\psi)$ is $\bar{\psi}$.  For a vanishing step size, i.e., $\lim_{t\to \infty}\alpha_t=0$, all regular nodes will asymptotically converge to $\bar{\psi}$ when following the distributed optimization dynamics \eqref{eqn:2do}. Since the functions $g_i$ are arbitrary and known only to
the nodes themselves, such  deceptions cannot be detected.
\end{rem}

The above fact is formally stated as follows.

\begin{thm}\cite{bahman,lilisu}
Suppose $\Gamma$ is a distributed algorithm that guarantees that all nodes calculate the global optimizer of \eqref{eqn:1do}. Then a single adversary can
cause all nodes to converge to any arbitrary value when they
run algorithm $\Gamma$, and furthermore, will remain undetected.
\label{thm:dj9w}
\end{thm}

Theorem \ref{thm:dj9w} indicates that it is impossible to develop an algorithm that always finds optimal solutions in the absence of adversaries and that is also resilient to carefully crafted attacks. Thus, the price that should be paid for resilient distributed optimization is a {\it loss in optimality}. In what follows, we describe  resilient consensus-based distributed optimization protocols.

Suppose the adversarial nodes form an $f$-local set. At each time step, every regular node gathers and sorts the states of all of its neighbors and, similar to the W-MSR algorithm in Section \ref{sec:rdistcons}, each agent disregards the highest $f$ and lowest  $f$ states from the gathered states, denoted by set ${\mathcal{R}_i[t]}$, and updates its state as 
\begin{equation}
\psi_i[t+1]=w_{ii}[t]\psi_i[t]+\sum_{j\in \mathcal{N}_i[t]\setminus{\mathcal{R}_i[t]}} w_{ij}[t]\psi_j[t]-\alpha_t d_i[t],
\label{eqn:lfdist}
\end{equation}
where $d_i[t], \alpha_t$ are the same as \eqref{eqn:2do}.

In \cite{bahman},  graph-theoretic conditions for agents to reach consensus in the presence of $f$-total malicious nodes under the linear filtering rule in \eqref{eqn:lfdist} are discussed.
The arguments rely on the fact that in $(2f+1)$-robust networks, the weight matrix $\boldsymbol W$ corresponding to the regular nodes is rooted at each time-step, i.e., there is a  node with a directed path to every other node in $\mathcal{G}$. Other sufficient conditions (similarly assuming the existence of rooted nodes in  the set of regular nodes $\mathcal{R}$) is presented in \cite{lilisu}.

\iffalse
The following two theorems from \cite{bahman} show   a graph-theoretic necessary and sufficient conditions for consensus in the presence of $f$-total malicious adversaries. 
\begin{thm} [\textbf{Reaching Consensus}]
Consider the network $\mathcal{G}=(\mathcal{V},\mathcal{E})$, with an $f$-total set of malicious nodes $\mathcal{F}$. Suppose the network is $(2f+1)$-robust, that the functions $g_i$, $i\in \mathcal{R}$, have subgradients bounded by some constant $L$, and
that the regular nodes run \eqref{eqn:lfdist}. Further suppose that $\lim_{t \to \infty} \alpha_t=0$. Then the regular nodes are guaranteed to reach consensus despite the actions of the adversaries, initial values, and local functions if and only if the graph is $(f+1,f+1)$-robust.
\end{thm}
\fi

The following result shows that \eqref{eqn:lfdist} provides a safety guarantee for distributed optimization, i.e., convergence to the convex hull of local minimizers, under additional conditions on the step size $\alpha_t$. 

\begin{thm} [\cite{bahman}]
Suppose that one of the following conditions
holds: 
\begin{itemize}
\item[(i)] The adversarial nodes are $f$-total malicious and the network is $(f+1,f+1)$-robust; or

\item[(ii)] The adversarial nodes are $f$-local Byzantine and the network is 
$(2f+1)$-robust.
\end{itemize}
Furthermore, for each node $i\in \mathcal{R}$, let
the local function $g_i(.)$ have  minimizer $m_i$. 
Define $\bar{M}=\max\{m_i|i\in \mathcal{R}\}$ and $\underbar{M}=\min\{m_i|i\in \mathcal{R}\}$. If stepsizes $\alpha_t$ satisfy $\sum \alpha_t=\infty$ and $\lim_{t \to \infty} \alpha_t=0$, then $\lim\sup_{t \to \infty} \psi_i[t]\leq \bar{M}$ and $\lim\inf_{t \to \infty} \psi_i[t]\geq \underbar{M}$ for all $i \in \mathcal{R}$, regardless of the
actions of the adversarial nodes and the initial values.
\end{thm}

\begin{rem}[\textbf{Lack of Convergence}]
Despite the fact that the above resilient distributed optimization technique guarantees that the optimizer remains in the convex hull of local optimizers, it does not guarantee  convergence to a {\it constant} value under certain type of adversarial actions and specific classes of step sizes. See examples in \cite{bahman}. 
\end{rem}

\subsection{Factors that Affect the Performance of Resilient Distributed Optimization Algorithms}

The following example from \cite{bahman} shows that under the  dynamics \eqref{eqn:lfdist}, the nature
of the individual optimization functions together with the
network topology determine how far away
the convergence point is from the minimizer of the average
of the regular nodes’ functions. 
\begin{example}
Consider network  $\mathcal{G}=\{\mathcal{V},\mathcal{E}\}$ which is $2f+1$ robust and let $\mathcal{T}\subset \mathcal{V}$ be an  $f$-local set.  Suppose
all nodes are regular. Pick an $a\in \mathbb{R}$ and let nodes in $\mathcal{T}$ have local functions $g_a(\psi)=(\psi-a)^2$ and the nodes in $\mathcal{V}\setminus{\mathcal{T}}$ have local functions $g_b(\psi)=\psi^2$ (both
functions can be modified to have their gradients capped at
sufficiently large values, so as to not affect the minimizer of any convex combination of the functions). Let $g(\psi)$ be the
average of all of the functions, with minimizer $\psi^*=\frac{|\mathcal{T}|}{n}a$. Then, under the local filtering dynamics with parameter $f$,
all nodes converge to the value $\bar{\psi}=0$ and thus $\bar{\psi}-\psi^*=\frac{|\mathcal{T}|}{n}a$ and $g(\bar{\psi})-g(\psi^*)=\frac{|\mathcal{T}|^2}{n^2}a^2$.
\end{example}

The above result shows that  if the network contains a large $f$-local set (in relation to the total
number of nodes) or the local functions have minimizers that
are very different (corresponding to a large $|a|$ in the above
result), then the value computed by \eqref{eqn:lfdist} will
have a greater divergence from the globally optimal solution.
Note that an $f$-local set in a graph will have size
at least equal to $f$ (since any set of size $f$ is $f$-local).

\subsection{Extension to Multi-Dimensional Functions}
In this subsection, we consider the case where the value for each agent is a $d$-dimensional vector. Thus, the local cost function for node $i$ becomes $g_i:\mathbb{R}^d\to \mathbb{R}$, and the  objective is to collaboratively
solve \eqref{eqn:1do} over  $\psi\in \mathbb{R}^d$. 

The extension to general multi-dimensional functions is a challenging problem as even the region containing the true minimizer of the functions is not easy to characterize. In particular, unlike the scalar case for which the minimizer of $g(\psi)$ lies within the convex hull of the minimizers of the  individual functions, for the multi-dimensional case, the true minimizer vector may lie outside the convex hull of the individual minimizer vectors \cite{kuwaranancharoen2018location}. %As an example, consider functions  $g_1(\psi_1,\psi_2)=\psi_1^2-\psi_1\psi_2+\frac{1}{2}\psi_2^2$ and $g_2(\psi_1,\psi_2)=\psi_1^2+\psi_1\psi_2+\frac{1}{2}\psi_2^2-4\psi_1-2\psi_2$ with minimizers $(0,0)$ and $(2,0)$ respectively, whose sum has minimizer $(1,1)$. 
However, there have been some recent attempts to address the resilient distributed multi-dimensional optimization problem, as we now briefly summarize. Assuming the malicious agents behave in a prescribed manner, \cite{Scagione} proposed an attack detection and isolation technique before the execution of the distributed optimization algorithm. In \cite{bajwa}, the authors  consider a resilient decentralized machine learning problem, and show that by utilizing a block coordinate descent method, the states of the regular agents will converge to the statistical minimizer with high probability. However, the analysis in \cite{bajwa} is restricted to i.i.d. training data across the network. %Moreover, no graph-theoretic condition for convergence is presented in \cite{bajwa}. 

In contrast to \cite{bajwa}, the authors in \cite{kananart} develop a two-step filtering technique and provide convergence guarantees that do not make any statistical assumptions on the agents' objective functions. For an $f$-local Byzantine attack model, under the assumption that $\mc{G}$ is $((2d+1)f+1)$-robust, the approach in \cite{kananart} guarantees asymptotic consensus of the states of all regular agents within a bounded region containing the global minimizer.\footnote{Note that the requirement on the network topology scales with the dimension $d$ of the parameter.} Simply applying the W-MSR algorithm to each coordinate of the parameter vector does not immediately lead to the above result. Instead, the approach in \cite{kananart} relies on a carefully designed second filtering step. 

\textbf{Exact Fault Tolerance:} Instead of settling for convergence to a \textit{proximity} of the global minimizer (as in \cite{bahman,kananart}), one may ask whether it is possible to converge \textit{exactly} to the minimizer of the sum of the objective functions of the regular agents, despite Byzantine attacks. As argued in \cite{bahman}, this is impossible unless additional assumptions are made on the agents' functions. In this context, the authors in \cite{nirupam} show that under a $2f$-redundancy assumption on the agents' objective functions, one can indeed achieve \textit{exact} convergence even in the multi-dimensional case based on a norm filter. In fact, such an assumption turns out to be necessary for guaranteeing exact convergence, as established in \cite{nirupam2}.

The analysis in \cite{nirupam} is carried out for a complete peer-to-peer network under an $F$-total Byzantine attack model. Extending the results in \cite{nirupam} to general networks remains an open direction of research. Moreover, investigating whether the graph-theoretic conditions in \cite{kananart} can be relaxed is also an interesting open problem.

\section{Resilient Distributed Estimation and Inference}
\label{sec:Res_Est}
Another canonical distributed problem involves estimating/tracking an unknown state of interest based on measurements that are collected by a network of sensors. Within this broad setting, there can be several variations: the unknown state may be static or may evolve based on a dynamical model; the measurements may be noise-free or may be corrupted by stochastic noise; and the goal could be to estimate the state asymptotically, or to derive finite-time guarantees. 

All of the above variations share a common unifying feature: \textit{new information flows into the network at every time-step.} This is a key distinction between the distributed estimation setup, and the consensus and optimization problems that we discussed earlier. Another important difference stems from the fact that agents typically have \textit{heterogeneous measurement/observation models} in an estimation problem. As a consequence, some agents may be more ``informative" than others. This disparity in information content across the network is another key feature that is absent in the standard consensus or distributed optimization formulations. As such, the algorithmic techniques and graph-theoretic conditions that we will cover in this section will differ significantly from those in Sections \ref{sec:rdistcons} and \ref{sec:rdist_opt}; see Remark \ref{rem:est_vs_cons} in this context. 

Before formally discussing resilient estimation algorithms, we outline two important considerations: the nature of the unknown quantity to be estimated, and the nature of the threat model. 

\textbf{Static Parameter Estimation vs. Dynamic State Estimation:} As the name suggests, in static parameter estimation, the goal is to estimate a static parameter $\theta^*$ based on noisy sensor observations acquired by the agents. In contrast, the task in (dynamic) state estimation is to track a state $\mathbf{x}[t]$ that evolves based on a dynamical system model (such as a linear time-invariant (LTI) model). Even in the absence of adversaries, tracking the state of an \textit{unstable} system based on dispersed measurements is a significantly challenging task. Thus, we will discuss the relatively simpler resilient distributed parameter estimation problem first, and then move on to the dynamic state estimation setting. 

\textbf{Sensor Attacks vs. Byzantine Attacks:} The works that we will review can also be broadly classified in terms of the threat model. In particular, there are two predominant attack models that are studied in the resilient distributed estimation literature: sensor attacks and Byzantine attacks. In the former case, measurement streams of certain agents are corrupted by an additive attack signal; these signals may or may not be bounded. However, \textit{all} agents behave normally, i.e., they follow the  prescribed protocol at all times. In contrast, a Byzantine agent can act \textit{arbitrarily}. Thus, the Byzantine attack model subsumes the sensor attack model. 
As we shall see, the nature of the threat model has significant implications for the graph-theoretic properties needed to combat attacks. 

\subsection{Parameter Estimation}
\label{subsec:rparam_est}
In the distributed parameter estimation problem, each agent $i\in\mathcal{V}$ receives measurements as follows:
\begin{equation}
    {y}_i[t]=\mathbf{H}_i \theta^* + w_i[t].
\label{eqn:param_model}
\end{equation}
Here, $\theta^* \in \mathbb{R}^{d}$ is the true unknown parameter, $y_i[t]\in \mathbb{R}^{r_i}$ is the measurement vector for agent $i$, $\mathbf{H}_i \in \mathbb{R}^{r_i \times d}$ is the local observation matrix for agent $i$, and ${w}_i[t]$ is the measurement noise that is typically assumed to independent and identically (i.i.d.) distributed over time, with zero mean and finite variance. Moreover, the noise sequences across different agents are assumed to be independent. 

\textbf{Objective:} In the non-adversarial setting, the goal is to design a \textit{consistent} distributed estimator, i.e., an estimator that ensures that the estimates of all agents converge to $\theta^*$ asymptotically almost surely. This is typically achieved by designing ``consensus +  innovations"-type estimators  \cite{karJSAC,karTIT,karSP,xieSG} that work under two standard assumptions: (i) the joint observation model is \textit{globally observable}, i.e., $\sum_{i\in\mathcal{V}} \mathbf{H}'_i \mathbf{H}_i$ is invertible, and (ii) the graph $\mathcal{G}$ is \textit{connected}.

In the resilient version of the above problem, a certain subset $\mathcal{A} \subseteq \mathcal{V}$ of the agents is corrupted either due to sensor attacks or due to Byzantine attacks. We now discuss the key algorithmic approaches to tackle such attacks. For each approach, we will focus on highlighting (i) the threat model; (ii) the main technique; (iii) the guarantees provided by the approach; and (iv) the assumptions on the observation model and the underlying graph needed to provide such guarantees.

\textbf{1) Methods based on Adversary Detection:} In \cite{chenkar1}, the authors consider a Byzantine attack model, and propose the \textit{Flag Raising Distributed Estimation} ($\mathcal{FRDE}$) algorithm where agents simultaneously perform parameter estimation and adversary detection. Specifically, for parameter estimation, the regular agents employ a consensus+innovations update rule, similar to those in \cite{karJSAC,karTIT,karSP,xieSG}. The consensus part of the update rule is based on a weighted average of neighbors' parameter estimates, while the innovation part processes the agent's own local measurements. For adversary detection, an agent computes the Euclidean distance between its own estimate and the estimates of its neighbors. If this distance exceeds a time-varying threshold, then an attack flag is raised. The design of this adaptive threshold constitutes the key part of the $\mathcal{FRDE}$ algorithm. 

The $\mathcal{FRDE}$ algorithm is analyzed under two main assumptions: (i) the joint observation model of the \textit{regular} agents is globally observable, i.e., $\sum_{i\in\mathcal{R}} \mathbf{H}'_i \mathbf{H}_i$ is invertible, where $\mathcal{R}=\mathcal{V}\setminus\mathcal{A}$; and (ii) the \textit{induced} sub-graph $\mathcal{G}_{\mathcal{R}}$ of the regular agents is connected. Under these assumptions, it is shown in \cite{chenkar1} that either all regular agents detect the presence of adversaries, or their local estimates converge to $\theta^*$ asymptotically almost surely. In other words, ``strong" attacks get detected while ``weak" attacks fail to disrupt the process of estimation. 

\textbf{Discussion:} The assumption that the parameter is globally observable w.r.t. the joint measurements of the regular agents is quite intuitive, and in fact necessary (under the Byzantine attack model). The necessity of the graph condition in \cite{chenkar1} is, however, an open question. Based on the $\mathcal{FRDE}$ algorithm, if the presence of adversaries is detected, the system needs to go through an external ``repair" phase; multiple such repair phases could potentially be quite expensive. An alternative is to thus design algorithms that \textit{always} allow the regular agents to estimate the true parameter, despite the presence of adversaries. We now discuss such methods. 

\textbf{2) Saturating Adaptive Gain Methods:} We will discuss this technique in some detail since it has recently been used in the context of resilient distributed state estimation as well \cite{xingkang1}. To isolate the core idea, we will review the simplest version of this method introduced in \cite{chenkar2} to tackle sensor attacks. For an agent $i\in\mathcal{A}$ under attack, its measurement model is as follows:
\begin{equation}
    y_i[t]=\theta^*+a_i[t],
\end{equation}
where $a_i[t]$ is the attack signal injected in the measurements of agent $i$. For agents whose measurements have not been corrupted, the attack signal is identically zero at all times. Note immediately that the measurement model is \textit{homogeneous} and \textit{noise-free}, and that all uncompromised agents can directly measure $\theta^*$. For this model, the authors in \cite{chenkar2} propose the \textit{Saturated Innovation Update} ($\mathcal{SIU}$) algorithm where all agents employ a consensus+innovations estimator with a time-varying gain applied to the local innovation term. Specifically, agent $i$'s estimate $x_i(t)$ of $\theta^*$ is iteratively updated as 
\begin{equation}
    \begin{aligned}
        x_i[t+1]&=x_i[t]-\beta_t \sum_{j\in\mathcal{N}_i} \left(x_i[t]-x_j[t]\right)\\
        &\hspace{2mm} +\alpha_t K_i[t] \left(y_i[t]-x_i[t]\right),
\label{eqn:SIU_update}
    \end{aligned}
\end{equation}
where $\alpha_t, \beta_t$, and $K_i[t]$ are strictly positive, scalar-valued design parameters. The  time-varying gain $K_i[t]$ is defined as
\begin{equation}
K_i[t] = \begin{cases}
1, &\text{ ${\Vert y_i[t]-x_i[t] \Vert}_2 \leq \gamma_t$}\\
\frac{\gamma_t}{{\Vert y_i[t]-x_i[t] \Vert}_2} &\text{otherwise,}
\end{cases}
\end{equation}
where $\gamma_t$ is an adaptive threshold. For a detailed description of how the parameter sequences $\{\alpha_t\}, \{\beta_t\}$, and $\{\gamma_t\}$ are designed, we refer the reader to \cite{chenkar2}. In what follows, we briefly explain why the design of $\gamma_t$ is a delicate matter. Indeed, if $\gamma_t$ is chosen to be too small, then the innovation gain $K_i[t]$ will limit the impact of adversaries; however, a very small innovation again may also prevent correct identification of $\theta^*$. On the other hand, if $\gamma_t$ is too large, then it may provide the adversaries with enough flexibility to direct the agents' estimates away from $\theta^*$. Thus, striking the right balance in the design of $\gamma_t$ is critical. The following theorem from \cite{chenkar2} characterizes the performance of the $\mathcal{SIU}$ algorithm. 

\begin{theorem} [\cite{chenkar2}]
\label{thm:SIU}
Suppose the following conditions hold. (i) The graph $\mathcal{G}$ is connected. (ii) The true parameter $\theta^*$ is bounded, i.e., ${\Vert \theta^* \Vert}_2 \leq \eta$, for some finite $\eta$ that is known a priori to all agents. (iii) Less than half of the agents are under sensor attack, i.e., $|\mathcal{A}|/|\mathcal{V}| < 1/2$. Then, the parameters $\alpha_t, \beta_t$, and $\gamma_t$ can be designed such that the update rule in Eq. \eqref{eqn:SIU_update} ensures
\begin{equation}
    \lim_{t\to\infty} {(t+1)}^{\tau_0} {\Vert x_i[t] - \theta^* \Vert}_2 =0, \forall i\in\mathcal{V},
\end{equation}
for all $\tau_0$ such that $0 \leq \tau_0 < \tau_1-\tau_2$,  where $\tau_1, \tau_2$ are design parameters satisfying $0 < \tau_2 < \tau_1 < 1$.\footnote{In \cite{chenkar2}, the adversarial set $\mathcal{A}$ is allowed to change over time.} 
\end{theorem}

The above theorem tells us that $\mathcal{SIU}$ is a consistent estimator, and that the rate of convergence is of the order of $1/t^{\tau_0}$, for any $\tau_0$ satisfying $0 \leq \tau_0 < \tau_1-\tau_2$. Building on the main idea of using an adaptive threshold to design the innovation gain, the authors in \cite{chenkar2} later generalized their results to account for heterogeneous measurement models corrupted by noise; see \cite{chenkar3} and \cite{chenkar4}. In \cite{chenkar5}, it was shown that the saturating adaptive gain idea is also effective in the context of resilient distributed field estimation under measurement attacks.

\textbf{Discussion:} We now highlight two subtle implications of the choice of threat model. First, note that the guarantee in Theorem \ref{thm:SIU} holds for \textit{all} agents, as opposed to just the regular agents. This is the typical guarantee one provides for measurement/sensor attack models. It is instructive to compare such a result with those for the Byzantine setting ({see, for instance, Theorem \ref{thm:res_dst_est}}) where the goal is to enable only the regular agents to estimate the unknown quantity of interest. 

The second key observation pertains to the graph condition in Theorem \ref{thm:SIU}. All that is needed is connectivity of the underlying network - the exact same condition even in the absence of adversaries. Thus, the main takeaway here is that \textit{the graph-theoretic conditions for solving the distributed parameter estimation problem are the same with and without sensor attacks.}  The main reason for this can be attributed to the fact that even if an agent's measurements are compromised, it does not try to actively disrupt the flow of information between regular agents; however, a Byzantine agent might. This necessitates much stronger graph-theoretic conditions to tackle Byzantine attacks, as we shall see in Section \ref{subsec:rstate_est}. 

\textbf{3) Methods based on Online Optimization:} Yet another way to approach the distributed parameter estimation problem is to view it from the lens of online optimization. This is precisely the method adopted in \cite{lilisu2}, where the authors consider a Byzantine attack model. To explain this method, for each agent $i\in\mathcal{V}$, define its local \textit{asymptotic} loss function $g_i:\mathbb{R}^d \rightarrow \mathbb{R}$ as
\begin{equation}
    g_i(x)=\frac{1}{2} \mathbb{E} \left[ {\Vert \mathbf{H}_i x - y_i\Vert}^2_2 \right],
\label{eqn:local_func}
\end{equation}
where $y_i$ is as in Eq. \eqref{eqn:param_model}, and the expectation is taken w.r.t. the measurement noise $w_i[t]$. Since the distribution of the noise sequence is unknown to agent $i$, it cannot access the above loss function. Nonetheless, agent $i$ can use all the measurements it has acquired up to each  time-step $t$ to compute an \textit{empirical approximation} of $g_i(x)$:
\begin{equation}
    g_{i,t}(x) = \frac{1}{2t} \sum\limits_{s=1}^{t} {\Vert \mathbf{H}_i x - y_i[s]\Vert}^2_2. 
\end{equation}
The algorithm in \cite{lilisu2} essentially combines local gradient descent on the above empirical loss functions, followed by coordinate-wise trimming to aggregate neighboring information; trimmed means are used to account for the presence of adversaries. This algorithm enables each regular agent to estimate the true parameter asymptotically almost surely. Moreover, as a departure from existing results on this problem, the authors provide \textit{finite-time concentration bounds} that hold with high probability. To arrive at the above results, the conditions imposed on the graph topology are the same as those for Byzantine-resilient scalar consensus \cite{vaidyacons}. In \cite{lilisu2}, certain additional assumptions are made on the observation model that may not be necessary. 

\textbf{Additional Results:} Before moving on to the state estimation setting, we briefly comment on a couple of related works. For a somewhat different observation model than in Eq. \eqref{eqn:param_model}, the authors in \cite{leblancparam}  provide guarantees against Byzantine attacks by drawing on the techniques and graph-conditions in \cite{Hogan}. To the best of our knowledge, this is the earliest work on resilient distributed parameter estimation.

Very recently, the authors in \cite{min_switch} proposed a \textit{min-switching} technique to account for the presence of Byzantine agents in the context of least-squares static estimation. The main idea behind the approach in \cite{min_switch} is to first construct an appropriate local Lyapunov function at each regular agent. The filtering technique then  comprises of using only those neighboring estimates that lead to maximum decrease of the Lyapunov function. It is shown that this method can help relax the graph-theoretic conditions in both \cite{lilisu2} and \cite{aritra}. 

\subsection{Dynamic State Estimation and Inference}
\label{subsec:rstate_est}
In a typical distributed state estimation problem, the goal is to track the state of a linear time-invariant system of the following form:
\begin{equation}
\boldsymbol{x}[t+1]=\boldsymbol A\boldsymbol{x}[t],
\label{eqn:1de}
\end{equation}
where $\boldsymbol{x}[t]\in \mathbb{R}^N$ is the state vector and $\boldsymbol A$ is the state transition matrix. The system is monitored by a network $\mathcal{G}=\{\mathcal{V},\mathcal{E}\}$ consisting of $n$ nodes. The measurement model of the $i$-th node is given by 
\begin{equation}
    y_i[t]=\boldsymbol C_i\boldsymbol{x}[t],
    \label{eqn:2de}
\end{equation}
where $y_i[t]\in \mathbb{R}^{r_i}$ and $\boldsymbol C_i\in  \mathbb{R}^{r_i\times N}$. We use $\boldsymbol C=[\boldsymbol C_1', \boldsymbol C_2', \cdots , \boldsymbol C_N']'$ to collect all the individual node observation matrices and $\boldsymbol{y}[t]=[y_1'[t], y_2'[t], \cdots , y_N'[t]]'$ to aggregate all the individual measurement vectors; accordingly,  $\boldsymbol{y}[t]=\boldsymbol C\boldsymbol{x}[t]$. Each node $i$ maintains an estimate $\hat{\bs{x}}_i[t]$ of the state $\bs{x}[t]$, and the goal is to ensure that these estimates converge to  $\bs{x}[t]$ asymptotically.

Even in the absence of adversaries, the distributed state estimation problem is quite challenging, and only recently were necessary and sufficient conditions discovered for this problem \cite{martins,mitraTAC,wang,han,rego,nozal,kimTAC}. The key technical challenge arises from the fact that $(\bs{A,C}_i)$ may not be detectable w.r.t. the measurements of any individual node $i$. This is precisely what necessitates communication between nodes in the graph. The difficulty of tracking an unstable dynamical process based on dispersed measurements only gets exacerbated in the presence of adversaries. Following the same style of exposition as in Section \ref{subsec:rparam_est}, we now discuss the main techniques for resilient distributed state estimation.

\textbf{1) Methods based on Observable Decompositions and Local Filtering:} We start by reviewing the approach developed in \cite{aritra} for solving the resilient distributed state estimation problem subject to an $f$-local Byzantine adversary model. In order to focus on the core ideas behind this approach, we assume that the system matrix $\boldsymbol A$ has real and simple eigenvalues; extensions to general spectra can be found in \cite{aritra}. As a first step, we diagonalize $\boldsymbol A$ using a coordinate transformation matrix $\boldsymbol V=[\boldsymbol v^1, \boldsymbol v^2, ..., \boldsymbol v^N]$ formed by $N$ linearly independent eigenvectors of $\boldsymbol A$. In the new coordinate system where 
 $\boldsymbol{z}[t]=\boldsymbol V^{-1}\boldsymbol{x}[t]$ is the state, the dynamics \eqref{eqn:1de} and \eqref{eqn:2de}  take the following form:
\begin{align}
\boldsymbol{z}[t+1]=\boldsymbol M\boldsymbol{z}[t],\nonumber\\
y_i[t]=\bar{\boldsymbol C}_i\boldsymbol{z}[t].
\label{eqn:r3de}
\end{align}
Here, $\boldsymbol M=\boldsymbol V^{-1}\boldsymbol A\boldsymbol V$ is a diagonal matrix with diagonal entries comprising of the distinct eigenvalues $\lambda_1, \lambda_2,..., \lambda_N$ of $\boldsymbol A$, and $\bar{\boldsymbol C}_i=\boldsymbol C_i\boldsymbol V$. Based on the above decomposition, each node $i$ can immediately identify (locally) the set of eigenvalues $\mathcal{O}_i$  that are detectable  w.r.t. its own measurements. The key observation made in \cite{aritra} is the following. Each node $i$ can estimate the components of the state vector $\boldsymbol{z}[t]$ corresponding to its detectable eigenvalues $\mathcal{O}_i$ without interacting with any neighbor. It needs to employ consensus only for estimating those components that correspond to its undetectable eigenvalues $\mc{UO}_i$. Specifically, every regular node $i$ employs the following scheme.

\begin{itemize}
\item [(i)] For each $\lambda_j \in \mc{O}_i$, node $i$ uses a standard Luenberger observer to estimate $z^{(j)}[t]$ - the component of $\bs{z}[t]$ corresponding to $\lambda_j$. 

\item [(ii)] For each $\lambda_j \in \mc{UO}_i$, node $i$ uses a ``local filtering'' technique to estimate $z^{(j)}[t]$. 
\end{itemize}

We now elaborate on item (ii). Let $\Lambda_{U}(\bs{M})$ represent the set of all unstable and marginally stable eigenvalues of $\bs{M}$. Moreover, for each $\lambda_j \in \Lambda_{U}(\bs{M})$, let $\mc{S}_j$ be the set of nodes that can detect $\lambda_j$. To enable the nodes in $\mc{V}\setminus\mc{S}_j$ to estimate $z^{(j)}[t]$, the following two requirements (stated loosely here, and more formally later) turn out to be critical. 

\begin{itemize}
    \item \textbf{Information redundancy:} The set $\mc{S}_j$ needs to be sufficiently large. Otherwise, if the sources of information for the mode $\lambda_j$ are corrupted by too many adversaries, then the other nodes cannot hope to recover the correct information about that mode.
    
    \item \textbf{Network-structure redundancy:} There must exist sufficient disjoint paths in $\mc{G}$ that link the nodes in  $\mc{S}_j$ to those in $\mc{V}\setminus\mc{S}_j$. Otherwise, the adversaries can form a bottleneck and disrupt the flow of information from $\mc{S}_j$ to $\mc{V}\setminus\mc{S}_j$. 
\end{itemize}

In \cite{aritra}, the authors introduce a graph-theoretic construct called the Mode Estimation Directed Acyclic Graph (MEDAG) to capture the above requirements. Essentially, a MEDAG $\mc{G}_j$ for mode $\lambda_j$ is a subgraph of $\mc{G}$ that provides a secure \textit{uni-directional} medium of information-flow from $\mc{S}_j$ to $\mc{V}\setminus\mc{S}_j$.\footnote{For a precise description of the properties of a MEDAG, see \cite{aritra} where a distributed algorithm is provided to construct such subgraphs.}  The uni-directional aspect is important to ensure stability of the estimation error dynamics; it has nothing to do with adversaries. Once a MEDAG $\mc{G}_j$ has been constructed for each $\lambda_j \in \Lambda_{U}(\bs{M})$, an agent $i\in\mc{V}\setminus\mc{S}_j$ uses the estimates of only its neighbors $\mc{N}^{(j)}_i$ in the MEDAG $\mc{G}_j$ to update its estimate $\hat{z}^{(j)}_i[t]$ of $z^{(j)}[t]$ as follows:

\begin{equation}
\hat{z}^{(j)}_i[t+1]=\lambda_j \sum_{\ell\in \mathcal{M}^{(j)}_i[t]}w_{i\ell}^{(j)}[t]\hat{z}^{(j)}_{\ell}[t]. 
\label{eqn:LFRE}
\end{equation}
In the above update rule,  $\mathcal{M}^{(j)}_i[t]\subset \mathcal{N}^{(j)}_i\subseteq \mathcal{N}_i$ is the set of those neighbors from whom node $i$ accepts estimates of 
$z^{(j)}[t]$ at time-step $t$, after removing the $f$ largest and $f$ smallest estimates of $z^{(j)}[t]$ from $\mathcal{N}^{(j)}_i$; the properties of a MEDAG $\mc{G}_j$ ensure that $\mathcal{M}^{(j)}_i[t]$ is always non-empty. 
The weights in \eqref{eqn:LFRE} are
non-negative and chosen to satisfy $\sum_{\ell\in \mathcal{M}^{(j)}_i[t]}w_{i\ell}^{(j)}[t]=1$. The overall approach we described above is called the Local-Filtering based Resilient Estimation (LFRE) algorithm in \cite{aritra}. 

To analyze the performance of the LFRE algorithm, we need to first understand when a given network $\mc{G}$ contains a MEDAG. The following graph-theoretic property is what we need in this context. 

\begin{definition}(\textbf{strongly} $r$-\textbf{robust graph} \textit{w.r.t.} $\mathcal{S}_j$) For $r \in \mathbb{N}_{+}$ and $\lambda_j \in \Lambda_{U}(\mathbf{A})$, a graph $\mathcal{G}=(\mathcal{V,E})$ is \textit{strongly $r$-robust w.r.t. the set of source nodes $\mathcal{S}_j$} if for any non-empty subset $\mathcal{C} \subseteq \mathcal{V}\setminus\mathcal{S}_j$, $\mathcal{C}$ is $r$-reachable.
\label{defn:strongrobust}
\end{definition}

\begin{figure}[t!]
\centering
\includegraphics[scale=.5]{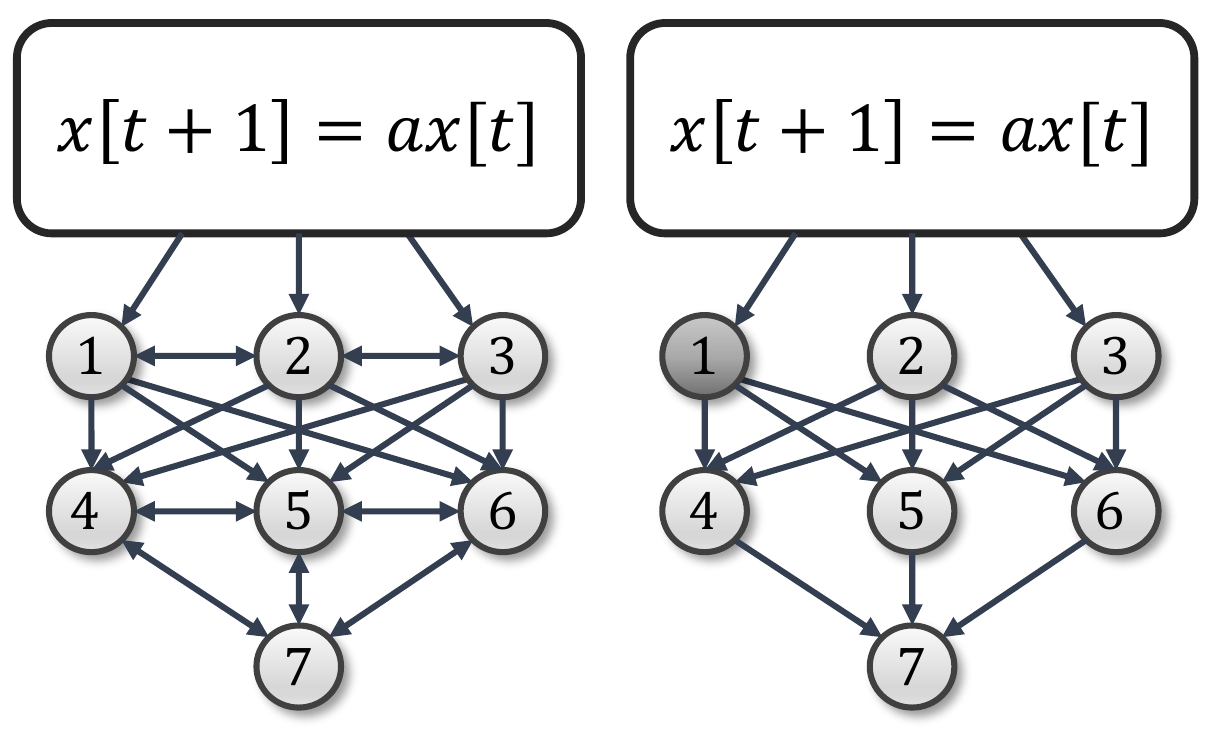}
\caption{(\textbf{Left}) A scalar unstable plant is monitored by a network
of 7 nodes. The network is strongly 3-robust w.r.t. the set of source nodes $\mc{S}=\{1,2,3\}$. (\textbf{Right}) A subgraph of the original graph satisfying
the properties of a MEDAG.}
\label{fig:chamedag0s82}
\end{figure}

For an illustration of the above definition, consider the setup in Fig. \ref{fig:chamedag0s82} (taken from \cite{aritra}) where a scalar unstable plant is monitored by a network of nodes. Nodes 1, 2, and 3 are the source nodes for this system, i.e., $\mc{S}=\{1,2,3\}$. The graph on the left in Fig. \ref{fig:chamedag0s82} is an example of a
network that is strongly 3-robust w.r.t. the set of source nodes $\mc{S}$.  Specifically, all subsets of $\{4,5,6,7\}$ are 3-reachable (i.e., each such subset has a node that has at least 3 neighbors outside that subset). The graph on the right is an example of a MEDAG. The next result exemplifies the role played by \textit{strong-robustness} in resilient distributed state estimation. 

 \begin{thm} [\cite{aritra}] 
 \label{thm:res_dst_est}
 Suppose $\mathcal{G}$ is strongly $(2f+1)$-robust w.r.t. $\mathcal{S}_j, \forall \lambda_j\in \Lambda_U(\boldsymbol A)$. Then, the LFRE algorithm guarantees $\lim_{t\to\infty} \|\hat{\boldsymbol x}_i[t]-\boldsymbol{x}[t]\|=0$ for every regular agent $i$, despite the presence of any $f$-locally bounded set of Byzantine adversaries. 
 \end{thm}
 
 \textbf{Discussion:} While the above result provides a sufficient condition for tolerating Byzantine adversaries in the context of distributed state estimation, separate necessary conditions are also identified in \cite{aritra}. The latter are a blend of requirements on the observation model and the network structure, and generalize the conditions for centralized state estimation subject to attacks \cite{fawzi,chong15}. 

 The techniques in \cite{aritra} were later generalized to account for time-varying networks in \cite{mitra_AR} and \cite{mitra_AOI}. Moreover, in \cite{mitra_TCNS}, the authors formally showed how one can incorporate the ideas of trust and diversity to relax some of the stringent redundancy requirements in \cite{aritra}. 
 
 One important takeaway is that the ideas of information redundancy and network-structure redundancy are quite general, and as such, applicable beyond the specific estimation problem we considered here. Indeed, we will later briefly comment on the fact that these ideas also turn out to be crucial in the context of resilient distributed hypothesis testing/statistical inference. 
 
 \begin{remark} 
 \label{rem:est_vs_cons}
 Unlike the $r$-robustness property that is coNP-complete to check \cite{hoganfatasundaram}, the strong $r$-robustness property in Definition \ref{defn:strongrobust} can be checked in polynomial-time. Aside from algorithmic differences, this highlights that the graph-theoretic condition required for resilient distributed state estimation (strong $r$-robustness) is fundamentally different from that for resilient consensus and optimization ($r$-robustness). 
 \end{remark}
 
\textbf{2) Methods based on Robust Control Theory:} In \cite{deghat}, the authors consider distributed state estimation of a continuous-time LTI system where both the state and measurement dynamics are subject to $L_2$-integrable disturbances. Given the nature of the disturbances, the authors build on the theory of distributed $H_{\infty}$ filters developed in \cite{ugrinov1,ugrinov2}. The threat model is that of a \textit{biasing attack} where an attacker injects an additive attack signal directly to the state estimator/observer dynamics of certain agents.

To tackle such biasing attacks, the authors in \cite{deghat} develop certain attack detection filters. Specifically, the attack detection filter at each agent $i$ takes as input two different innovation signals. The first innovation signal is the gap between the actual measurement $y_i[t]$ and the predicted sensor measurement $C_i\hat{x}_i[t]$. Intuitively, if agent $i$'s observer dynamics is compromised, then one should expect $C_i\hat{x}_i[t]$ to be biased, leading to a significant deviation from $y_i[t]$.\footnote{As in \cite{deghat}, the innovation signal $(y_i[t]-H_i\hat{x}_i[t])$ plays a key role in the saturating adaptive gain method of \cite{chenkar2}. However, while $y_i[t]$ is accurate and $C_i\hat{x}_i[t]$ is biased in \cite{deghat}, the situation is exactly the opposite in \cite{chenkar2},  where the measurements are biased but the agents' estimators are not.} With a similar motivation, a second innovation signal is computed based on the deviations of agent $i$'s state estimate from those of its neighbors. It is shown that the problem of designing the parameters of the above detector can be recast as the problem of stabilizing a distributed dynamical system via output injection. The latter problem is addressed by drawing on ideas from vector dissipativity theory \cite{ugrinov1,ugrinov2}. Since the overall design procedure is quite intricate, we refer the reader to \cite{deghat} for details.

As their main result, the authors in \cite{deghat} show that if certain LMI's are feasible, then their approach ensures attack-detection and guarantees a desired level of $H_{\infty}$ disturbance attenuation. The requirements on the network structure are implicitly captured by the LMI's.

\textbf{Discussion:} It is instructive to compare the results in \cite{aritra} with those in \cite{deghat}. In \cite{aritra}, accounting for a worst-case Byzantine attack model necessitates the requirement of sufficient redundancy in the underlying network, as captured by the ``strong robustness'' condition in Definition~\ref{defn:strongrobust}.  Moreover, for the problem to be meaningful in \cite{aritra}, only a subset of the agents can be adversarial. In contrast, since a specific class of biasing attacks is considered in \cite{deghat}, the network requirements are relatively less stringent, and it is plausible for the entire network to be under attack. 

Following up on \cite{deghat}, more general biasing attacks are considered in \cite{Ugrinovskii} where an attacker is allowed to bias both the state observer and the attack detector at an agent. Moreover, unlike \cite{deghat} where the design of the filter gains involves solving LMI's that are coupled across agents, the design procedure is carried out locally at every agent in \cite{Ugrinovskii}. 

\textbf{3) Methods based on Saturating Gains and Attack Detection:} Recently, in \cite{xingkang1}, the authors developed secure distributed filters for tackling measurement attacks. Their main approach relies on a saturating adaptive gain technique, similar to \cite{chenkar2}. There are, however, considerable differences with \cite{chenkar2} that stem from the fact that \cite{xingkang1} considers dynamical state estimation while \cite{chenkar2} studies static parameter estimation. One such difference is that \cite{xingkang1} employs a two-time-scale estimation technique: between two consecutive time-steps of the dynamics, the agents are allowed to perform multiple consensus steps to bridge the gap between their estimates. Under reasonable assumptions on observability, and connectivity of the graph, the estimation error is shown to be uniformly bounded. 

\textbf{Discussion:} Notably, the analysis in \cite{xingkang1} applies to \textit{time-varying adversarial sets}. When the adversarial set is fixed, an attack detection algorithm is further developed in \cite{xingkang1} that leads to tighter error bounds. It should be noted, however, that the two-time-scale approach is crucial to the stability of the distributed filter in \cite{xingkang1}.  

\begin{remark}
In some very recent work, the authors in \cite{mao2} have studied the resilient distributed state estimation problem from a dynamic average consensus perspective. In \cite{mao1}, the setting where some of the communication links can also be corrupted (in addition to node attacks) has been explored. 
\end{remark}
\subsection{Hypothesis Testing and Statistical Inference}
\label{sec:res_hyp_test}
In this subsection, we will briefly discuss an approach for tackling Byzantine attacks in the context of distributed hypothesis testing \cite{jad1,jad2,liu,shahin,nedic,lalitha,su1,uribe,mitraACC19,mitra2019new} - a problem similar in flavor to the static parameter estimation setting we considered in Section \ref{subsec:rparam_est}. In this problem, each agent in a network receives a sequence of stochastic measurements generated by a common underlying distribution that is parameterized by an  $\textit{unknown, static}$ state $\theta^*$. Each agent is equipped with a local likelihood model, and is aware that $\theta^*$ belongs to a finite set $\Theta=\{\theta_1, \ldots, \theta_m\}$ of $m$ candidate hypotheses. The goal is for the agents to collaboratively identify $\theta^*$ from $\Theta$. 
The challenge arises from the fact that $\theta^*$ may not be locally identifiable w.r.t. the likelihood model of any specific agent. In other words, no one agent can, in general, eliminate every false hypothesis on its own. Instead, we assume \textit{global identifiability} of the joint observation model, i.e., $\theta^*$ can be uniquely identified based on the collective observations of the agents.\footnote{Global identifiability for distributed hypothesis testing is the exact analogue of global observability for distributed state estimation.} 

In a typical approach to solving the above problem, each agent maintains a belief vector $\boldsymbol{\mu}_{i,t}$ which is a distribution over $\Theta$. Formally, the objective is to design belief-update and propagation rules that ensure $\mu_{i,t}(\theta^*) \rightarrow 1, \forall i\in \mathcal{V}$, almost surely. The predominant approach is to employ some form of ``belief-averaging" protocol to update the belief vectors \cite{jad1,jad2,liu,shahin,nedic,lalitha,su1,uribe}. In a departure from these algorithms, a min-rule was recently developed in \cite{mitra2019new}; the asymptotic learning rate of this rule is strictly better than those based on averaging. However, all of the above approaches are vulnerable in the face of attacks: a single malicious agent can essentially cause all good agents to eliminate the true hypothesis. 

\textbf{The Min-Rule:} A desirable feature of the min-rule in \cite{mitra2019new} is that it admits a simple, computationally-efficient extension that is robust to worst-case attacks. We first describe the basic min-rule, and then its adversarial extension. Each agent $i\in \mc{V}$ maintains an auxiliary local belief vector $\boldsymbol{\pi}_{i,t}$ that is updated in a Bayesian manner based on just the observations of agent $i$. For every false hypothesis $\theta \in \Theta \setminus \{\theta^*\}$ that agent $i$ can eliminate on its own, we will have $\pi_{i,t}(\theta) \rightarrow 0$ almost surely. Thus, agent $i$ only needs to interact with neighbors for eliminating those false hypotheses that it cannot rule out on its own. Let $\mc{S}(\theta^*,\theta)$ be those agents that can distinguish between $\theta^*$ and $\theta$, i.e., these agents can eliminate $\theta$ individually. The main idea is to transmit low beliefs on $\theta$ from agents in $\mc{S}(\theta^*,\theta)$ to the rest of the network. This is achieved via the following rule at each agent $i$:

\begin{equation}
\mu_{i,t+1}(\theta) \propto \min\{\{\mu_{j,t}(\theta)\}_{{j\in\mathcal{N}_i\cup\{i\}}},\pi_{i,t+1}(\theta)\}.
\label{eqn:min_rule}
\end{equation}

The above beliefs are normalized to ensure that $\boldsymbol{\mu}_{i,t}$ is a valid distribution at every time-step. For the adversarial setting, the approach is very similar, except that the min-rule is applied to a set of \textit{moderate beliefs}. Specifically, each regular agent $i$ updates its belief on a state $\theta$ by first rejecting the highest $f$ and lowest $f$ beliefs on $\theta$ received from $\mc{N}_i$, and then employing:
\begin{equation}
\mu_{i,t+1}(\theta) \propto \min\{\{\mu_{j,t}(\theta)\}_{j\in\mathcal{M}^{\theta}_{i,t}},\pi_{i,t+1}(\theta)\}, 
\label{eqn:res_min_rule}
\end{equation}
where $\mathcal{M}^{\theta}_{i,t}$ are those agents that do not get rejected in the above filtering step. This is known as the Local-filtering based Resilient Hypothesis Elimination (LFRHE) algorithm. The correctness of the LFRHE algorithm once again rests on the two key ingredients we identified in Section \ref{subsec:rparam_est}, namely, information-redundancy and network-structure redundancy. In particular, for every pair $\theta_p, \theta_q$, we need $\mc{S}(\theta_p,\theta_q)$ to be large enough, and we also need a sufficient number of disjoint paths from  $\mc{S}(\theta_p,\theta_q)$ to $\mc{V}\setminus \mc{S}(\theta_p,\theta_q)$. These requirements are succinctly captured in the following theorem.

\begin{theorem} [\cite{mitra2019new}]
Suppose that for every pair of hypotheses $\theta_p,\theta_q\in\Theta$, the graph $\mathcal{G}$ is strongly $(2f+1)$-robust w.r.t. the  source set $\mathcal{S}(\theta_p,\theta_q)$. Moreover, suppose each regular agent $i$  has a non-zero prior belief on every hypothesis, i.e., $\pi_{i,0}(\theta) > 0$ and $\mu_{i,0}(\theta) > 0$, $\forall \theta\in\Theta$. Then, the LFRHE algorithm guarantees that $\mu_{i,t}(\theta^*) \rightarrow 1$ almost surely for every regular agent $i$, despite the actions of any $f$-local set of Byzantine adversaries.
\end{theorem}

\textbf{Discussion:} One of the main takeaways from the above result is that just like the resilient distributed estimation problem, the strong-robustness property in Definition \ref{defn:strongrobust} ends up playing a crucial role when it comes to tolerating Byzantine attacks for distributed hypothesis-testing as well. We conjecture that this graph-theoretic property will prove to be useful for other distributed learning problems where information is diffused across the network. A recent work that studies resilient distributed best-arm identification for stochastic multi-armed bandits supports this conjecture \cite{Fed_bandits}. 

\textbf{Additional Results:} The algorithm in \cite{mitra2019new} was later extended in \cite{wuACC} to account for time-varying networks. In \cite{su1}, the authors proposed an alternate approach to tackling adversaries by building on the log-linear belief-update rule in \cite{shahin, nedic, lalitha}. Their approach requires the agents to compute Tverberg partitions (see Section \ref{subsec:resilient_vector_consensus}); however, as discussed in that section, there is no known algorithm that can compute an exact Tverberg partition in polynomial time for a general $d$-dimensional finite point set \cite{Mulzer}.

\section{Attack Detection and Identification Over Networks}
\label{sec:attackdetection}
In the preceding sections, we covered several techniques for solving a variety of distributed information-processing problems subject to attacks (e.g., consensus, optimization, and estimation).  As discussed in those sections, depending on the attack model and nature of the information available to the nodes, detection and identification of (worst-case) adversarial behavior may be impossible in general.  In particular, the ``local-filtering'' algorithms discussed in those sections  did not explicitly rely on detection/identification of adversarial behavior.  However, in other settings, detection and identification of adversarial behavior may indeed be possible.  The purpose of this section is to briefly summarize algorithms for such settings.

Since the precise nature of the attack detection algorithm is usually dictated by the specific distributed task at hand, we will not be able to cover all such detection mechanisms here. Instead, we will primarily restrict our attention to the attack model in Eq. \eqref{eqn::OAEGI0} of Section \ref{subsec:func_calc} that we studied in the context of distributed function calculation. Using this model, we will discuss graph-theoretic requirements for detecting and identifying attacks in a network. In particular, we will demonstrate how \textit{structured systems theory} plays a key role in this context. Before delving into the technical details, we remind the reader that the $f$-total attack model under consideration involves a set of malicious nodes $\mathcal{F}$, where $|\mathcal{F}|\leq f$ for a known $f\geq 0$. 

\subsection{Attack Detection}
Centralized and distributed detection techniques can be used to detect attacks, see \cite{Pasqualettiii}. Here, we characterize graph-theoretic  conditions for detecting attacks. We start by considering that  the initial states, $\boldsymbol\psi[0]$, are known.
An attack $\boldsymbol \zeta$ is called {\it undetectable} or {\it perfect} if $\boldsymbol y(\boldsymbol\psi[0],\boldsymbol\zeta, t)=\boldsymbol y(\boldsymbol\psi[0],0, t)$ for all $t\geq 0$, i.e., the measurement is the same as the case of no attack. The notion of a perfect attack has an equivalent algebraic condition, which is based on the following definition.

\begin{definition}\label{def:gnr}
The generic normal rank (gnr) of the matrix pencil of dynamics \eqref{eqn::OAEGI0} 
$$\boldsymbol{P}(z)= \begin{bmatrix}
      \boldsymbol{W}- z\mathbf{I}_{n} & \boldsymbol B_{\mathcal{F}}    \\[0.3em]
     \boldsymbol C & \mathbf{0}
     \end{bmatrix},$$
is the maximum rank of the matrix over all choices of free (nonzero)
parameters in $(\boldsymbol W,\boldsymbol B_{\mathcal{F}} ,\boldsymbol C)$ and $z\in \mathbb{C}$.
\end{definition}

It is shown in \cite{Dion} that having a perfect attack and the generic normal rank of $\boldsymbol{P}(z)$ being less than $n+|\mathcal{F}|$ are equivalent .
Recalling the input set $\mathcal{U}$ and measurement set $\mathcal{Y}$ for a structured system from Section \ref{sec:str}, the following result  interprets the generic normal rank of $\boldsymbol{P}(z)$ in terms of the disjoint paths in the graph of structured system $\mathcal{G}$.

\begin{lem} [\cite{Dion2}]
The generic normal rank of the matrix pencil $\boldsymbol{P}(z)$
is equal to $n+r$, where $r$ is the size of the largest linking in $\mathcal{G}$ from the
input vertices, $\mathcal{U}$, to the output vertices, $\mathcal{Y}$.
\label{lem:qwkdn}
\end{lem}

Note that the generic normal rank of the matrix pencil is at least $n$, since the matrix $\boldsymbol W-z\mathbf{I}_{n}$ will have generic rank $n$ for any choice of parameters in $\boldsymbol W$ and any $z$ that is
not an eigenvalue of $\boldsymbol W$. Lemma \ref{lem:qwkdn} implies that to prevent perfect attacks,  parameter  $r$ has to be equal to the number of attacks, i.e., $r=|\mathcal{F}|$. This implicitly indicates that the number of sensors must be at least $|\mathcal{F}|$. Lemma \ref{lem:qwkdn} along with  Menger's theorem and Expansion lemma, c.f. \cite{west}, yield the following graph-theoretic result on attack detectability.

\begin{thm} 
Suppose that dynamics \eqref{eqn::OAEGI0} with measurement \eqref{eqn:measuremetn} is subject to a set of $f$ attacked nodes and that the initial states, $\boldsymbol \psi[0]$, are known.  To prevent a perfect attack, it is sufficient for graph $\mathcal{G}$ to be $(f+1)$-connected.
\label{thm:qwks8dn}
\end{thm}

\begin{rem}
As the number of sensors is often limited and the underlying network  may be sparse, detecting all attacked nodes may not be always possible. An alternative approach is to place the available sensors on key nodes in the network in order to maximize  $r$ in Lemma \ref{lem:qwkdn}, i.e., detecting maximum number of attacks. The sensor placement problem for optimal attack detection is discussed in Section \ref{sec:nc08w2}. 
\end{rem}

For the cases where the initial condition of the system is unknown, an undetectable attack $\boldsymbol{\zeta}$ is characterized by the existence of a pair of initial states $\boldsymbol\psi_1[0]$ and $\boldsymbol\psi_2[0]$ such that $\boldsymbol{y}(\boldsymbol{\psi}_1[0],0,t)=\boldsymbol{y}(\boldsymbol{\psi}_2[0],\boldsymbol{\zeta},t)$ for all $t\geq 0$. In such cases, one needs to first recover the initial conditions of the system in order to detect the attack. This demands the system to be strongly observable. Recall from Section \ref{sec:distfunccalc} and  Theorem \ref{thm:1disfunccalc} that   system \eqref{eqn::OAEGI0} is strongly observable if the graph is $(2f+1)$-connected. This extra level of graph connectivity, compared to Theorem \ref{thm:qwks8dn}, is a price paid for the lack of the knowledge of the initial states. 

\begin{thm} 
Suppose that dynamics \eqref{eqn::OAEGI0} with measurement \eqref{eqn:measuremetn} is subject to a set of $f$ attacked nodes and unknown initial conditions.  To prevent a perfect attack, it is sufficient for graph $\mathcal{G}$ to be $(2f+1)$-connected.
\label{thm:qwksaf18}
\end{thm}

\subsection{Attack Identification Procedure}
In order to identify the attacked nodes, first, node $i$  must find the true initial value of all other nodes (we assume that it satisfies the condition in Theorem \ref{thm:1disfunccalc}). After obtaining the vector of initial states, $\boldsymbol{\psi}[0]$,  and assuming that the interaction matrix, $\boldsymbol{W}$, is known to $i$, it can apply dynamics \eqref{eqn::OAEGI0} to obtain $\boldsymbol{\psi}[1]-\boldsymbol{W}\boldsymbol{\psi}[0]=\boldsymbol B_{\mathcal{F}}\boldsymbol{\zeta}[0]$. Every nonzero component in the vector on the left hand side of this equation indicates an additive error injected by the corresponding node. Thus, every node that is malicious during time-step $0$ can be identified by this method. The same process can be repeated to find all nodes that were
malicious during the first $L$ time steps from the transmitted values $y_i[0:L]$ in \eqref{eqn:oaisbv}. Note that using iteration policy \eqref{eqn::OAEGI0} to identify the attacks requires the system to be free of noise or external disturbances, as otherwise, they mislead us in the above procedure.  Further details on centralized and distributed attack detection and identification  techniques can be found in \cite{Pasqualettiii}.

\subsection{Other Attack Detection and Identification Approaches}
Our discussion in this section has thus far focused on the attack model in Eq. \eqref{eqn::OAEGI0} for distributed function calculation. Before closing this section, we briefly summarize certain other attack detection mechanisms that are relevant in other contexts. For resilient distributed parameter estimation, attack detection constitutes a key component of the $\mathcal{FRDE}$ algorithm in \cite{chenkar1} that we discussed in Section \ref{subsec:rparam_est}. For state estimation, agents under attack are identified based on robust control techniques in \cite{deghat}, and a saturating adaptive gain method in \cite{xingkang1}; see Section \ref{subsec:rstate_est}. At a high-level, the approaches in \cite{chenkar1}, \cite{deghat}, and \cite{xingkang1} share a common principle for detecting whether an agent $i$ is under attack: they involve computing an appropriate innovation signal that captures the extent to which the estimates/measurements of agent $i$ differ from those of its neighbors. For distributed optimization, \cite{Scagione} proposed a heuristic gradient based method to detect the misbehaving agents. The insight  is that the attackers’ biasing actions can result in a large gradient value. Hence, a regular agent can attempt to detect such attacks by approximating the gradient of each neighbor and track it over time relative to the
mean of the gradients of the remaining neighbors. Finally,  we note that in the context of multi-robot coordination, the authors in \cite{spoof1,spoof2} propose methods to tackle the so called ``Sybil attack”, where an attacker spoofs or impersonates the identities
of existing agents to gain a disproportionate advantage in the
network. The key idea in these works is to detect such spoofing attacks by exploiting the physics of wireless signals. For specific details, we refer the reader to the respective papers.

\section{Graph-Theoretic Interpretations  of the Attack Impact}
\label{sec:quant}
In the previous sections, we discussed how to use network connectivity to 
withstand the adversarial actions in several distributed algorithms.  In some cases, however, it is not possible to completely nullify the attacker's actions. In those cases, an alternative approach is to mitigate its impact (as was done in the local filtering approaches for consensus and optimization). In this section, we quantify the attacker's impact on networked control systems in terms of the topology of the underlying network. 

\subsection{Controllability of Networks Under Attack}
\label{sec:nd08}
One way to define the attacker's impact is via the largest subset of nodes which can be controlled (or reached) by the set of the attacked nodes.  There is a vast literature that studies the controllability of networks with a limited number of actuators \cite{Rahmani, alex, Dion, shreyasstructue, Pequitooo}. In this section, our focus is on structural controllability as a discrete measure of controllability, and Gramian-based methods as  continuous measures.

\subsubsection{Structural and Strong Structural Controllability}

As mentioned in Section \ref{sec:str}, a structured system is said to be controllable if this
property holds for at least one numerical choice of free (nonzero) parameters in the system. Theorem \ref{thm:strcon}  provided graph-theoretic conditions for structural controllability. Condition (i) in Theorem \ref{thm:strcon} is called the reachability condition. The reachability, by itself, can be a measure of the attack impact. In particular, when the attacker's goal is to disseminate a signal throughout the network and infect as many nodes as possible, maximizing reachability is beneficial to the attacker. On the other hand, when the attacker's objective is to steer the states towards its desired direction, controllability is the appropriate measure.
We present the following definition. 

\begin{definition}
The  reachable  set,  $\mathcal{S}^r(\mathcal{F})$ (respectively controllable set, $\mathcal{S}^c(\mathcal{F})$) for a set of attacked nodes $\mathcal{F}$, is a subset of nodes in graph $\mathcal{G}=\{\mathcal{V},\mathcal{E}\}$ whose elements can all be reached (controlled) by the set of attacked nodes $\mathcal{F}$ and is {\it maximal} in the sense that no additional node can be added to $\mathcal{S}^r(\mathcal{F})$ ($\mathcal{S}^c(\mathcal{F})$) without breaking this property.  Elements of the attacked  set,  $\mathcal{F}$, are called the source nodes of $\mathcal{S}^r(\mathcal{F})$ { (or of $\mathcal{S}^c(\mathcal{F})$)}.% The minimum number of attacked nodes which make the whole graph $\mathcal{G}$ reachable (respectively controllable)  is denoted by $\mathcal{K}_r$ (respectively $\mathcal{K}_c$).
\label{def:reachable}
\end{definition}

\begin{figure}[t!]
\centering
\includegraphics[scale=.6]{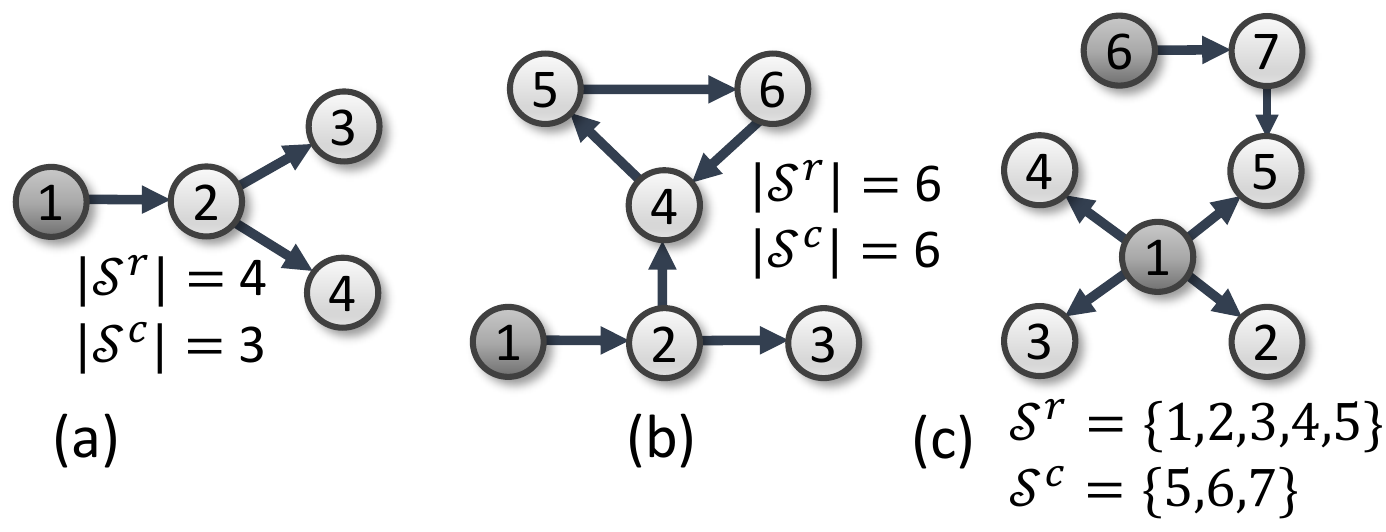}
\caption{Graphs and their largest reachable and controllable subsets.}
\label{fig:1}
\end{figure}

Based on Theorem \ref{thm:strcon}, it is clear that $|\mathcal{S}^c(\mathcal{F})|\leq |\mathcal{S}^r(\mathcal{F})|$. Fig.~\ref{fig:1} (a) shows a graph whose largest reachable set, for $|\mathcal{F}|=1$, is larger than the largest controllable set. In graph (b) the largest reachable and controllable sets are identical. Note that $\mathcal{S}^c(\mathcal{F})$ is not necessarily a subset of $\mathcal{S}^r(\mathcal{F})$ as shown in graph (c). The source nodes are shown with a darker color. 
Due to resource constraints, the attacker naturally tries to solve either of the following problems: 
\begin{itemize}
    \item [(i)] Controlling (or reaching) the {\it largest} possible subset of nodes in the network with a given number of attacked nodes, or
    \item[(ii)] Controlling (or reaching) the whole network with {\it minimum} number of attacked nodes.
\end{itemize}
 The latter has been investigated under the context of minimal structural controllability problems \cite{Sergiocomplex}. It is shown that the problem of finding the minimum number of control input (attacked) nodes is in general NP hard and in specific cases (such as dedicated inputs) can be solved in polynomial time. When reachability is the objective, the problem is related to estimating the reachable set of nodes which can be solved in polynomial time \cite{aji}.

A system is called strong structurally controllable if ${\rm rank}(\boldsymbol W,\boldsymbol B)=n$ for { {\it all} (nonzero)} choice of free parameters in $\boldsymbol W$ and $\boldsymbol B$. The dimension of the strong structurally controllable subspace is the minimum rank of the controllablility matrix, $\mathcal{C}\left(\boldsymbol W,\boldsymbol B\right)$. There are graph-theoretic bounds on this quantity for consensus dynamics as stated below.

\begin{thm}[{\cite{yasin}}]
For set $\mathcal{F}$ chosen by the attacker, the dimension of the strong structurally controllable subspace is lower bounded by 
\begin{equation}
{\rm rank}\hspace{1mm}\mathcal{C}\left(\boldsymbol W,\boldsymbol B\right)\geq \max_{i\in \mathcal{V},j\in \mathcal{F}}{\rm dist}(i,j)+1,
\end{equation}
\label{thm:lowercontrol}
\end{thm}
where ${\rm dist}(i,j)$ is the shortest distance between nodes $i$ and $j$. Tighter lower bounds for the dimension of the strong structurally controllable subspace can be found in \cite{yasin}.  According to Theorem \ref{thm:lowercontrol}, from the attacker's perspective, the optimal decision is to select a node with maximum distance from the rest of the nodes in the graph. Another interpretation from the above result, which had been discussed before in \cite{Rahmani}, is the reverse effect of network connectivity on the controllability. Specifically, sparse networks may contain pairs of nodes that are far apart, and  consequently may have large controllability subspaces, while well-connected networks may have smaller controllability subspaces.

\subsubsection{Gramian-Based Controllability}

Unlike the (discrete) structural or rank-based controllability measures of dynamical systems \cite{kalman}, the  controllability Gramian provides a continuous measure for this property, in the form of the energy required to drive the dynamical system towards specific directions in the state space.  The $T$-step controllability Gramian is defined as
\begin{equation}
    \mathcal{W}_{\mathcal{F},T}\triangleq \sum_{\tau=0}^{T-1}\boldsymbol W^{\tau}\boldsymbol B_{\mathcal{F}}\boldsymbol B_{\mathcal{F}}'(\boldsymbol W')^{\tau}
\end{equation}
where $\boldsymbol {B}_{\mathcal{F}}$ corresponds to the set of attacked nodes and defined in \eqref{eqn::OAEGI0}. The controllability Gramian $\mathcal{W}_{\mathcal{F},T}$ is
positive definite if and only if the system is controllable in $T$ steps \cite{kailath}.   %The spectra of the controllability Gramian determines the energy needed for a control system to reach certain states. 
However, even if a system is controllable, certain directions of the state space may be hard to reach \cite{sun}. The smallest eigenvalue of the Gramian, $\lambda_1(\mathcal{W}_{\mathcal{F},T})$,  is inversely related to the amount of energy required to move the system in the direction
that is the most difficult to control, i.e., the eigenvector corresponding to $\lambda_1(\mathcal{W}_{\mathcal{F},T})$. Other controllability metrics, such as ${\rm trace}(\mathcal{W}_{\mathcal{F},T})$ and ${\rm trace}(\mathcal{W}_{\mathcal{F},T}^{-1})$, quantify the energy needed on average to move the system around on the state space. From the attacker's perspective, the system should be easily controllable. Thus, it targets nodes in which one of the above mentioned spectra is optimized. In particular, the attacker attempts to minimize its effort to steer the system by maximizing $\lambda_1(\mathcal{W}_{\mathcal{F},T})$ and ${\rm trace}(\mathcal{W}_{\mathcal{F},T})$ or minimizing ${\rm trace}(\mathcal{W}_{\mathcal{F},T}^{-1})$.

The selection of control nodes to optimize the spectrum of the Gramian does not generally admit a closed-form solution. The exception is ${\rm trace}(\mathcal{W}_{\mathcal{F},T})$ for which  \cite{Pasqualetti}
$$
{\rm trace}(\mathcal{W}_{\mathcal{F},T})=\sum_{i\in\mathcal{F}}\bigg(\sum_{\tau=0}^{T-1}\boldsymbol W^{2\tau}\bigg)_{ii}.
$$
When $\boldsymbol W$ is Schur stable, then $\sum_{\tau=0}^{\infty}\boldsymbol W^{2\tau}=(I-\boldsymbol W^2)^{-1}$. For continuous time systems, if $\boldsymbol W$ is Hurwitz, this closed form solution becomes ${\rm trace}(\mathcal{W}_{\mathcal{F}})=\sum_{i\in\mathcal{F}}\big(\boldsymbol W^{-1}\big)_{ii}$. For other metrics, selecting optimal nodes, from the attacker's perspective, is a combinatorial problem and (in general) hard to solve. However, recent studies on submodularity and monotonicity of some of those metrics indicate that greedy algorithms for selecting the control nodes result in a sub-optimal solution with a guaranteed performance bound \cite{tyler, alexx, Clarkbook, Summers}. 

In the following subsection, we outline a graph-theoretic interpretation of the attacker's strategy to optimize ${\rm trace}(\mathcal{W}_{\mathcal{F},T})$ in consensus dynamics.

\subsubsection{Case Study: Consensus Dynamics}
\label{sec:h9c8h}
We consider  Gramian-based controllability on two types of consensus dynamics  on undirected graphs, namely average consensus and leader-following consensus. 

\begin{figure}[t!]
\centering
\includegraphics[scale=.5]{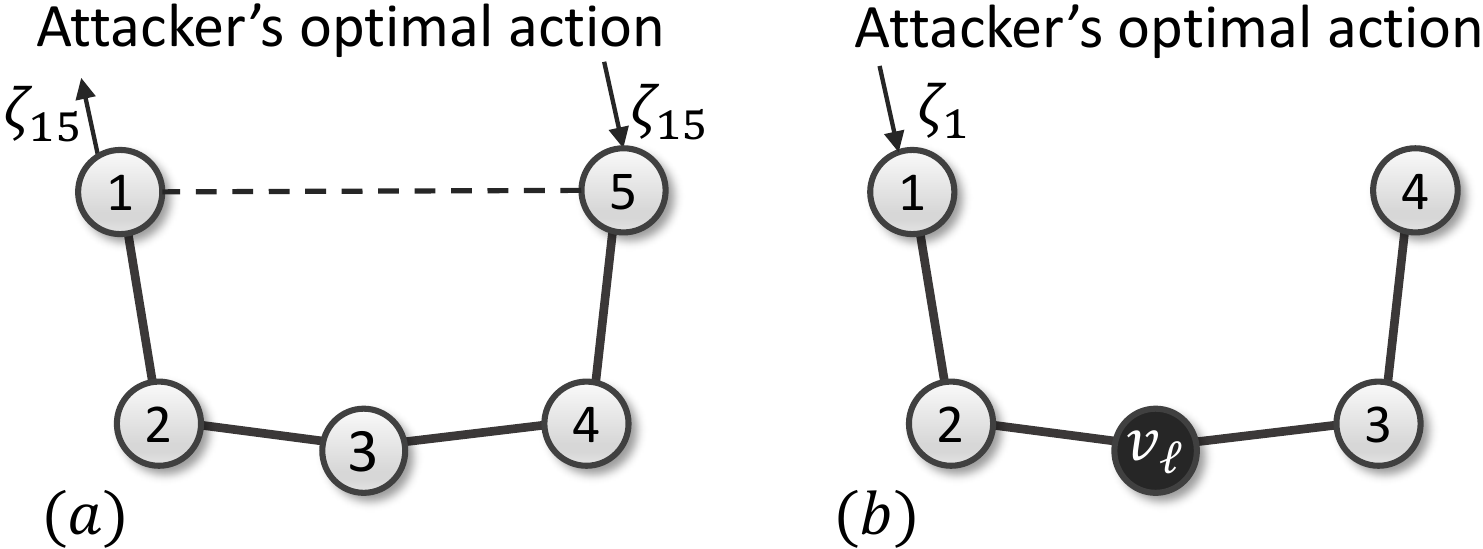}
\caption{(a) Attacker's optimal decisions based on \eqref{eqn:traceclosedform}, (b) attacker's optimal decision based on \eqref{eqn:n0d9sn}.}
\label{fig:1214}
\end{figure}

\textbf{1. Edge Attack in Average Consensus:} Assume that the attack happens in the form of a flow which enters one node and exists from another node. In particular, we say that $ik\in \mathcal{F}$ if the pair $i, k$ are chosen by the attacker and their dynamics are 
\begin{align}
\dot{\psi}_i&=\sum_{j\in \mathcal{N}_i^{\rm in}}(\psi_j-\psi_i)+\zeta_{ik},\nonumber\\
\dot{\psi}_k&=\sum_{j\in \mathcal{N}_k^{\rm in}}(\psi_j-\psi_k)-\zeta_{ik},
\end{align}
where $\zeta_{ik}$ is the attack flow.
This type of input signal, as schematically shown in Fig.~\ref{fig:1214} (a), happens in  power systems (DC input links) and distribution networks \cite{piranipower, porthamiltonian}. If we write the dynamics in vector form, it becomes 
\begin{align}\label{eqn:nc0a8gh2}
\dot{\boldsymbol\psi}&=-L\boldsymbol\psi+\mathcal{B}\boldsymbol\zeta,
\end{align}
{ where $\mathcal{B}$ is the incidence matrix of the graph induced by the attacked edges.} The controllability Gramian is 
\begin{equation}
\mathcal{W}_c=\int_0^{\infty} \mathbf{e}^{-L\tau}\mathcal{B}\mathcal{B}'\mathbf{e}^{-L'\tau}d\tau. \label{eqn:efn}
\end{equation}
Since $L$ is marginally stable, the infinite integral does not exist. However, the eigenvector of the marginally stable eigenvalue belongs to the subspace corresponding to the consensus value, which is of little interest to the attacker (as otherwise no attack would be needed). We remove this subspace by grounding one node (removing a row and column corresponding to that node) which makes the Laplacian non-singular. The {\it grounded Laplacian} matrix induced by the grounded node $v$ is denoted by $L_{fv}$ (or simply $L_f$). Then we have 
\begin{align}
{\rm tr}(\mathcal{W}_c)={\rm tr}\left( \mathcal{B}'\int_0^{\infty}\mathbf{e}^{-2L_f\tau}d\tau \mathcal{B}\right) &=\frac{1}{2}{\rm tr} \left(\mathcal{B}'L^{-1}_f\mathcal{B}\right).
\end{align}
It was shown in \cite{Boyd} that the above value is independent of the choice of the grounded node, i.e., we have 
\begin{align}
{\rm tr}\left(\mathcal{B}'L^{-1}_f\mathcal{B}\right)={\rm tr}\left(\mathcal{B}'L^{\dagger}\mathcal{B}\right)=\sum_{ij\in \mathcal{F}}\mathfrak{R}_{ij},\label{eqn:traceclosedform}
\end{align}
{ where $L^{\dagger}$ is the Moore–Penrose inverse of $L$} and $\mathfrak{R}_{ij}$ is the effective resistance between nodes $i$ and $j$. 
 Equation \eqref{eqn:traceclosedform} indicates that the trace of the controllability Gramian is the summation of effective resistances between node pairs chosen by the attacker. Thus, if the attacker seeks to maximize \eqref{eqn:traceclosedform}, i.e., minimize the attack energy,  by choosing $m$ node pairs, it should choose $m$ pairs with {\it the largest effective resistance} in the graph.  Figure~\ref{fig:1214} (a) is an example of an optimal attack. 

\textbf{2. Node Attack in Leader-Follower Consensus:}
We consider a leader-follower dynamical system on undirected graphs which is widely studied in formation control problems \cite{Rahmani, ArxiveRobutness, haobarroah}. Based on this model, there is a leader, which has access to the control input or determines the set-point,  and a set of followers, which follow the state of the leader. The dynamics of leader and followers are given by
\begin{equation}
\begin{bmatrix}
      \dot{\boldsymbol{\psi}}_{f}(t)          \\[0.3em]
       \dot{\boldsymbol{\psi}}_{\ell}(t) 
     \end{bmatrix}=-\underbrace{ { \begin{bmatrix}
      L_{f} & L_{\ell f}           \\[0.3em]
       L_{f\ell} & L_{\ell}           
     \end{bmatrix}}}_{L}\begin{bmatrix}
      {\boldsymbol{\psi}}_f(t)          \\[0.3em]
       {\boldsymbol{\psi}}_{\ell}(t) 
     \end{bmatrix}+\begin{bmatrix}
      \boldsymbol B_{\mathcal{F}}       \\[0.3em]
       0
     \end{bmatrix}\boldsymbol{\zeta}_{\mathcal{F}}(t),
\label{eqn:mat}
\end{equation}
where $\boldsymbol \psi_f(t), \boldsymbol \psi_{\ell}(t)$ and $\boldsymbol \zeta_{\mathcal{F}}(t)$ are the state of followers, the state of the leader, and the attack vector, respectively. The leader is not affected by communication attacks. It also keeps its state constant and does not incorporate the states of the followers, i.e.,  $\dot{\boldsymbol\psi}_{\ell}(t)=0$. Thus, we have $ L_{f\ell} = L_{\ell}=0$.  Matrix $\boldsymbol B_{\mathcal{F}}$, { formed from indicator vectors as in \eqref{eqn::OAEGI0}}, determines the nodes that are under attack  and $L_{\ell f}$ determines the connection  of the leader to the followers. Matrix $L_f$ is the grounded Laplacian matrix discussed before. One of the key properties of this matrix, which is used in this analysis, is that $[L_f^{-1}]_{ii}=\mathfrak{R}_{i\ell}$, where $\mathfrak{R}_{i\ell}$ is the effective resistance between node $i$ and the leader. 
Similar to the case of edge attack, the objective of the attacker is to maximize the trace of the controllability Gramian. Calculating the Gramian integral yields
\begin{equation}
    {\rm trace}(\mathcal{W}_{\mathcal{F}})=\sum_{i\in\mathcal{F}}\big(L_f^{-1}\big)_{ii}=\sum_{i\in \mathcal{F}}\mathfrak{R}_{i\ell}. 
    \label{eqn:n0d9sn}
\end{equation}
 Based on \eqref{eqn:n0d9sn},  if the attacker wants to minimize the average energy by attacking $m$ nodes, it must select $m$ nodes with the largest effective resistance from  $ {\ell}$.  Figure~\ref{fig:1214} (b) is an example of an optimal attack.

\subsection{System Norm Approaches}
\label{sec:robust}

Another way to quantify the attacker's impact is through the use of system norms from the attack signal to the output of interest, e.g., the state of the agents. Two widely used system norms in robust control are $\mathcal{H}_2$ and $\mathcal{H}_{\infty}$ norms. 
Since these methods were initially developed on continuous time systems, we focus on such systems in this section.

Suppose that the evolution of a network of agents is described by \eqref{eqn:struccont}
where $\boldsymbol D=0$ and $\boldsymbol B=\boldsymbol{B}_{\mathcal{F}}$ which corresponds to the set of attacked nodes defined in \eqref{eqn::OAEGI0}. The transfer function of the error dynamics from attack input $\boldsymbol{\zeta}_{\mathcal{F}}(t)$ to output $\boldsymbol{y}(t)$ in the Laplace domain is $G(s)\triangleq \boldsymbol C(sI-\boldsymbol W)^{-1}\boldsymbol B_{\mathcal{F}}$. The system $\mathcal{H}_2$ and $\mathcal{H}_{\infty}$ norms are defined as
\begin{align}\label{eqn:defsys}
&\|G\|_2 \triangleq \left( \frac{1}{2\pi}{\rm trace}\int_0^{\infty}G^*(j\omega)G(j\omega)d\omega \right)^{\frac{1}{2}},\nonumber \\
&\|G\|_{\infty} \triangleq \sup_{\omega\in \mathbb{R}}{\sigma_{\rm max}(G(j\omega))},
\end{align}
where $\sigma_{\rm max}(.)$ is the maximum singular value of a matrix. 
The system $\mathcal{H}_2$ norm can also be calculated based on the observability Gramian $\mathcal{W}_o$, which is the solution of the following Lyapunov equation
\begin{align}\label{eqn:lyap}
    \boldsymbol W'\mathcal{W}_o+\mathcal{W}_o\boldsymbol W=-\boldsymbol C'\boldsymbol C,\nonumber\\
    \|G\|_2={\rm trace}(\boldsymbol B_{\mathcal{F}}'\mathcal{W}_o\boldsymbol B_{\mathcal{F}}).
\end{align}
Unlike the approach of the structured system theory to resilient NCSs, which was based on nonzero patterns of  system matrices and not the magnitude of the elements, the system norm approach  depends on the exact value of the matrix elements. Thus, having knowledge about  the nonzero patterns of dynamic matrices is not sufficient and one has to specify the type of  matrices  which describe the interactions between agents. We revisit the leader-follower consensus dynamics discussed in Section \ref{sec:revisiting} to further explain this fact.

\begin{remark}\textbf{($\mathcal{H}_2$ versus  $\mathcal{H}_{\infty}$ norm)}
From a security perspective, either the system  $\mathcal{H}_2$ or  $\mathcal{H}_{\infty}$ norm can be used to quantify the attack impact. If the frequency content of the attack signal is unknown, using a $\mathcal{H}_2$ norm is a more reasonable choice as it is calculated over all frequencies. However, if the objective is to find the worst-case attack impact over all frequencies, the system $\mathcal{H}_{\infty}$ norm is an appropriate choice. 
\end{remark}

\iffalse

From the attacker's perspective, there are two types of outputs: (i) the set of nodes which are equipped with detection sensors, $\mathcal{D}$, as discussed in Section XX, in which the attacker tries to have small impact on to be less visible. (ii) The set of target nodes nodes, $\mathcal{T}$, chosen by the attacker which it tries to have large impact on. These two sets can have nodes in common. The output equation in \eqref{eqn:mat23} for cases (i) and (ii) can be written with $C=C_{s}$ and $C=C_{t}$, respectively. The visibility and the impact of the attacker  is determined by some system norms from the attack signal to the corresponding sets, i.e., {\tb Remove visibility. Only Impact.}
\begin{align}
&{\rm Visibility:}  &&\mathfrak{V}\triangleq\|G(A_{ra},C_{s})\|_p,\nonumber\\
&{\rm Impact:}  &&\mathfrak{I}\triangleq\|G(A_{ra},C_{t})\|_p,\nonumber\\
&{\rm Attacker's Objective:}  &&J(A_{ra},C_{s},C_{t})=\mathfrak{V}+\lambda \mathfrak{I}
\end{align}
where $p$ determines an appropriate system norm. In cases where the impact on all nodes is desired, i.e., all nodes are target nodes, we have $C_{t}=I$. Here, $\mathfrak{V}$ is a continuous measure of the attacker visibility, compared the rank of the matrix pencil which is a discrete measure. Parameter $\lambda \geq 0$ determines the dominance of visibility to impact from the attacker's perspective.

{\tb Indicate that this method can handle malicious and Byzantine agents as it is an input-output approach. }

{\tb Show that in an informative figure, showing sensor nodes and target nodes.}

\fi

\subsubsection{Case Study: Consensus Dynamics Revisited} \label{sec:revisiting}
\iffalse
Consider the leader-follower dynamics on undirected graphs \eqref{eqn:mat}.  We study two scenarios
\newline
\textbf{1. Attack impact on all nodes:} In the first scenario, the attack impact is modeled as the $\mathcal{H}_{2}$ norm from the attack signal $\boldsymbol{\zeta}(t)$ to the state of followers $\boldsymbol{\psi}_f(t)$, i.e., $C=I$. Since the system $\mathcal{H}_2$ norm is obtained by integrating the transfer function over all frequencies, as in \eqref{eqn:defsys}, it is an appropriate measure for the cases where the frequency content of the attack signal is unknown. Solving the Lyapunov equation \eqref{eqn:lyap} yields
\begin{equation}
\|G\|_2=\frac{1}{2}{\rm trace}(\boldsymbol B_{\mathcal{F}}'L_f^{-1}\boldsymbol B_{\mathcal{F}})=\frac{1}{2}\sum_{i\in \mathcal{F}}\mathfrak{R}_{i\ell},
\end{equation}
which is the same as trace of the controllability Gramian discussed in \eqref{eqn:n0d9sn}. Hence, from the attacker's perspective, optimizing the system $\mathcal{H}_2$ norm and the energy for controlling the system results in the same set of attacked nodes. 
\fi

\textbf{1. Edge Attack in Average Consensus:} Consider the consensus dynamics under an attack flow as in \eqref{eqn:nc0a8gh2}. Suppose that the attacker measures the difference between node values under attack, i.e., $\boldsymbol y=\mathcal{B}'\boldsymbol\psi$. Dynamics \eqref{eqn:nc0a8gh2} together with this measurement form a symmetric system, i.e., $L$ is symmetric and $\boldsymbol B=\boldsymbol C'$. Thus, the $\mathcal{H}_{\infty}$ norm is equal to the DC gain of the system \cite{tan}, i.e., for the transfer function from $\boldsymbol\zeta $ to $\boldsymbol y$ we have
\begin{equation}\label{eqn:cvan9082}
   { \|G\|_{\infty}}=\sigma_{\rm max}\big( \mathcal{B}'L^{\dagger}\mathcal{B} \big)
\end{equation}
Unlike the trace of   $\mathcal{B}'L^{\dagger}\mathcal{B}$, in \eqref{eqn:traceclosedform}, interpreting its largest singular value is hard. We consider the simple case where only one node pair, $i$ and $j$, is under attack, i.e., $\mathcal{B}=\mathbf{e}_{ij}$. In this case, \eqref{eqn:cvan9082} becomes scalar and we have $\|G\|_{\infty}=\sigma_{\rm max}\big(\mathbf{e}_{ij}'L^{\dagger}\mathbf{e}_{ij}\big)=\mathfrak{R}_{ij}$. Thus, in order to have a large impact, the attacker must choose node pairs with the largest effective resistance in the network.

\textbf{2. Node Attack in Leader-Follower Consensus:}  Here, a single attacker $i$ targets a set of nodes and $\boldsymbol B_{\mathcal{F}}=C'$. As before, the $\mathcal{H}_{\infty}$ norm is equal to the DC gain of the system  { \cite{Farina}} and we have  $\|G\|_{\infty}=\sigma_{\rm max}\big(\boldsymbol B_{\mathcal{F}}'L_f^{-1}\boldsymbol B_{\mathcal{F}}\big)$. For the case of a single node under attack, we have $\boldsymbol B_{\mathcal{F}}=\mathbf{e}_i$ and   $\|G\|_{\infty}=\mathbf{e}_i'L_f^{-1}\mathbf{e}_i=\mathfrak{R}_{i\ell}$. Hence, to have a large impact on a target node, the attacker must choose a node in the network with the largest effective resistance from the leader.

\section{Related Problems in Resilient Networked Control Systems}

In this section, we briefly discuss other problems  on the resilience of networked control systems which use graph theory as a tool in the analysis.

\subsection{Resilience to Actuator/Sensor and Link Removals}

So far, we discussed the case where the agents or the communications between the agents are under attack.  In that context, we assume that the control inputs are not affected by the attacker. 
In some situations, however, the attacker may choose to remove certain sensors, actuators, or communication links entirely. In those cases, the main concern is to retain the controllability of the system. This can be written in terms of a {\it robust structural controllability} problem.  In particular, the objective is to maintain a system's controllability despite the removal of a subset of actuators.

Consider the linear time invariant system \eqref{eqn:struc} where a subset of control inputs are removed (potentially due to adversarial actions). In this case, the minimum number of actuators which retain the controllability of the system is determined by the following problem 
\begin{equation}
    \begin{aligned}
        &\underset{\boldsymbol B\in \mathbb{R}^{n\times m}}{\text{arg min}}&& \|\boldsymbol B\|_0\\
        &\text{such that} 
        && (\boldsymbol W, {\boldsymbol B}_{\mathcal{U}\setminus{\mathcal{F}}})  {\text{ is struc. cont. $ \forall \mathcal{F} \subset \mathcal{U}$}},\\
         &&&\text{$|\mathcal{F}|\leq m$,} 
    \end{aligned}
       \label{eqn:po1}
\end{equation}
where  ${\boldsymbol B}_{\mathcal{U}\setminus{\mathcal{F}}}$ corresponds to the structure of the input matrix $\boldsymbol B$ whose columns corresponding to set $\mathcal{F}$ are removed (i.e., actuators are failed). The number of actuator faults are upper bounded by $m$. It is shown that the above  problem is NP-hard and polynomial time algorithms to approximate the solution of those problems have been proposed \cite{Xiaofei, sergiorobust}.

In other set of problems, the attacker targets a set of links in the network to remove. Several performance measures may be affected by such an action.  When robust controllability (or observability) is of interest, the problem can be written as follows \cite{Xiaofei}
\begin{equation}
    \begin{aligned}
        &\underset{\boldsymbol B\in \mathbb{R}^{n\times m}}{\text{arg min}}&& \|\boldsymbol B\|_0\\
        &\text{such that} 
        && (\boldsymbol W_{\mathcal{E}_{\mathcal{X},\mathcal{X}}\setminus{\mathcal{E}_{\mathcal{F}}}}, {\boldsymbol B}) {\text{ is struc. cont. $\forall \mathcal{E}_{\mathcal{F}} \subset \mathcal{E}_{\mathcal{X},\mathcal{X}}$}},\\
        &&&\text{$|\mathcal{E}_\mathcal{F}|\leq \bar{m}$,} 
    \end{aligned}
    \label{eqn:optim}
\end{equation}
where $\mathcal{E}_{\mathcal{F}}$ is the set of edges affected by the attacker, upper bounded by $\bar{m}$. 
Similar to \eqref{eqn:po1}, the above problem is NP-hard  in general.

\subsection{Strategic Sensor and Actuator Placement on  Graphs}
\label{sec:nc08w2}
Our focus in the previous sections was to find graph  conditions which ensure resiliency to attacks for certain distributed algorithms. 
However, in many situations, the underlying network topology is sparse and cannot be changed. Furthermore,  the number of sensors/actuators is limited. In these cases, an alternative approach is to place those limited number of sensors (or actuators) on specific nodes in the network in order to optimally detect the attack or mitigate its impact.

There is a vast literature on sensor (or actuator) placement to enhance the observability (or controllability) in terms of the rank of the observability (or controllability) matrix, Gramian-based metrics (as discussed previously),  or the error variance of the Kalman filters \cite{Ayoub, ye2020complexity}. In all these problems, there is a single decision maker which deploys sensors (or place actuators) on a set of nodes.
However, in security problems, the adversary plays as another decision maker which tries to optimize its own cost function, e.g., maximize impact or minimize visibility. This introduces a {\it strategic} sensor (or actuator) placement problem, taking the attacker's actions into account. In this direction, game theory can be used as a powerful tool to address this set of problems; see  \cite{zhubasar, Manshaie, Alpcan, Basarbook,ye2020resilient} and references therein.

Strategic sensor placement in the network to detect cyber-attacks has been recently studied \cite{jezdimirr, aminn, Piranigame, piranisinopoli}. In this setting, the attacker seeks to apply attack inputs while being stealthy and the detector tries to detect the attack.
Several approaches have been adopted to characterize the equilibria of the security games. Nash equilibrium is used to model simultaneous decision making and Stackelberg game model is used for the case where the defender must act before the attacker. In design problems, the Stackelberg game is a popular approach to defend against cyber-attacks. In particular, the detector acts as the game leader and places sensors on nodes considering the worst case attack strategies. The applicability of each method, based on the nature of the attack and the structure of the cyber physical system, is discussed in \cite{Manshaie}. 

In addition to the strategic attack detection, a defense mechanism can help  mitigate the impact of the attack via certain control actions \cite{Piranigameecc, Qhe, Gueye, zhubasar}. The attacker's impact can be quantified by either of the methods discussed in Section \ref{sec:quant}. A comprehensive defense strategy must include both strategic detection and mitigation mechanisms. This is a potential avenue for further research.

\subsection{Network Coherence as a Measure of Resiliency}

In the network control systems literature, the notion of network coherence is used to quantify the ability of a network to reject communication disturbances while performing a formation control or a consensus algorithm in large scale systems \cite{Bamieh2}. It is usually described in terms of system $\mathcal{H}_2$ or $\mathcal{H}_{\infty}$ norms from the disturbance signal to the output of interest, e.g., position (phase) or velocity (frequency) \cite{Emma, CDCPower, Poola}. Interpreting the disturbances as attack inputs, some of the approaches and the results in this line of research, e.g., scalability of algorithms and leader selection schemes, can be readily used in resilient distributed  algorithms, as we will briefly discuss in the following paragraphs.

\subsubsection{Leader Selection}
The objective is to choose the optimal nodes in the network as leaders, i.e., nodes which receive the control signal, such that the network coherence is maximized, i.e.,  system $\mathcal{H}_2$ or $\mathcal{H}_{\infty}$ norms from the disturbance input to the output of interest are minimized. It is shown in \cite{Leonard2} that the optimal leader to minimize the $\mathcal{H}_2$ norm in consensus dynamics is the {\it information central} node in the network: a node in which the summation of effective resistances to the rest of the nodes in the network is minimized. It is also shown that the leader which optimizes the $\mathcal{H}_{\infty}$ norm  is  not necessarily the graph's information center \cite{ArxiveRobutness}. Moreover, a graph-theoretic condition for the leader to co-optimize both metrics is discussed in \cite{ArxiveRobutness}.
 
 \subsubsection{Scalability}
Another problem of interest is the scalability of the  network coherence in graphs with various structures. In \cite{Bamieh2}, it is discussed that how the network coherence scales with the network size for regular lattices in 1, 2 and higher dimensions. The scalability of these metrics is also discussed for random graphs. In particular, for Erd\H{o}s-R$\acute{e}$nyi Random Graphs and random regular graphs, tight characterizations of the network coherence are discussed  in \cite{ArxiveRobutness, PiraniSundaramArxiv}.

%\subsubsection{Attacker-Defender Games}

%\subsubsection{Optimal Sensor Placements for Distributed Filtering/Estimation}

%\subsubsection{Optimal Sensor Placement for Attack Detection}

%\subsubsection{Leader Selection for Optimal Network Coherence}

%\subsection{Game-Theoretic Frameworks}

\section{Resilience in Certain Classes of Graphs}
\label{sec:nca0sc8ag0}
In this section, we summarize some pertinent topological properties of certain classes of graphs. We focus on the properties of graph connectivity and graph robustness  provided in Section~\ref{sec:connectivity}, since they are particularly relevant to the resilience of distributed algorithms against adversarial actions. Our focus is on undirected networks, unless indicated. We start from simpler cases. 

\subsection{Paths, Cycles, Trees, and Complete Graphs}
An undirected path of length $n$ is the simplest connected graph with connectivity $\kappa=1$ and robustness $r=1$. A cycle is a path of length $n$ whose start and end nodes are connected. For a cycle graph, the connectivity is $\kappa=2$ and the robustness is $r=1$. Trees are connected acyclic graphs with connectivity $\kappa=1$ and robustness $r=1$.  A complete graph is $(n-1)$-connected and $\lceil \frac{n}{2} \rceil$-robust.

%An undirected path of length $n$ is the simplest connected graph with connectivity $\kappa=1$ and robustness $r=1$. Cycle is a path of length $n$ whose start and end nodes are connected. For a cycle graph, the connectivity is $\kappa=2$ and the robustness is $r=1$. Trees are connected acyclic graphs with connectivity $\kappa=1$ and robustness $r=1$, and isoperimetric constant at most $i(\mathcal{G})=1$, for star graphs, and at least $i(\mathcal{G})=\frac{2}{n}$, for path graphs. A complete graph is formed when $(i,j)\in \mathcal{E}$ for all $i, j\in \mathcal{V}$ and $i\neq j$. A complete graph is $(n-1)$-connected and $\lceil \frac{n}{2} \rceil$-robust. The isoperimetric constant of a complete graph is  $\frac{n}{2}$ ($n$ even). 

\subsection{Circulant Networks, $k$-Nearest Neighbor Paths, and 1-D Random Geometric Graphs}

\begin{figure}[t!]
\centering
\includegraphics[scale=.5]{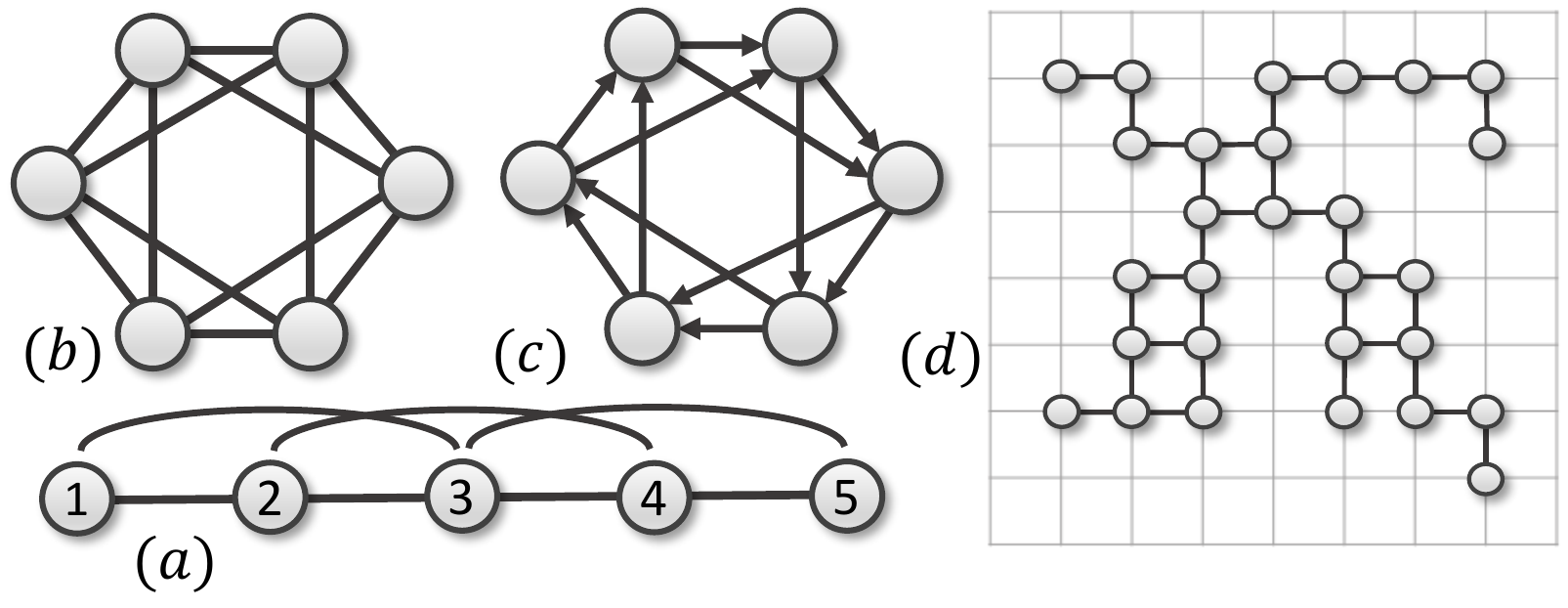}
\caption{(a) A 2-nearest neighbor path (b,c) an undirected and a directed circulant graph, (d) A connected formation on a 2-dimensional lattice.}
\label{fig:sampleclasses}
\end{figure}

\begin{definition}[\textbf{$k$-Nearest Neighbor Paths}]
 A $k$-nearest neighbor path,
$\mathcal{P}(n,k)$, is a network comprised of $n$ nodes in a path, where the nodes are labeled as $ 1,  2,..., n$ from one end of the path to the other and each node $i$ can communicate with its $k$ nearest neighbors behind it and $k$ nearest neighbors ahead of it, i.e.,  $ {i-k} ,  {i-k+1},..., {i-1},  {i+1},  {i+2},..., {i+k}$, for some
$k\in \mathbb{N}$. An example of $\mathcal{P}(n,k)$ is shown in Fig.~\ref{fig:sampleclasses} (a).
\end{definition}

Such $k$-nearest neighbor paths are relevant for modelling vehicle platoons, due to the limited sensing and communication range for each vehicle.
It is shown in \cite{piranitvt} that a $k$-nearest neighbor path, $\mathcal{P}(n,k)$, is a $k$-connected graph. We will discuss the robustness of $\mathcal{P}(n,k)$ later in this section. A similar structure to  $\mathcal{P}(n,k)$ is a 1-dimensional geometric random graph which captures edges between nodes that are in close (spatial) proximity to each other. 
\begin{definition}[\textbf{Geometric Random Graphs}]
A geometric random graph $\mathcal{G}_{n,\rho, l}^d=\{\mathcal{V},\mathcal{E}\}$ is an undirected graph
generated by first placing $n$ nodes (according to some mechanism) in a $d$-dimensional region $\Omega_d=[0,l]^d$, where $d\in \mathbb{Z}_{\geq 1}$. We denote the position of node $i\in \mathcal{V}$ by $x_i\in \Omega_d$. Nodes $i,j\in \mathcal{V}$ are connected
by an edge if and only if $\|x_i-x_j\|\leq \rho$ for some threshold $\rho$, where $\|.\|$ is some appropriate norm (often taken to be the standard Euclidean norm). When the node positions are generated randomly (e.g., uniformly and independently) in the region, one obtains a geometric random graph.
\label{def:cjd09}
\end{definition}

In the more general models of $\mathcal{G}_{n,\rho, l}^d$, the length $l$ is also allowed to increase and the density $\frac{n}{l^d}$ can converge to some constant, making it suitable for capturing both dense and sparse random networks. { The following result holds for 1-dimensional geometric random graphs.}

\begin{prop}
In $\Omega_1=[0,l]$ with fixed $l$, if { $\mathcal{G}_{n,\rho, l}^1$} is $k$-connected, then it is at least $\lfloor \frac{k}{2}\rfloor$-robust. 
\label{prop:cnsa02}
\end{prop}

Based on Definition \ref{def:cjd09}, the $k$-nearest neighbor path can be seen as a geometric graph $\mathcal{G}_{n,\rho, l}^d$ with $\rho=\frac{lk}{n-1}$ and placing the nodes as follows: the first node, $ 1$, is placed on one end of the line and the $i$-th node is placed at distance $\frac{(i-1)l }{n-1}$ from $ 1$. Thus, based on Proposition \ref{prop:cnsa02} and the fact that the $k$-nearest neighbor path $\mathcal{P}(n,k)$ is $k$-connected, we conclude that it is at least $\lfloor \frac{k}{2}\rfloor$-robust.

%{\tb Seroiusly work on the two properties: 1- Connectivity of directed $k$-nn platoon and 2- Robustness of this network. }

\begin{definition}[\textbf{Circulant Graphs}]
An undirected graph of $n$ nodes is called circulant if the $n$ vertices of the graph can be numbered from 0 to $n-1$ in such a way that if some two vertices numbered $x$ and $(x \pm d)$ $\rm mod$ $n$ are adjacent, then every two vertices numbered $z$ and $(z \pm d)$  $\rm mod$ $n$ are adjacent. A directed graph is circulant   (with the above labeling) if some two vertices numbered $x$ and $(x + d)$  $\rm mod$ $n$ are adjacent, then every two vertices numbered $z$ and $(z + d)$ mod $n$ are adjacent.
\end{definition}

Informally speaking, an undirected $k$-circulant graph is a $k$-nearest neighbor cycle graph. Thus, with the same reasoning, an undirected 
$k$-circulant graph is $2k$-connected and at least $\lfloor \frac{k}{2}\rfloor$-robust. The result is extended to directed circulant graphs where it is shown that it is at least $\lfloor \frac{k+2}{4}\rfloor$-robust \cite{Usevitch}. An example of an undirected and a directed circulant graph is shown in Fig.~\ref{fig:sampleclasses} (b,c).

\subsection{Formation Graphs on 2-Dimensional Lattice}
 A specific type of geometric graph is the two dimensional lattice which has been widely used in 
formation control of autonomous robots \cite{kumar, kumar2}. 
  A lattice is a set of linear combinations with integer coefficients of the elements of a basis of $\mathbb{R}^2$.  The elements of
the set are lattice points. Let $\boldsymbol v_1$ and $\boldsymbol v_2$ be bases of a 2-dimensional lattice with $\|\boldsymbol v_1\|=\|\boldsymbol v_2\|=\ell$ where $\ell$ is the lattice length. Every point $ x$ on the lattice can be described by $ x=a_i\boldsymbol v_1+b_j\boldsymbol v_2$ where $a_i,b_j\in \mathbb{Z}$. In a lattice, two nodes $i$ and $j$ are connected if $\|x_i-x_j\|\leq \ell$ where $\|.\|$ is the Euclidean norm. Given a set of nodes $\mathcal{V}$ and a distance $\ell$, the graph $\mathcal{G}_{\ell}=\{\mathcal{V},\mathcal{E}_{\ell}\}$ with edge set $\mathcal{E}_{\ell}\triangleq \{(i,j)|\|x_i-x_j\|\leq \ell\}$ is called the proximity graph of the set $\mathcal{V}$. We describe the communication range by a
function $R:\mathbb{Z}_{\geq 1} \to \mathbb{R}$ that maps the number of robots $m$ to a distance where $m$ robots are ensured to be reached. A formation of $n$ robots
is said to be connected if its associated proximity graph $\mathcal{G}_{\ell}$ is
connected. An example of a lattice and a connected formation is shown in Fig.~\ref{fig:sampleclasses} (d). { It was shown in \cite{kumar} that in a connected formation of $n$ robots, every robot has at least $1\leq m\leq n-1$ robots within a distance $m\ell$. Based on this, one can compute a minimum communication range  for the robots in a formation to guarantee resilience in the communication network.} 
\begin{prop}[\cite{kumar}]
Given a set $\mathcal{V}$ of $4f+1$ nodes in a connected
formation, if the communication range of every node satisfies
$R\geq 3f\ell$, then the associated  graph of the formation is $(2f+1)$-robust.
\end{prop}

\subsection{ Random Graphs}

A common approach to modelling complex networks is via the framework of {\it random graphs}, i.e., by drawing a graph from a certain probability distribution over the set of all possible graphs on a given set of nodes.  Such random graph models have diverse applications \cite{Bollobs, newmann}, including in modeling  cascading failures in large scale systems  \cite{Crucitti, yagan1}.  Here, we summarize the connectivity and robustness properties of certain commonly studied random graph models.

\subsubsection{Erd\H{o}s-R$\acute{e}$nyi Random Graphs}

An Erd\H{o}s-R$\acute{e}$nyi (ER) random graph $\mathcal{G}(n,p)$ is a graph on $n$ nodes, where each edge between two distinct nodes is presented independently with probability $p$ (which could be a function of $n$).  We say that a graph property holds {\it asymptotically almost surely} if the probability of drawing a graph with that property goes to $1$ as $n \rightarrow \infty$. 
\iffalse
\begin{definition}
Let $\Omega_n$ be the set of all  graphs on $n$ nodes.  For a given graph function $f: \Omega_n \rightarrow \mathbb{R}_{\ge 0}$ and another function $g: \mathbb{N} \rightarrow \mathbb{R}_{\ge 0}$, we say $f(\mathcal{G}(n,p)) \le (1+o(1))g(n)$ asymptotically almost surely if there exists some function $h(n) \in o(1)$ such that $f(\mathcal{G}(n,p)) \le (1+h(n))g(n)$ with probability tending to $1$ as $n \rightarrow \infty$. An essentially identical definition holds for lower bounds of the above form.  Consider function $t(n)=\frac{g(n)}{n}$ where $g(n)\to \infty$ as $n\to \infty$ and a function $x=o(g(n))$ satisfying $x\to \infty$ as $n\to \infty$. We say that $t(n)$ is a threshold function for graph property P if $p(n)=\frac{g(n)+x}{n}$ implies that almost all $\mathcal{G}\in \mathcal{G}(n,p)$  have property P and  $p(n)=\frac{g(n)-x}{n}$ implies that almost no $\mathcal{G}\in \mathcal{G}(n,p)$  has property P.
\end{definition}
\fi
The following theorem shows the probability threshold for which a graph $\mathcal{G}\in \mathcal{G}(n,p)$ is $r$-connected and $r$-robust.
\begin{thm}[\cite{hoganfatasundaram}]
For any constant $r\in \mathbb{Z}_{\geq 1}$, $$t(n)=\frac{\ln n + (r-1)\ln\ln n}{n}$$ is a threshold function for the ER random graph $\mathcal{G}$ to have minimum degree $r$,  to be $r$-connected, and to be $r$-robust.
\label{thm:svdin}
\end{thm}

According to the example graph in Fig.~\ref{fig:by9qwfqdyz} (a), graph robustness is a much stronger property than the graph connectivity and the minimum degree. However, Theorem \ref{thm:svdin} indicates that the above threshold function for $r$-connectivity (and minimum degree $r$) is also a threshold function for the
stronger property of $r$-robustness in ER random graphs.

\iffalse

Here, we discuss the network coherence in ER random graphs.

\begin{thm} 
Consider a random graph $\mathcal{G}(n,p)$ with  $p(n) \ge \frac{c\ln{n}}{n}$, for constant $c>1$.  Let $\mathcal{S} \subset \mathcal{V}$ be a set of grounded nodes chosen uniformly at random with $|\mathcal{S}| = o(\sqrt{np})$. Then for the $\mathcal{H}_{\infty}$ disorder we have 
\begin{equation}
(1 - o(1))\frac{1}{|\mathcal{S}|p} \le \|G\|_{\infty} \le(1 + o(1))\frac{1}{|\mathcal{S}|p},
\label{eqn:per4}
\end{equation}
asymptotically almost surely. Moreover, for the $\mathcal{H}_2$ norm we have 
\begin{equation}
(1 - o(1))\frac{|\mathcal{S}|+1}{2|\mathcal{S}|p} \le \|G\|_2 \le(1 + o(1))\frac{|\mathcal{S}|+1}{2|\mathcal{S}|p},
\label{eqn:per44}
\end{equation}
asymptotically almost surely. 
\label{thm:cohh}
\end{thm}
{\tb Show a simulation result for $H_2$ and $H_{\infty}$ of ER graphs, like what I  did for TCNS.}
\fi

\subsubsection{Random Regular Graphs}

Let $\Omega_{n,d}$ be the set of all undirected graphs on $n$ nodes where every node has degree $d$ (note that this assumes that $nd$ is even).  A {\it random $d$-regular graph} (RRG), denoted $\mathcal{G}_{n,d}$ is a graph drawn uniformly at random from $\Omega_{n,d}$.  For $d\geq 3$, it is shown that $\mathcal{G}_{n,d}$ is asymptotically almost surely $d$-connected \cite{Bollobs}.
Based on \cite{Friedman}, for any $\epsilon > 0$, the algebraic connectivity of a random $d$-regular graph satisfies
\begin{equation}
\lambda_2(L) \geq d-2\sqrt{d-1}-\epsilon,
\label{eqn:vcan972ebi}
\end{equation}
asymptotically almost surely. 
As  discussed in \cite{kumar2},  if the algebraic connectivity of a graph is bigger than $r-1$, then the network is at least $\lfloor\frac{r}{2}\rfloor$-robust.
Hence, according to \eqref{eqn:vcan972ebi}, an RRG  is at least $\lfloor \frac{d-2\sqrt{d-1}}{2}\rfloor $-robust asymptotically almost surely.

\subsubsection{Random Interdependent Networks}

An interdependent network $\mathcal{G}$  is denoted by a tuple $\mathcal{G}=(\mathcal{G}_1, \mathcal{G}_2, \allowbreak \ldots, \mathcal{G}_k, \mathcal{G}_p)$ where $\mathcal{G}_l=(\mathcal{V}_l, \mathcal{E}_l)$ for $l=1,2,\ldots, k$ are called the {\it subnetworks} of the network $\mathcal{G}$, and $\mathcal{G}_p=(\mathcal{V}_1\cup \mathcal{V}_2 \cup \ldots \cup \mathcal{V}_k, \mathcal{E}_p)$ is a $k$-partite network with $\mathcal{E}_p \subseteq \cup_{l \neq t} \mathcal{V}_l \times \mathcal{V}_t$ specifying the interconnection (or inter-network) topology. Applications of interdependent networks in modelling communication networks and power grid are discussed in \cite{marzieh}. Define the sample space $\Omega_n$ to  consist of all possible interdependent networks $(\mathcal{G}_1, \mathcal{G}_2, \ldots, \mathcal{G}_k, \mathcal{G}_p)$ and the index  $n \in \mathbb{N}$ denotes the number of nodes in each subnetwork.
%Define the probability space $(\Omega_n, \mathcal{F}_n, \mathbb{P}_n)$, where the sample space $\Omega_n$ consists of all possible interdependent networks $(\mathcal{G}_1, \mathcal{G}_2, \ldots, \mathcal{G}_k, \mathcal{G}_p)$ and the index $n \in \mathbb{N}$ denotes the number of nodes in each subnetwork. %The $\sigma$-algebra $\mathcal{F}_n$ is the power set of $\Omega_n$ and the probability measure $\mathbb{P}_n$ associates a probability $\mathbb{P}(\mathcal{G}_1, \mathcal{G}_2, \ldots, \mathcal{G}_k, \mathcal{G}_p)$ to each network $\mathcal{G}=(\mathcal{G}_1, \mathcal{G}_2, \ldots, \mathcal{G}_k, \mathcal{G}_p)$. 
A random interdependent network is a network $\mathcal{G}=(\mathcal{G}_1, \mathcal{G}_2, \ldots, \mathcal{G}_k, \mathcal{G}_p)$ drawn from $\Omega_n$ according to a given probability distribution. 

We assume that $|\mathcal{V}_1|=|\mathcal{V}_2|=\cdots=|\mathcal{V}_k|=n$ and that the number of subnetworks $k$ is at least 2.  Similar to Theorem \ref{thm:svdin} for ER random graphs, there exists a sharp threshold for connectivity and robustness of random interdependent networks. 

\begin{thm}[\cite{Automaticapirani}]
\label{thm:rob_main}
Consider a random interdependent network  $\mathcal{G}=(\mathcal{G}_1, \mathcal{G}_2, \ldots, \mathcal{G}_{k}, \mathcal{G}_p)$. Then, for any positive integers $r$ and $k \ge 2$,
$$
t(n)=\frac{\ln n +(r-1) \ln \ln n}{(k-1)n}
$$
is a threshold for $r$-connectivity and $r$-robustness of $\mathcal{G}$.
\end{thm}

\subsubsection{Random Intersection Networks}
Random intersection graphs belong to class of random graphs for which every node is assigned a set of objects selected by some random mechanism. 
They have applications in wireless sensor networks, frequency hopping spread spectrum, spread of epidemics, and social networks \cite{Cohen}.

Given a node set $\mathcal{V}=\{ 1, 2,..., n\}$, each node $i$ is assigned  an object set $S_i$ from an object pool $\mathcal{P}$ consisting of $P_n$ distinct objects, where $P_n$ is a function of $n$. Each object $S_i$ is constructed using the following two-step procedure: (i) The size of $S_i$, $|S_i|$, is determined according to some probability distribution $\mathcal{D}: \{1,2,...,P_n\}\to [0,1]$ in which $\sum_{x=1}^{P_n}\mathbb{P}\left(|S_i|=x\right)=1$. (ii) Conditioning on $|S_i| = s_i$, set $S_i$ is chosen uniformly among all $s_i$-size subsets of $\mathcal{P}$. Finally, an undirected edge is assigned between two nodes if and only if their corresponding object sets have at least one object in common. There are variations of the general random intersection graph such as binomial random intersection graphs and  uniform random intersection graphs, each of which focuses on a certain probability distribution $\mathcal{D}$. The following theorem discusses the connectivity and robustness of random intersection graphs.

\begin{thm}[\cite{yagan}]
Consider a general random intersection graph $\mathcal{G}(n,P_n,\mathcal{D})$. Let $X$ be a random variable following probability distribution $\mathcal{D}$. With a sequence $\alpha_n$ for all $n$ defined through $\frac{\{\mathbb{E}[X]\}^2}{P_n}=\frac{\ln n+(r-1)\ln\ln n+\alpha_n}{n}$, if $\mathbb{E}[X]=\Omega(\sqrt{\ln n})$, $Var[X]=o(\frac{\{\mathbb{E}[X]\}^2}{n(\ln n)^2})$ and $\alpha_n=o(\ln n)$, and $\lim_{n\to \infty}\alpha_n=\infty$, then  the graph is asymptotically almost surely $r$-connected and $r$-robust.
\end{thm}

\iffalse

\begin{thm}[\cite{yagan}]
Consider a general random intersection graph $\mathcal{G}(n,P_n,\mathcal{D})$. Let $X$ be a random variable following probability distribution $\mathcal{D}$. With a sequence $\alpha_n$ for all $n$ defined through $\frac{\{\mathbb{E}[X]\}^2}{P_n}=\frac{\ln n+(k-1)\ln\ln n+\alpha_n}{n}$, if $\mathbb{E}[X]=\Omega(\sqrt{\ln n})$, $Var[X]=o(\frac{\{\mathbb{E}[X]\}^2}{n(\ln n)^2})$ and $\alpha_n=o(\ln n)$, and $\lim_{n\to \infty}\alpha_n=\infty$, then
\begin{align*}
   & \lim_{n\to \infty}\mathbb{P}(\mathcal{G}(n,P_n,\mathcal{D})\hspace{1mm} {\rm is}\hspace{1mm} k-{\rm connected})\\
 &=   \begin{cases}
    0      & \quad  {\rm if} \hspace{1mm} \lim_{n\to \infty}\alpha_n=-\infty,\\
   1 & \quad  {\rm if} \hspace{1mm} \lim_{n\to \infty}\alpha_n=\infty,\\
   e^{-\frac{e^{-\alpha^*}}{(k-1)!}} & \quad {\rm if} \hspace{1mm} \lim_{n\to \infty}\alpha_n=\alpha^*\in (\infty , \infty).\\
  \end{cases}
\end{align*}
Using the same assumptions except that $\mathbb{E}[X]=\Omega\big((\ln n)^3\big)$, we have 
\begin{align*}
   & \lim_{n\to \infty}\mathbb{P}(\mathcal{G}(n,P_n,\mathcal{D})\hspace{1mm} {\rm is}\hspace{1mm} k-{\rm robust})\\
 &=   \begin{cases}
    0      & \quad  {\rm if} \hspace{1mm} \lim_{n\to \infty}\alpha_n=-\infty,\\
   1 & \quad  {\rm if} \hspace{1mm} \lim_{n\to \infty}\alpha_n=\infty.\\
  \end{cases}
\end{align*}
\end{thm}

\fi

\section{Future Directions} 
\label{sec:future}
This paper provided an overview of the existing graph-theoretic tools which can be used to analyze the resilience of distributed algorithms in networked control systems. Comparing with system-theoretic approaches to the robustness and fault tolerance of control systems, graph-theoretic approaches are relatively new and demand more development, primarily in the following three directions:  (i)  Developing graph-theoretic methods to facilitate analyzing a wider range of distributed algorithms and more complex adversarial actions. 
 (ii) Reinterpreting the known system-theoretic notions of resilience and robustness of dynamical systems from a graph-theoretic perspective.  (iii) Investigating the resilience of a wider range of distributed algorithms using available graph-theoretic tools. 
 
Here, we propose a few research avenues which are worth  investigating in the future.  

\textbf{Spectral Approach to Network Structures:}
One of the necessary steps towards reconciling system-theoretic approaches and graph theory is to find algebraic interpretations of certain network structures. Algebraic graph theory is an active topic of research in mathematics \cite{Godsil}. However, specific structural properties of networks that are widely used in analyzing the resilience of networked control systems, e.g., network robustness, are quite new notions defined within the field of systems and control. Hence, their algebraic interpretations have not been well studied. 

An example is a work on the relation of the algebraic connectivity and the network robustness. Defining the {\it edge-boundary} of a set of nodes $S \subset \mathcal{V}$ is given by $\partial{S} = \{(i,j) \in \mathcal{E} \mid i \in S, j \in \mathcal{V}\setminus{S}\}$.  The {\it isoperimetric constant} of $\mathcal{G}$ is defined as \cite{ChungSpectral}
 \begin{equation}
 i(\mathcal{G})\triangleq \min_{S \subset \mathcal{V}, |S| \le \frac{n}{2}}\frac{|\partial S|}{|S|}.
 \label{eqn:iso}
 \end{equation}
Based on the above definition and the definition of the network robustness we conclude that if $i(\mathcal{G})> r-1$, then the graph is at least $r$-robust. Moreover, we have $\lambda_2(L) \leq 2i(\mathcal{G})$ \cite{ChungSpectral}. Based on this, if $\lambda_2(L)>r-1$, then the network is at least $\lfloor \frac{r}{2}\rfloor$-robust \cite{Automaticapirani, kumar2}. However, $\lfloor \frac{r}{2}\rfloor$ provides a loose lower bound for the network robustness. An example is a star graph which is 1-robust with $\lambda_2(L)=1$.  Further research is needed to be done in this direction.

\textbf{Resilience with Minimum Communication:} Due to the fact that communications between nodes can be costly in many applications, the problem of reaching a certain level of resilience with minimum communication, i.e., edges between nodes, is worth investigating.  In specific dynamical systems, e.g., consensus dynamics, adding edges may also degrade the controllability of the system.  Few recent works have focused on minimizing the number of edges while reaching a
 certain level of security \cite{shan} or maximizing the connectivity without violating the controllability of the system  \cite{abbaswaseem}. While reaching a certain level of connectivity with the minimum number of edges has been well-studied in the literature, the same problem for network robustness remains open.

\textbf{Graph-Theoretic Approach to Attack Energy:} In Section \ref{sec:quant} some graph-theoretic interpretations of the attack's impact was discussed. Among those, the attack energy has not been well studied. The objective of the attacker, other than impact and detectability, can be to access the system with minimum energy. One way to quantify the attack energy is via using the spectra of the controllability Gramian $\mathcal{W}_{\mathcal{F}}$.  An interesting research avenue is to design the network to maximize the attack energy using an appropriate spectrum of the controllability Gramian.

\textbf{Resilience of other Classes of Random Graphs:}
The study of the connectivity and network robustness of random graphs with the various probability distribution of edge formation is another important future research line. Among them are scale-free networks in which their degree distributions follow a power-law. Many networks have been reported to be scale-free, including preferential attachment and the fitness model \cite{Clauset}. Analyzing the structural properties of these classes of networks enables us to quantify the resilience of distributed algorithms on various natural and man-made large-scale systems.

\textbf{Resilience of Networks with Nonlinear Interactions:}
Throughout this survey paper, the focus was on NCSs in which the interactions between agents are linear. In some cases of NCSs, the local interactions are nonlinear, e.g., synchronization of Kuramoto Oscillators with applications to power systems \cite{florian} and attraction-repulsion functions in swarm robotics \cite{veysel}. Extension of the methods discussed in this survey to those classes of nonlinear systems requires further investigations.

\textbf{Resilience of Multi-Agent Reinforcement Learning Algorithms:} Another emerging area of interest pertains to enabling a team of agents to cooperatively learn  optimal policies for interacting with their environment, particularly when the dynamics of the environment are not initially known to the agents.  Such {\it multi-agent reinforcement learning} settings share some common features with the distributed consensus, optimization, and estimation problems we described earlier in the paper, in that adversarial agents can provide incorrect information about their observations and costs to the other agents.  However, these problems also introduce additional lines of complexity for resilience, in that adversarial agents can also affect the underlying system with their inputs.  There have been recent initial explorations of resilient algorithms in such settings  \cite{xie21, lin2020toward, Wu21, figura2021adversarial}, but much work remains to be done to understand how to mitigate adversaries that can not only send incorrect information but can also take destructive actions.

\bibliographystyle{unsrt}        
\bibliography{autosam}         % and a bib file to produce the 
                                 % bibliography (preferred). The
                                 % correct style is generated by
                                 % Elsevier at the time of printing.

%\begin{thebibliography}{99}     % Otherwise use the  
                                 % thebibliography environment.
                                 % Insert the full references here.
                                 % See a recent issue of Automatica 
                                 % for the style.
%  \bibitem[Heritage, 1992]{Heritage:92}
%     (1992) {\it The American Heritage. 
%     Dictionary of the American Language.}
%     Houghton Mifflin Company.
%  \bibitem[Able, 1956]{Abl:56}
%     B.~C.~Able (1956). Nucleic acid content of macroscope. 
%     {\it Nature 2}, 7--9. 
%  \bibitem[Able {\em et al.}, 1954]{AbTaRu:54}   
%     B.~C. Able, R.~A. Tagg, and M.~Rush (1954).
%     Enzyme-catalyzed cellular transanimations.
%     In A.~F.~Round, editor, 
%     {\it Advances in Enzymology Vol. 2} (125--247). 
%     New York, Academic Press.
%  \bibitem[R.~Keohane, 1958]{Keo:58}
%     R.~Keohane (1958).
%     {\it Power and Interdependence: 
%     World Politics in Transition.}
%     Boston, Little, Brown \& Co.
%  \bibitem[Powers, 1985]{Pow:85}
%     T.~Powers (1985).
%     Is there a way out?
%     {\it Harpers, June 1985}, 35--47.

%\end{thebibliography}

%\appendix
%\section{A summary of Latin grammar}    % Each appendix must have a short title.
                                        % in the appendices.
\end{document}